\documentclass[12pt,letterpaper]{article}
\usepackage[margin=1in]{geometry}

\usepackage{amssymb,amsmath,amsfonts,eurosym,enumitem,graphicx,caption,color,comment,footmisc,array,longtable,rotating,float,xcolor}
\usepackage[normalem]{ulem}
\usepackage{makecell}
\usepackage{multirow}
\usepackage{booktabs, adjustbox}
\usepackage{threeparttable}
\usepackage{subcaption}
\usepackage{colortbl}

\usepackage{mathrsfs}
\usepackage{bm}
\usepackage{bbm}

\usepackage[numbers]{natbib}

\usepackage[T1]{fontenc}
\usepackage{libertinus}

\usepackage{xspace}

\newcommand{\Suppl}{Supplementary Material\xspace}

\usepackage{microtype}   

\usepackage{siunitx}
\definecolor{burgundy}{RGB}{153, 30, 61}
\usepackage[
  colorlinks=true,
  linkcolor=burgundy,
  citecolor=burgundy,
  urlcolor=burgundy
]{hyperref}

\newcolumntype{L}[1]{>{\raggedright\let\newline\\arraybackslash\hspace{0pt}}m{#1}}
\newcolumntype{C}[1]{>{\centering\let\newline\\arraybackslash\hspace{0pt}}m{#1}}
\newcolumntype{R}[1]{>{\raggedleft\let\newline\\arraybackslash\hspace{0pt}}m{#1}}

\usepackage[affil-it]{authblk}

\begin{document}

\begin{titlepage}
\title{Challenging Partisan Expectations Reduces Political Polarization}  

\author[1,*]{Do Won Kim}
\author[2]{Ozgur Can Seckin}
\author[1]{Saumya Bhadani}
\author[2]{Alessandro Flammini}
\author[1]{Giovanni Luca Ciampaglia}
\author[2,3]{Bao Tran Truong}

\affil[1]{University of Maryland, United States}
\affil[2]{Observatory on Social Media, Indiana University Bloomington, United States}
\affil[3]{Center Synergy of Systems, Dresden University of Technology, Germany}
\affil[*]{dowonkim@umd.edu}

\maketitle

\begin{abstract}
\noindent Political conversations are often proposed as a remedy for political polarization, yet their effectiveness remains inconsistent. We argue that this inconsistency partly reflects a neglected feature of political contact: the expectations partisans bring to these encounters. 
We hypothesize that conversations should reduce political polarization the most when they violate the expected link between partisan identity and issue position. We test this hypothesis in a 2$\times{}$2 experiment in which $1{,}983$ U.S. adults engaged in structured conversations with an AI chatbot whose presented partisan identity and policy stance were independently manipulated. We find that expectation-challenging conversations in which participants talk with a disagreeing ingroup member or an agreeing outgroup member are effective in reducing affective and issue polarization. Although these effects emerge without meaningful shifts in participants' own policy positions, a follow-up survey shows that most effects disappear over one month. Interestingly, these conversations maintain or improve objective measures of deliberation but are experienced as less satisfying by participants. Our findings identify expectation violation as an underexplored depolarization mechanism. Our results also demonstrate the promises and limitations of how conversational AI can serve as a scalable method for experimentally studying interventions to mitigating partisan divides.

\vspace{1em}


\end{abstract}
\setcounter{page}{0}
\thispagestyle{empty}
\end{titlepage}
\pagebreak \newpage

\pdfbookmark[1]{Introduction}{intro}
\section*{Introduction}
\label{sec:introduction}

Political polarization has risen over the past decades in many countries~\cite{boxell2024cross}, with well-documented repercussions for those democracies and their intergroup relations~\cite{graham2020democracy, piazza2023political, iyengar2015fear, EndersEtAl2019}. Especially in the United States, the growing partisan divide extends beyond actual policy disagreement into affective polarization, whereby citizens feel warmer toward co-partisans and colder toward out-partisans, and into  perceived issue polarization, whereby citizens hold distorted perceptions of where each side stands~\cite{mason2015disrespectfully, dellaposta2015liberals}. 

Theories of intergroup contact maintain that dialogue is a long-standing response to such divisions. Under favorable conditions, the theory posits that interactions with members of opposing groups can reduce prejudice about them~\cite{allport1954nature}. This perspective has shaped many recent depolarization efforts that encourage citizens to engage across political divides through various forms of contact, including dialogue, media exposure, or interpersonal contact~\cite{voelkel2024strengthening}. Meta reviews suggest that contact typically reduces prejudice~\cite{pettigrew2006meta,paluck2019contact}, however, causal evidence for scalable interventions remains limited~\cite{paluck2021prejudice} and the conditions distinguishing successful from ineffective interventions are still debated~\cite{blattner2023does,rossiter2024cross,paolini2010negative,graf2016investigating}. 

We propose expectation violation as a lens for clarifying when cross-partisan contact reduces polarization. 
Citizens enter political conversations with expectations about others. Party labels carry assumptions about policy disagreement, moral character, interpersonal hostility, and democratic commitments~\cite{rahn1993effect,Ahler2018,goggin2019goes,dias2022nature}. 
We conceptualize cross-partisan dialogue as a setting in which these expectations encounter interactional evidence. Contact is most likely to matter when the encounter gives people a reason to disrupt the usual link between partisan identity and political judgment. Importantly, expectation violations can take multiple forms. An out-partisan may violate expectations by agreeing on an issue, but also by appearing more reasonable than expected, by reasoning carefully, behaving warmly, acknowledging shared values, or expressing democratic commitments that contradict partisan stereotypes.
Social-cognitive accounts suggest that such violations matter because they can shift how people process political information~\cite{fiske1990continuum,jussim1987nature}. People often rely on category-based judgment, e.g., ``this is what people from that party are like,'' when another person behaves in line with a familiar partisan stereotype, but shift toward more individuated processing when the encounter does not fit. 
This may prompt citizens to reconsider the individual, the group, or the assumed link between party identity and issue positions. 

The same logic can also apply to expectations about the ingroup such as shared commonalities among co-partisans, or people who are otherwise socially similar. Expectation violations can thus occur not only in cross-partisan contact, but in political contact more generally. For example, an in-partisan who disagrees may challenge the expectation that shared partisan identity implies shared political judgment. Evidence that exposure to similar people with differing political views can reduce polarization is consistent with this broader logic~\cite{balietti2021reducing}. In both cases, the encounter weakens the perceived correspondence between social identity and political opinion.

By contrast, when partisans disagree in predictable ways, they are more likely to be interpreted through existing partisan schemas. Such encounters provide little information beyond what citizens already believe about the social structure of politics, and are therefore less likely to change attitudes. This logic helps explain why cross-partisan conversations most effectively bridge divides when they focus on areas of agreement rather than disagreement~\cite{santoro2022promise,deJong2024} --- disagreement from outgroup members is already expected. 

Conversational AI provides a useful tool to experimentally test this mechanism. 
A central challenge in studying the effects of contact is that the key features shaping contact are typically entangled in naturally occurring political conversations~\cite{pettigrew1998intergroup}. For example, many political issues are ``party-branded,'' such that learning about someone's position often makes their party affiliation predictable, and vice-versa~\cite{dias2022nature}. As a result, a conversation with an out-partisan is usually also a conversation with someone expected to disagree, while a conversation with a co-partisan is usually expected to involve agreement.

This entanglement makes it difficult to isolate whether the effects of contact on polarization arise from either group identity, policy disagreement, or topic characteristics. Real interpersonal exchanges make it difficult to manipulate these dimensions independently of each other. 
A common way to control for these factors is to match participants by disagreement level measured with pre-conversation survey responses. But partner matching is costly and often leaves a substantial subset of participants unmatched~\cite{blattner2023does}. Even when matching succeeds, researchers cannot fully control how the interaction unfolds, as individuals vary in warmth, receptiveness, the willingness to listen and perceived common ground.

AI chatbots can be assigned a partisan identity and a policy stance independently, decoupling these factors while holding other features of the interaction relatively constant, something that is difficult to achieve with human participants. 
The value of AI in this context is therefore neither as a substitute for political conversation with humans nor a delivery method for depolarization interventions, but rather as a methodological tool for isolating a mechanism that is difficult to study in human exchanges. 

We hypothesize that political contact is most consequential when it violates partisan expectations. To test this hypothesis, we conducted a $2\times{}2$ randomized survey experiment where U.S. participants engage in short, structured conversations with AI chatbots (see Fig.~\ref{fig:schema}(I)). Because many political preferences are strongly sorted by party~\cite{dias2022nature, Ahler2018, huddy2015expressive, mason2018uncivil, barber2019does, kinder2017neither, bartels2002beyond}, we operationalize a direct form of expectation violation: issue agreement from an out-partisan or issue disagreement from an in-partisan. 
Specifically, the partisan identity and opinion alignment of the chatbot are independently manipulated to be presented as either an ingroup- or outgroup-partisan, and as either agreeing or disagreeing with the participant's position on a self-selected political issue. This produces four different conditions: two that confirm expected partisanship-opinion alignment (Ingroup Agree, Outgroup Disagree), and two that challenge it (Ingroup Disagree, Outgroup Agree). 
We estimate the effects of these expectation-violating conversations on affective polarization and perceived issue polarization. Beyond these dimensions, we also examine attitude polarization, defined as the extent to which partisans hold extreme positions on contested policy issues~\cite{stagnaro2025factual, velez_confronting_2024}. 

We find that conversations with AI chatbots reduce affective and perceived issue polarization when they challenge partisan expectations (see Fig.~\ref{fig:schema}(II)). These effects are observed across political topics of different controversiality. 
The way these reductions emerge differs depending on which expectation is challenged. When participants talked with an ingroup AI that disagreed with them, polarization declined primarily because they felt less warm toward their own partisan group and perceived a larger gap between their own issue positions and those of ingroup members. In contrast, when participants talked with an outgroup AI that agreed with them, polarization declined primarily because they felt warmer toward the opposing partisan group and perceived a smaller gap between their own issue positions and those of outgroup members. Thus, expectation-challenging interactions can reduce polarization through two distinct pathways: by making partisans feel less attached to, and less politically similar to, their own partisan group, or by making them feel warmer toward, and more politically similar to, the opposing partisan group. 

Importantly, these effects occur without substantial shifts in participants' own issue attitudes, indicating that how people feel toward partisans and perceive the ideological divide can change even when their own views do not --- reducing polarization need not require persuasion. Overall, these findings cast AI under a different light, as a configurable political corrective, whose impact depends on the social signals it conveys. As conversational AI systems become routine participants in democratic discourse~\cite{schroeder2026how,lin2025persuading,velez2025chatbot}, understanding and governing the dynamics of these interactions is critical.

\pdfbookmark[1]{Experimental Design}{design}
\section*{Experimental Design}
\label{sec:study-design}

\begin{figure}
\centering
\includegraphics[width=0.6\linewidth]{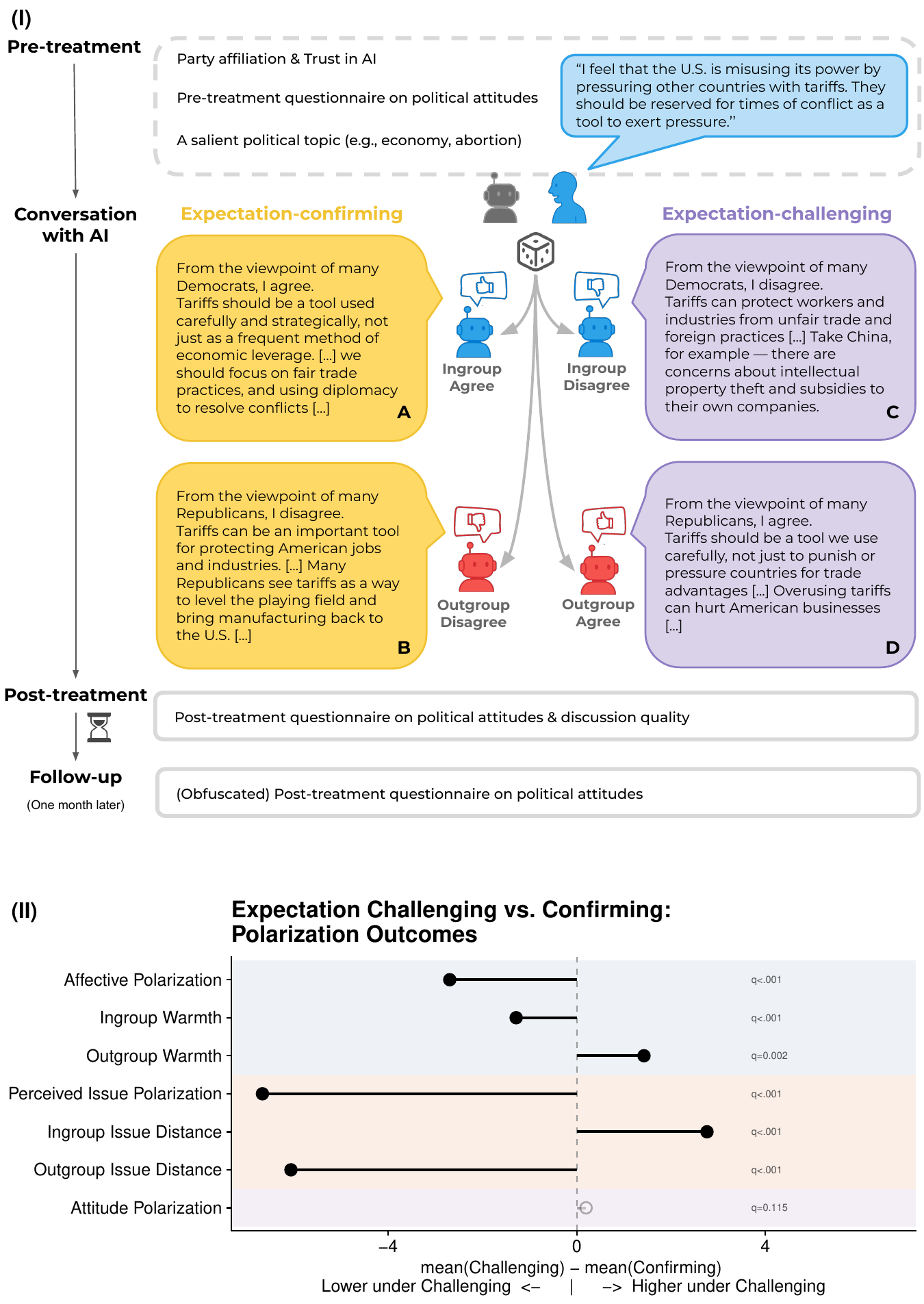}
\caption{
\textbf{(I) Study design.} We conduct two waves of survey. The first wave included a pre-treatment questionnaire, random assignment to the treatment --- an AI conversation that was either expectation-confirming (Ingroup Agree and Outgroup Disagree) or expectation-challenging (Ingroup Disagree and Outgroup Agree) --- and a post-treatment questionnaire measuring the main outcomes of the study.
After a minimum period of one month, participants were invited to a follow-up survey measuring again only the main outcomes of the study. The figure shows an example of the elicited issue of one of the participants assigned to the Ingroup Agree condition, the actual response from the chatbot (A), and as a counterfactual, what the chatbot would have responded in the other conditions ((B)--(D)). \\
\textbf{(II) Expectation-challenging versus expectation-confirming conditions.} The figure shows linear contrasts comparing expectation-challenging conditions (Ingroup Disagree and Outgroup Agree) with expectation-confirming conditions (Ingroup Agree and Outgroup Disagree) for polarization outcomes. Points indicate contrast estimates comparing expectation-challenging and expectation-confirming conditions (see \Suppl{}, section~\ref{sec:si-contrast}). Negative estimates indicate lower values under expectation-challenging conditions; positive estimates indicate higher values under expectation-challenging conditions. Filled markers denote contrasts with BKY sharpened two-stage FDR-adjusted $q \le .05$, whereas hollow markers denote $q > .05$. Reported $q$-values are shown on the right. This analysis is exploratory and was not preregistered.}
\label{fig:schema}
\end{figure}

We conducted a two-wave, preregistered experiment to examine how political conversations with an AI chatbot reshape how partisans feel about and perceive one another, whether these shifts extend to participants' own issue positions, and whether such conversations improve or impair the quality of democratic discourse itself (Fig.~\ref{fig:schema}). In particular, we test whether conversations that challenge partisan expectations, either through ingroup disagreement or outgroup agreement, reduce political polarization (affective polarization, perceived issue polarization, and attitude polarization) and enhance discussion-related outcomes (satisfaction, willingness to engage in future discussions, and discussion quality).

The first wave of the experiment was conducted in September 2025, in which U.S. adults identifying as Democrats or Republicans completed a set of pre-treatment survey items that measured party identification, political ideology, affective polarization, and trust in AI (see \Suppl~\ref{sec:si-questionnaire} for full item wording).
Participants were then asked to select an issue of personal salience from a list of political topics, and describe their position in an open-ended statement of at least $100$ characters. This response was then summarized by a Large Language Model (LLM), GPT-4.1 (see \Suppl~\ref{sec:si-llm-prompts} for prompt templates). To ensure the summaries accurately capture the stated positions, participants were then given the opportunity to review and edit them as needed. Participants were then asked to rate how much they agreed with their validated summaries and how much they thought Democrats and Republicans would agree with them (see~\nameref{sec:methods} for more details). 

Participants were then randomly assigned to one of four conditions in a $2\times{}2$ factorial design: Ingroup Agree, Ingroup Disagree, Outgroup Agree, and Outgroup Disagree. The partisan identity and stance of a GPT-4.1 chatbot were prompted to match the assigned treatment according to the participant’s political identity and position on the selected issue (see \Suppl~\ref{sec:si-llm-prompts} for the prompts). In making these assignments, we used block randomization based on party affiliation and pre-treatment AI trust. Pre-treatment characteristics were fully balanced (see \Suppl~\ref{sec:si-covar-balance}).

The chatbot started the conversation and responded to participant's non-summarized opinion, taking a partisan and issue stance according to the assigned condition.
For example, Fig.~\ref{fig:schema}(d) shows the response of our chatbot to a participant who identified as a Democrat and opposed U.S. policy on tariffs in the counterfactual scenario in which they had been assigned to the and assigned to the Outgroup Agree condition. In this case, the chatbot would have framed the response from the perspective of a Republican who endorses the same position as that of the participant (see \Suppl~\ref{sec:si-llm-prompts} for prompt templates and validation procedures). The conversation continued for up to four total exchanges in a structured chat interface (\Suppl~\ref{sec:si-interface}). 

After the conversation, participants completed a post-treatment questionnaire asking them to again rate agreement with their initial summarized issue statement, and political polarization outcomes, such as their perceptions of how much Democrats and Republicans would agree with it, and their level of in/outgroup warmth toward both parties. The questionnaire also included a number of discussion-related outcomes, such as satisfaction with the discussion and their willingness to engage in future conversations with the group represented by the AI chatbot. Finally, participants also answered questions about anti-democratic attitudes and their post-conversation trust in AI (see \Suppl~\ref{sec:si-questionnaire} for the full item wording). In total, $N=1{,}983$ participants completed the first wave of the study.

The second wave measured the same set of political polarization outcomes, while being framed as an unrelated study, to assess whether the effects persisted over time. Participants were invited to complete a second survey at least four weeks after the first-wave survey. The follow-up survey included items about affective polarization, perceived issue polarization, participants' own issue stance, and anti-democratic attitudes. In total, $N=1{,}627$ ($82\%$ retention rate) completed the follow-up survey. There was no evidence of differential attrition across experimental conditions ($p = 0.165$; see Table~\ref{tab:differential_attrition}) or selective attrition on any preregistered demographic variables across waves (all $p > 0.05$ for political affiliation, race, sex, and education level; see Table~\ref{tab:selective_attrition}).

When reporting our results, we use the Ingroup Agree condition as the baseline and estimate treatment effects for the other conditions relative to it. We treat this condition as the status quo of political discussion, as citizens most frequently engage with co-partisans who share their views, both online~\cite{cinelli2021echo} and in everyday life~\cite{carlson2020talking, mcpherson2001birds}.

\pdfbookmark[1]{Results}{results}
\section*{Results}

\pdfbookmark[2]{Political Polarization}{results-polarization}
\subsection*{Political Polarization}

\begin{figure}[!ht]
    \centering
    \includegraphics[width=1\linewidth]{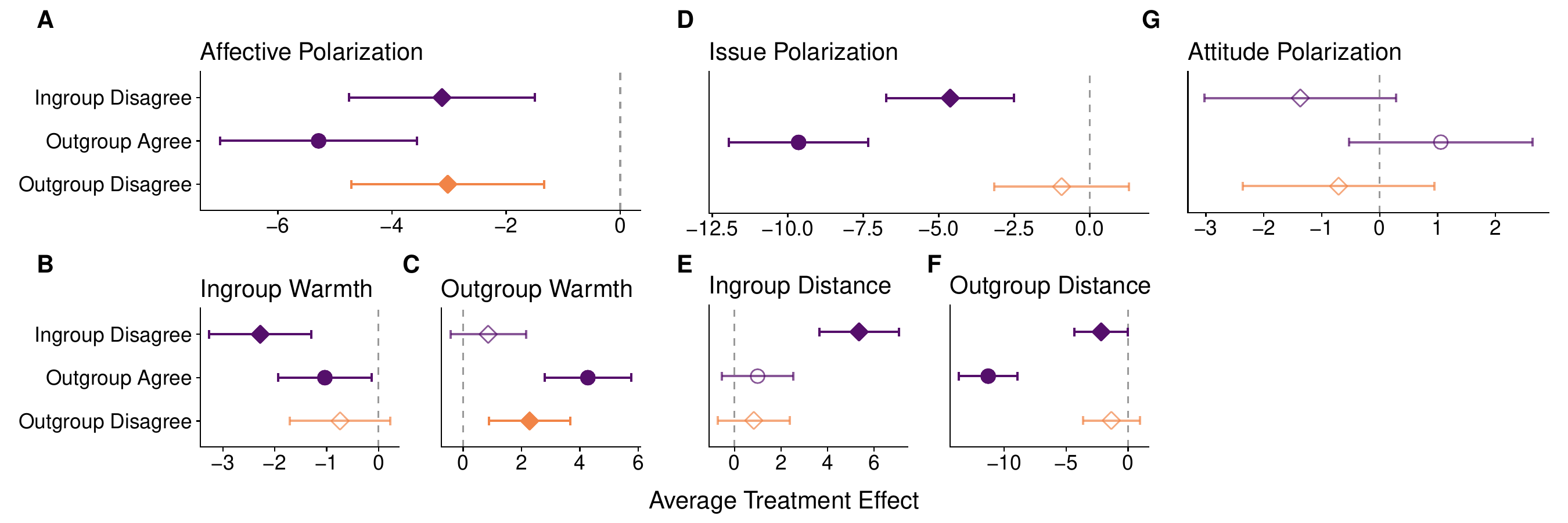}
    \caption{\textbf{Effects of AI conversations on political polarization.}
    Each plot shows average treatment effects for the Outgroup Agree (purple circle), Ingroup Disagree (purple diamond) and Outgroup Disagree (orange diamond), relative to the Ingroup Agree condition (gray dashed line), with 95\% confidence intervals. Non-significant results ($q > .05$) are indicated by lighter colors and unfilled markers.
    (A)~Expectation-challenging conversations (Outgroup Agree and Ingroup Disagree) and conversations in the Outgroup Disagree condition reduce affective polarization. 
    (B,C)~The reduction in affective polarization in the Outgroup Agree and Outgroup Disagree conditions is primarily driven by warmer feelings toward the outgroup. In the Ingroup Disagree condition, the decrease is driven by colder feelings toward the ingroup.
    (D)~Expectation-challenging conversations also reduce perceived issue polarization.
    (E,F)~The reduction in perceived issue polarization in the Outgroup Agree condition is driven by a decrease in perceived outgroup distance, while in the Ingroup Disagree condition it is largely driven by a increase in perceived ingroup distance.
    (G)~We find no evidence that challenging partisan expectations persuade participants either by moderating or strengthening their previous attitudes on the issue.}
\label{fig:main_1}
\end{figure}

Fig.~\ref{fig:main_1}(A) shows that expectation-challenging interactions significantly reduce affective polarization relative to the Ingroup Agree condition ($N=1{,}982$). 
In the Ingroup Disagree condition ($n=501$), affective polarization decreases (beta coefficient $(b) = -3.12$, $95\%~CI=[-4.751, -1.496]$, $p < .001$, false discovery rate-adjusted p-value $(q) < .001$), corresponding to a $6.57\%$ reduction relative to pre-treatment mean ($M = 47.5$). This reduction is driven primarily by a substantial decrease in ingroup warmth (Fig.~\ref{fig:main_1}(B); $b = -2.28~[-3.261, -1.297]$, $p < .001$, $q < .001$).  
In the Outgroup Agree condition ($n=496$), affective polarization decreases ($b = -5.29~[-7.01, -3.56]$, $p < .001$, $q < .001$), equivalent to an $11.14\%$ reduction relative to pre-treatment mean. Here, the effect is driven primarily by a significant increase in outgroup warmth (Fig.~\ref{fig:main_1}(C); $b = 4.27~[2.788, 5.757]$, $p < .001$, $q < .001$), accompanied by a modest but significant decrease in ingroup warmth (Fig.~\ref{fig:main_1}(B); $b = -1.03~[-1.938, -0.131]$, $p = .025$, $q = .026$).

Even the Outgroup Disagree condition ($n=497$) shows a significant reduction in affective polarization relative to the Ingroup Agree baseline ($n=488$) (Fig.~\ref{fig:main_1}(A); $b = -3.02~[-4.711, -1.332]$, $p < .001$, $q < .001$). This reduction was driven by a significant increase in outgroup warmth (Fig.~\ref{fig:main_1}(C); $b = 2.28~[0.884, 3.675]$, $p = .001$, $q = .002$), suggesting that interacting with an outgroup AI, even a disagreeing one, may improve feelings toward the outgroup more than interacting with an agreeing ingroup AI. However, this effect disappears in subsequent robustness checks where we restricted the sample to participants who correctly identified the AI's assigned stance, suggesting that this particular effect may partly reflect the chatbot's instruction to keep the conversation civil even when disagreeing from the outgroup perspective. See \Suppl~\ref{sec:si-robustness-check} for more information.

We therefore interpret the Outgroup Disagree finding cautiously. At the same time, this result does not by itself contradict our broader theoretical claim that expectation-challenging interactions reduce polarization more than expectation-confirming interactions. To assess this claim at the level of the theorized condition classes, we estimated an exploratory linear contrast from the same OLS specification used in the primary analyses, comparing expectation-challenging conditions (Ingroup Disagree and Outgroup Agree) with expectation-confirming conditions (Ingroup Agree and Outgroup Disagree). This contrast showed that expectation-challenging conditions produced significantly greater reductions in affective polarization ($C = -2.69$, $\text{SE} = 0.61$, 95\% CI $[-3.89, -1.49]$, $p < .001$, $q < .001$) and perceived issue polarization ($C = -6.67$, $\text{SE} = 0.82$, 95\% CI $[-8.27, -5.06]$, $p < .001$, $q < .001$), relative to expectation-confirming conditions. This pattern is consistent with our theoretical expectation, although the contrast was exploratory and not preregistered. See SI Section~\ref{sec:si-contrast} (Figure~\ref{fig:si-contrast} and Table~\ref{tab:si-contrast}) for full results.

A similar pattern emerges for perceived issue polarization ($N=1{,}982$). Talking with an AI chatbot representing disagreeing ingroup members (Ingroup Disagree; $n=501$) reduced perceived issue polarization (Fig.~\ref{fig:main_1}(D); $b = -4.62~[-6.738, -2.504]$, $p < .001$, $q < .001$), corresponding to a $9.33\%$ reduction from the pre-treatment mean ($M = 49.5$). This reduction is driven primarily by a significant increase in perceived ingroup issue distance ($n=502$; Fig.~\ref{fig:main_1}(E); $b = 5.36~[3.651, 7.062]$, $p < .001$, $q < .001$), while the decrease in perceived outgroup issue distance is smaller and only marginally significant ( $n=501$; Fig.~\ref{fig:main_1}(F); $b = -2.18~[-4.331, -0.032]$, $p = .047$, $q = .036$). 
 
In the Outgroup Agree condition ($n=496$), the reduction in perceived issue polarization is substantially larger ($b = -9.64~[-11.954, -7.332]$, $p < .001$, $q < .001$), equivalent to a $19.47\%$ decrease from the pre-treatment mean. Here, the effect is driven almost entirely by a pronounced decrease in perceived outgroup issue distance (Fig.~\ref{fig:main_1}(F); $b = -11.30~[-13.658, -8.938]$, $p < .001$, $q < .001$).

While affective and perceived issue polarizations decrease, we observe no corresponding treatment effects in attitude polarization (Fig.~\ref{fig:main_1}(G)). The reduction therefore results from shifting perceptions of others, not from participants changing their own minds on the issue.

None of the effects discussed here persisted over time; by Wave 2, the differences between conditions were no longer statistically significant (see \Suppl~\ref{sec:si-results}), with one exception: the Ingroup Disagree condition ($n=418$) continued to show increased perceived ingroup issue distance ($N=1{,}626$; $b = 3.89~[1.948, 5.832]$, $p < .001$, $q = .002$), relative to the Ingroup Agree baseline ($n=401$).

\pdfbookmark[2]{Discussion-related Outcomes}{results-discussion}
\subsection*{Discussion-related Outcomes}

\begin{figure}[!ht]
    \centering
    \includegraphics[width=\linewidth]{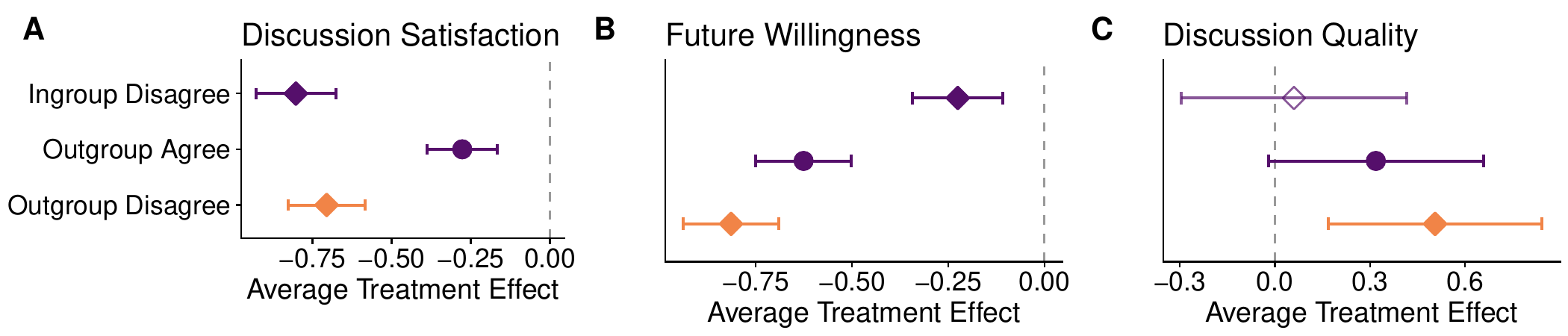}
    \caption{\textbf{Effects of AI conversations on discussion-related outcomes} 
    Each plot shows average treatment effects relative to the Ingroup Agree condition (gray dashed line), with 95\% confidence intervals. Non-significant results ($q > .05$) are indicated by lighter colors and unfilled markers.
    Expectation-challenging conversations (Outgroup Agree and Ingroup Disagree) and conversations in the Outgroup Disagree condition all reduce (A)~discussion satisfaction, and (B)~future willingness to discuss politics with the group represented by the AI. 
    (C)~We find no evidence that expectation-challenging conditions (Ingroup Disagree and Outgroup Agree) reduce overall discussion quality; in fact, we observe an increase in discussion quality in the Outgroup Agree condition after accounting for multiple comparisons.}
    \label{fig:H3}
\end{figure}

Challenging partisan expectations may produce discomfort, deterring participants from engaging in such conversations. For the interventions to be successful, it is important to assess not only their effects on political attitudes but also how participants experience these conversations. We examine how participants experience these conversations through a set of exploratory outcomes capturing the experience and quality of the conversations.  
We report participant answers to two subjective survey measures: satisfaction with the chatbot discussion (Fig.~\ref{fig:H3}(A); $N=1{,}982$) and future willingness to discuss politics with others (Fig.~\ref{fig:H3}(B); $N=1{,}982$) (See \Suppl~\ref{sec:si-questionnaire} for the wording of these questions). 
We also evaluate the objective quality of the conversations, observable through participant responses to the chatbot. Specifically, we extracted from the responses linguistic features such as responsivity and elaboration. These features are then aggregated to construct a Discussion Quality Index (Fig.~\ref{fig:H3}(C); $N=1{,}982$) (see Subsection~\ref{sec:si-discussion-quality} for details).

In general, participants found conversations most satisfying and were more willing to have similar discussions in the future when talking to an ingroup chatbot that agreed with them (Ingroup Agree; $n=488$). Participants in all other conditions rated their conversations less positively both in terms of discussion satisfaction in Figure~\ref{fig:H3}(A) (Ingroup Disagree ($n=501$): $b = -0.80$ $[-0.929, -0.677]$, $p < .001$, $q < .001$; Outgroup Agree ($n=496$): $b = -0.28$ $[-0.389, -0.165]$, $p < .001$, $q < .001$; Outgroup Disagree ($n=497$): $b = -0.71~[-0.826, -0.585]$, $p < .001$, $q < .001$) and willingness to engage in future discussions in Figure~\ref{fig:H3}(B) (Ingroup Disagree ($n=501$): $b = -0.23~[-0.343, -0.108]$, $p < .001$, $q < .001$; Outgroup Agree ($n=496$): $b = -0.63$ $[-0.750, -0.502]$, $p < .001$, $q < .001$; Outgroup Disagree ($n=497$): $b = -0.82~[-0.938, -0.691]$, $p < .001$, $q < .001$), though the drop in willingness for participants in the Ingroup Disagree was not as pronounced as in the other two conditions (Figure~\ref{fig:H3}(B)). See \Suppl~\ref{si:tab:SI_H3} for full regression tables.

These somewhat lower satisfaction and willingness do not affect the participants' adherence to deliberative norms. As shown in Figure~\ref{fig:H3}(C), compared to the baseline ($n=488$), discussion quality showed no significant difference in the Ingroup Disagree condition ($n=501$; $b = 0.06$ $[-0.296, 0.416]$, $p < .742$, $q = .09$) and was marginally higher in the Outgroup Agree condition ($n=496$). 
Our chatbot was designed to maintain a civil tone (see \nameref{sec:methods}), and participants generally responded in kind. Interestingly, the Outgroup Disagree exhibited the highest discussion quality, driven mostly by high responsivity and elaboration (see \Suppl \nameref{sec:si-discussion-outcomes} and Figure~\ref{fig:si-discussion_quality} for a detailed breakdown of the individual components of the Discussion Quality Index).

\pdfbookmark[2]{Manipulation Checks, Robustness Checks, and Heterogeneous Treatment Effects}{results-others}
\subsection*{Manipulation Checks, Robustness Checks, and Heterogeneous Treatment Effects}
The validity of our results depends on whether participants paid attention and received the treatment as intended. To assess whether participants meaningfully paid attention to the conversation, we examined participant response times to chatbot messages across conversational turns using interaction log data. Both reading and writing times remained relatively long and consistent throughout the interactions, suggesting that participants consistently took time to read and formulate their responses. The median is approximately 30 seconds for reading time and 30--45 seconds for writing time. This yields a speed per character consistent to a previous eye-tracking study of reading speed~\cite{franken2015eye} (\Suppl Section~\ref{sec:si-attention-check}). 

We also examined, by directly asking the participants, their perceptions of the issue stance and partisanship of the chatbot. Some participants did not perceive the chatbot's issue stance as defined by their assigned condition (\Suppl Section~\ref{sec:si-mani-check}). We thus assessed the robustness of our results to this ambiguity by re-running models restricting the analysis to participants who passed the manipulation check. Restricting to manipulation-check passers ($N = 1{,}456$) yielded robust results with somewhat larger effect sizes, suggesting that our main reported effects are conservative estimates (Section~\ref{sec:si-robustness-check}). We also conducted two additional robustness checks, namely, excluding pilot-phase participants, and re-estimating models using linear mixed-effects specifications for pre--post outcomes. Our main findings remained stable across these checks (Section~\ref{sec:si-robustness-check}).

In addition, we also estimated heterogeneous treatment effects by discussion topic and partisan identity as exploratory analyses. Existing work on intergroup contact shows mixed evidence on whether exposure to
outgroup disagreement can reduce political polarization. A key moderating factor appears to be the topic, where discussions yield depolarizing effects only when conversations do not touch contentious political issues that could lead to disagreement~\cite{santoro2022promise}. In our sample, the `Economy' was the most frequently chosen topic by participants, while many of the other options available to them corresponded to politically charged topics such as gun policy and abortion (see
\Suppl Figure~\ref{fig:topic_dist}). In comparison, economic issues are typically easier to talk about~\cite{pew2019topic}. Therefore, we tested whether our results were driven largely by topic selection (economy vs. non-economy). We find that the effects of expectation-challenging conversations depend on both the type of expectation being violated and the politicization of the issue (\Suppl Section~\ref{sec:si-hte}). While the effects of Ingroup Disagree only emerge for contentious non-economy topics, the effects of Outgroup Agree are strong and consistent across topics. 
We also find a pronounced partisan asymmetry: among Democrats ($n=1{,}037$), both Ingroup Disagree ($n=253$) and Outgroup Agree ($n=263$) (i.e., expectation-challenging conditions) reduce affective and perceived issue polarization. In contrast, only Outgroup Agree ($n=233$) produces statistically significant effects among Republicans ($n=945$) (\Suppl Section~\ref{sec:si-hte}). Despite these heterogeneities, the effects of Outgroup Agree is relatively large, robust and consistent across topics and partisan groups.

\pdfbookmark[1]{Discussion}{discussion}
\section*{Discussion}
\label{sec:discussion}

Our study shows that when partisans engage in political conversations with an AI, challenging partisan expectations can mitigate affective and perceived issue polarization without reinforcing or moderating issue attitudes. Instead of persuading someone to change their attitude on a given issue, these conversations mitigate polarization by weakening the perceived correspondence between partisan identity and political judgment. 

Given the well-documented difficulty of depolarization~\cite{holliday2025depolarization}, these findings are particularly noteworthy. Participants discussed salient policy issues of their own choosing where stakes may be higher and attitudes more resistant to change. Yet we still observed meaningful shifts. The largest effect we observe (for the Outgroup Agree condition) is substantial: roughly five-point reduction in affective polarization. For comparison, a two-point shift in this metric corresponds to reversing approximately three years of aggregate partisan animosity in the United States~\cite{piccardi2025reranking}. 

Our second contribution to the literature is identifying the mechanisms through which the depolarizing effect of political conversations may operate. This is an area that has received limited attention in the literature. Prior depolarization research has examined the impact of either group membership (ingroup vs. outgroup)~\cite{you2024ingroupnorms} or issue alignment (agreement vs. disagreement)~\cite{braghieri2024talking, fang2025person} in conversations, but in isolation of each other. We show that when participants interact with an ingroup chatbot that unexpectedly disagrees with them (Ingroup Disagree), polarization declines primarily due to colder feelings toward the ingroup and greater perceived issue distance from co-partisans. In contrast, when participants engage with an outgroup chatbot that unexpectedly agrees with them (Outgroup Agree), polarization declines due to warmer feelings toward the outgroup and reduced perceived distance from them. Overall, the same depolarizing outcome can emerge through distinct mechanisms, depending on how partisan expectations are violated. 

Several theoretical explanations are consistent with such observed effects. One possibility is the correction of misperceived group norms. Individuals frequently overestimate the ideological extremity and homogeneity of both their own group and the opposing group~\cite{levendusky2016mis, lees2020inaccurate, you2024ingroupnorms}. Conversational encounters that violate partisan expectations -- such as discovering that the outgroup is less extreme or that the ingroup holds a variety of views -- may therefore reduce perceived intergroup distance and mitigate hostility. Furthermore, cognitive dissonance and social balance theories suggest that expectation-violating interactions can prompt individuals to reassess their beliefs about partisan group boundaries~\cite{festinger1957theory, heider2013psychology}. 

Although expectation-challenging conversations with AI chatbots can reduce political polarization, they introduce a practical tradeoff for real-world deployment of this type of interventions. Participants in our study reported lower satisfaction and reduced willingness to engage in future conversations across all conditions relative to the Ingroup Agree baseline. This tension reflects the reality of online political discussions, which are typically embedded in ingroup settings and characterized by agreement. As a result, while expectation-challenging interactions can be effective at reducing political polarization, they may run counter to users’ conversational preferences, posing a barrier to their sustained use in practice.

Our findings also highlight the risk of AI sycophancy in political conversations. When models prioritize alignment with partisans' views, they can generate a self-reinforcing feedback loop in which partisans seek validation and AI systems reinforce existing expectations, thereby amplifying partisan biases~\cite{sharma2023towards, glickman2025human, rathje2025sycophantic, schroeder2026how}. Approaches to mitigate these risks focus for the most part on constraining model behavior through enhancing political neutrality~\cite{fisher2025political} or prompting models to adhere to deliberative norms~\cite{dryzek2019crisis, strandberg2019discussions}. Instead, our results point to a different, unexplored lever: strategically disrupting partisan expectations. In other words, rather than aligning with users, AI systems could reduce polarization by introducing expectation-violating interactions. Notably, in our study, interactions with an outgroup AI that agreed not only attenuated polarization; they also improved discussion quality, an effect likely driven by greater elaboration and responsiveness. This suggests that breaking partisan expectations can simultaneously weaken polarization and foster more substantive cross-partisan dialogue, consistent with prior work on deliberative norms~\cite{caluwaerts2023deliberation}.

As suggested by our findings on discussion-related outcomes, such expectation-challenging interactions may be difficult to sustain in human-to-human settings. Everyday political discourse is structured by homophily and typically unfolds within like-minded groups~\cite{huber2017political}. When outgroup interactions do occur, they can backfire without shared conversational norms~\cite{barlow2012contact}. Thus, expectation-challenging exchanges are unlikely to arise organically and instead require deliberate intervention. Encouraging partisans to engage with such interventions remains a practical challenge for future work.

We contribute to the practice of using AI for social science research by demonstrating how AI can be leveraged to actively structure expectation-challenging exchanges that would otherwise be rare or unstable in real-world settings. What makes AI particularly well-suited for this role is that it relaxes a key constraint of human interaction: while interpersonal disagreement often carries social risks that hinder engagement, AI can facilitate such exchanges in a scalable and controllable manner without those risks~\cite{argyle2023leveraging, tessler2024ai}. Crucially, this advantage does not come at the expense of effectiveness, as suggested by prior work demonstrating that AI--human conversations produce effects comparable to those observed in human-to-human conversations~\cite{boissin2025aiperceived}.

Our study has several important limitations.
First, the external validity of our findings depended on the willingness of our participants to engage with an AI chatbot, in particular one that challenges partisan expectations. Partisans, especially those with entrenched beliefs about other political actors, may be unlikely to voluntarily seek out such interactions outside controlled settings. As a result, the effectiveness of this approach in naturalistic environments remains an open question. Future work should examine how to design AI systems that not only induce depolarization but also sustain user engagement.

Second, when we followed up with participants, we did not observe persistent effects in our main outcomes. While a single interaction was sufficient to produce short-term changes, these effects did not last over time. Our follow-up period extended beyond that of prior interventions~\cite{voelkel2024strengthening,piccardi2025reranking}, which often examine shorter time horizons (i.e., two weeks), suggesting that the effects we study here are not merely transient. At the same time, this pattern indicates that one-off interactions may be insufficient to generate lasting change. We expect that repeated or sustained exposure to expectation-challenging interactions may be necessary to produce durable reductions in polarization, representing an important direction for future research.

One might worry that participants inferred the study’s purpose and responded in ways intended to satisfy the researchers. While acknowledging the possibility of such demand effects, we address this concern in several ways.

First, although participants may have inferred that the study concerned depolarization, our design made condition-specific demand effects less likely. Every participant, regardless of condition, engaged in a multi-turn conversation with an AI chatbot representing a partisan group's view on a salient policy issue; the surface-level task was identical across all four cells of the \{Ingroup/Outgroup\} $\times$ \{Agree/Disagree\} design. No participant completed a qualitatively different task, such as a survey with no conversation, that would clearly mark one condition as the focal treatment. This concern is more acute in designs in which a treatment group converses with a chatbot while a control group merely completes a survey, because the asymmetry in experience more readily signals the study's intent.

Second, even if some participants inferred the study's broad purpose, such awareness would not by itself threaten our between-condition estimates. For demand effects to differentially bias these estimates, participants would have had to infer not merely that the study concerned political polarization, but that the researchers expected the interaction to reduce polarization along a particular experimental contrast---ingroup versus outgroup, agreement versus disagreement, or their interaction---and in a particular direction. They would then have had to adjust their post-treatment responses selectively across conditions in line with that inferred hypothesis. Moreover, recent evidence indicates that even when the purpose of an experiment is explicitly revealed to participants, treatment effects are largely unchanged, and that financial incentives to respond in line with researcher expectations fail to consistently induce demand effects~\cite{mummolo2019demand}. Measuring polarization before treatment is a much weaker cue than either explicit disclosure or financial inducement, making strong condition-specific demand effects less plausible in our setting.

Of course, none of these points rule out demand effects entirely. 
Because our outcomes are measured immediately after the conversation, we cannot fully separate genuine attitude change from short-term responses to the conversational context, and the attenuation of most effects by Wave~2 is consistent with either short-lived genuine change or responses shaped by the immediate post-treatment setting. We therefore interpret the immediate post-treatment estimates as upper bounds on durable change, and we regard the design features noted above as reasons treatment effects are unlikely to be driven entirely by demand artifacts. Future work could address this more directly by moving beyond the survey-experimental setting altogether---for example, through field experiments that observe attitudes and behavior in naturalistic contexts, where cues to the study's purpose are less salient.

Another potential limitation is related to how participants perceived the specific condition the chatbot represented. Lower perceived representational accuracy in the expectation-challenging conditions suggests that participants viewed these chatbots as less consistent with group-based expectations. One might therefore worry that participants discounted expectation-challenging AI as unrepresentative of real partisans. If so, they may have engaged less seriously with the chatbot's arguments, thereby attenuating the treatment effects we estimate. We cannot entirely rule out this possibility. Critically, however, to the extent that this concern operates through reduced engagement with expectation-challenging AI, it implies that the effects reported here would be conservative with respect to those that would be observed under fuller uptake of the chatbot's position. Our robustness analysis restricted to participants who passed the manipulation check---that is, those who perceived the chatbot's stance as intended---yields treatment effects that are substantially larger than in the full sample (Supplementary Information, Section~\ref{sec:si-robustness-manip}), consistent with attenuation from imperfect uptake or engagement.  

Our design operationalized expectation violation through issue agreement and disagreement because these dimensions are strongly sorted by partisanship and can be cleanly manipulated. But expectation violations in political contact can take many forms. An out-partisan may violate expectations by being unusually warm, reasonable, democratically committed, or open to learning. An in-partisan may violate expectations by expressing ambivalence, acknowledging heterogeneity, or disagreeing respectfully. Future work should examine which kinds of expectation violations are most effective, when they generalize beyond the immediate interaction, and how they interact with conversational quality and perceived representativeness.

In summary, our findings show that structured political conversations can reduce polarization when they challenge the expected alignment between partisan identity and political judgment. Expectation violation is not the only pathway through which contact may matter, but it offers a useful framework for explaining why some encounters reduce polarization while others do not. By decoupling partisan identity from issue agreement, conversational AI makes it possible to study this mechanism directly. As AI systems become more common participants in democratic discourse, understanding how their social signals shape political perception will be increasingly important.

\pdfbookmark[1]{Methods}{methods}
\section*{Methods}
\label{sec:methods}

The study was approved by the Institutional Review Board at the University of Maryland (package 2326731-1) and Indiana University Bloomington (protocol 27968) in August 2025. 

\pdfbookmark[2]{Recruitment and Sample}{recruitment-sample}
\subsection*{Recruitment and Sample}

We recruited participants via CloudResearch Connect using the platform's prescreening filters. To be eligible, participants had to reside in the U.S., be of age 18 or older, and self-identify as Democrats or Republicans.
In total, 2,050 individuals entered Wave 1, including 100 participants recruited as part of a pilot sample. 
Partisanship was re-verified at the beginning of Wave 1, and 10 participants were excluded for not identifying as or leaning toward either the Democratic or Republican Party. An additional 2 participants were excluded for not completing the survey.
Finally, we excluded 65 participants due to technical failures with the LLM API: 50 whose initial opinion summaries failed to generate and 15 who experienced errors during the LLM conversation. These failures were unlikely to be systematic, showing no significant 
associations with participant characteristics or treatment assignment (see \Suppl \ref{sec:si-technical}).

The final sample for Wave 1 consisted of $N = 1{,}983$ eligible participants. Analyses of long-term effects were restricted to the subset of participants who completed both Wave 1 and Wave 2 ($N = 1{,}627$). Table~\ref{tab:si-descriptive-stat} presents descriptive statistics for both samples.

\pdfbookmark[2]{Experimental Design}{experimental-design}
\subsection*{Experimental Design}

Participants were randomly assigned to one of four conditions in a 2×2 factorial design that varies partisan identity (ingroup vs. outgroup) and opinion alignment (agreement vs. disagreement) of an AI chatbot. 
The randomization procedure was effective: participants assigned to each condition exhibited no systematic differences in pre-treatment covariates beyond what would be expected by chance. See Supplementary Information Section~\ref{sec:si-covar-balance} for further details.

\pdfbookmark[2]{Outcome Measures}{outcome-measures}
\subsection*{Outcome Measures}
\label{measures}

Our primary outcomes include three different measures of political polarization.
First, affective polarization reflects how much warmer one feels toward the ingroup relative to the outgroup, operationalized as ingroup warmth minus outgroup warmth. Ingroup- and outgroup warmth are each measured via feeling thermometer scores on a 0--100 scale, with higher values indicating greater warmth toward the respective group. 

Second, perceived issue polarization reflects how far apart on policy issues the ingroup and outgroup are perceived to be, operationalized as the absolute difference between the perceived issue stances of ingroup and outgroup members, measured on a 0--100 scale. Its two sub-components reflect perceived issue distance from the in- and outgroup members, respectively. Distances were measured as the absolute difference between one's own issue stance and the perceived issue stance of the in- or outgroup members, respectively, such that higher values reflect greater perceived distance to that group.  

Finally, attitude polarization reflects the degree to which one's issue position moved away from the scale midpoint as a result of the conversation. It was computed as the change in distance from the neutral midpoint of 50 before and after the conversation: as a pre--post difference when the pre-treatment issue position was below 50, and as a post--pre difference when it was above 50, such that positive values indicate attitude reinforcement (i.e., polarization) and negative values indicate attitude moderation or depolarization.

As our secondary outcomes, we instead have three different discussion-related measures. First, satisfaction was calculated as the average of two items, each measured on a 5-point Likert scale, where higher values indicate greater satisfaction with the AI chatbot discussion. 
Second, the future willingness to engage in political conversations was measured on a 5-point Likert scale where higher values indicate greater willingness to engage in political discussions with others.
Third and final, we defined a Discussion Quality Index to aggregate multiple conversation-level metrics signaling civility and substantive engagement. As the AI was already prompted to maintain civility across treatment groups, the index was calculated only for participant responses. Linguistic features were first computed at the message level and then averaged across all of a participant's turns to yield conversation-level scores. For instance, the ratio of gratitude terms was calculated per message, then averaged to produce a single conversation-level gratitude score. The conversation-level metrics were then standardized across the full sample to have a mean of zero and a standard deviation of one. The resulting composite index is the sum of the standardized scores of seven metrics: gratitude, hedge usage, question-asking, toxicity, responsivity, elaboration, and polarity (see \Suppl Section~\ref{sec:si-discussion-quality} for more details).
 
\pdfbookmark[2]{Statistical Analysis}{statistical-analysis}
\subsection*{Statistical Analysis}

Participants were randomly assigned to one of four experimental conditions following a 2 $\times$ 2 factorial design. These conditions are represented with binary indicator variables: (1) Outgroup = 1 if the participant was assigned to an outgroup condition, 0 otherwise; and (2) Disagreement = 1 if the participant was assigned to a disagreement condition, 0 otherwise.
 
Specifically, we estimate
\[
Y_i = \beta_0 + \beta_1\text{Ingroup--Disagree}_i + \beta_2\text{Outgroup--Agree}_i + \beta_3\text{Outgroup--Disagree}_i + \mathbf{X}_i\boldsymbol{\gamma} + \varepsilon_i,
\]
where the three treatment indicators are dummy-coded with the \textit{Ingroup--Agree} condition as the reference category, and $\beta_1$, $\beta_2$, $\beta_3$ represent the average treatment effect of each condition relative to baseline. $\mathbf{X}_i$ denotes a vector of prognostic covariates selected via the lasso procedure, plus the block indicator (partisanship $\times$ AI trust). 
The set of candidate variables that we select from includes age (18 and above), sex (male, female), race (white, nonwhite), ethnicity (non-Hispanic, Hispanic), college graduate (yes, no), ideology (1--7, where 1 indicates ``very liberal'' and 7 indicates ``very conservative''). 
The block indicator, which was used in the randomization procedure, is not subject to lasso selection and is included in all models. We use two-sided tests with $\alpha = .05$ and report HC2 robust standard errors. Within each conceptual family of outcomes, raw $p$-values are adjusted using the Benjamini–Krieger–Yekutieli (BKY) two-stage false discovery rate (FDR) procedure~\cite{benjamini2006adaptive}; both raw $p$-values and $q$-values are reported.

The analysis described deviates from the preregistered plan for ease of interpretation. Specifically, we re-parameterized the model to directly estimate pairwise contrasts relative to the baseline condition (see \Suppl Section~\ref{sec:si-replication} for details). 

\pdfbookmark[2]{Reporting Summary}{reporting-summary}
\subsection*{Reporting Summary}
Further information on research design is available in the Nature Portfolio Reporting Summary linked to this article.

\pdfbookmark[1]{Data Availability}{data-availability}
\section*{Data Availability}
The replication data are available on GitHub at~\url{https://github.com/DO-WON/AI_Conv}. The repository contains anonymized data, all variables derived from conversation logs, as well as all codes and protocols necessary for replication.  

\pdfbookmark[1]{Code Availability}{code-availability}
\section*{Code Availability}
Code is available on GitHub at~\url{https://github.com/DO-WON/AI_Conv}. 

\pdfbookmark[1]{Acknowledgments}{acknowledgments}
\section*{Acknowledgments}
This material is based on work supported in part by the Institute for Trustworthy AI in Law and Society (TRAILS), which is supported by the National Science Foundation under Award No. 2229885. Any opinions, findings, and conclusions or recommendations expressed in this material are those of the author(s) and do not necessarily reflect the views of the National Science Foundation or the National Institute of Standards and Technology. GLC acknowledges support from the NSF under CAREER grant no.~2239194 and from the University of Maryland Social Data Science Center. 
DWK, SB, BTT, and OCS received support from the Institute for Humane Studies under grant no. IHS018624, IHS018625, and IHS018278.
BTT received support from the Swiss National Science Foundation (Sinergia grant CRSII5\_209250) and the Knight Foundation.
OCS received support from Volkswagen Foundation under the Bots Building Bridges grant.
The authors wish to acknowledge the UC Berkeley Center for Human-Compatible AI and Jonathan Stray for partial support through the final prize of the Prosocial Ranking Challenge. 
This work used the IU JetStream 2 computational infrastructure through allocation CIS240118 from the Advanced Cyberinfrastructure Coordination Ecosystem: Services \& Support (ACCESS) program, which is supported by National Science Foundation grants \#2138259, \#2138286, \#2138307, \#2137603, and \#2138296. 
The authors wish to acknowledge Filippo Menczer and Yeeun Kim for helpful discussions, and Pasan Kamburugamuwa, Nick Liu, Rivado Edah and Janielle Jackson for their help with software development.

\pdfbookmark[1]{Author Contributions}{author-contributions}
\section*{Author Contributions}
DWK, BTT, OCS, SB, and GLC designed the study. 
DWK and BTT programmed the survey. 
SB, DWK, BTT, and OCS developed the chatbot. 
OCS processed the conversation data. 
BTT processed the metadata. 
DWK processed the data and performed the analysis. 
All authors wrote and approved the manuscript. 

\pdfbookmark[1]{Competing Interests}{competing-interests}
\section*{Competing Interests}
The authors declare no competing interests.

\pdfbookmark[1]{Materials \& Correspondence}{materials-correspondence}
\section*{Materials \& Correspondence}
Correspondence and requests for materials should be addressed to Do Won Kim.

\clearpage

\setcounter{page}{1}
\renewcommand{\thepage}{S\arabic{page}}

\setcounter{section}{0}
\setcounter{subsection}{0}
\setcounter{subsubsection}{0}

\setcounter{figure}{0}
\setcounter{table}{0}
\setcounter{equation}{0}

\renewcommand{\thesection}{S\arabic{section}}
\renewcommand{\thesubsection}{S\arabic{section}.\arabic{subsection}}
\renewcommand{\thesubsubsection}{S\arabic{section}.\arabic{subsection}.\arabic{subsubsection}}

\renewcommand{\thefigure}{S\arabic{figure}}
\renewcommand{\thetable}{S\arabic{table}}
\renewcommand{\theequation}{S\arabic{equation}}

\begin{center}
{\Large \textbf{Challenging Partisan Expectations Reduces Political Polarization --- Supplementary Information}}

\vspace{1em}

Do Won Kim$^{1,*}$,
Ozgur Can Seckin$^{2}$,
Saumya Bhadani$^{1}$,
Alessandro Flammini$^{2}$,
Giovanni Luca Ciampaglia$^{1}$,
and Bao Tran Truong$^{2,3}$

\vspace{1em}

{\small
$^{1}$University of Maryland, United States\\
$^{2}$Luddy School of Informatics, Computing, and Engineering, Indiana University, United States\\
$^{3}$Center Synergy of Systems, Dresden University of Technology, Germany\\
$^{*}$dowonkim@umd.edu
}
\end{center}

\tableofcontents

\clearpage

\section{Supplemental Results}
\label{sec:si-results}

\subsection{Descriptive Statistics} \label{sec:si-descriptive-stat}

Table~\ref{tab:si-descriptive-stat} shows descriptive statistics for the sample of Wave 1 completers and that of completers  of both Wave 1 and Wave 2.

\begin{table} 
\centering
\caption{Descriptive Statistics}
\label{tab:si-descriptive-stat}
\centering
\resizebox{\ifdim\width>\linewidth\linewidth\else\width\fi}{!}{
\fontsize{9}{11}\selectfont
\begin{threeparttable}
\begin{tabular}[t]{lcccc}
\toprule
\multicolumn{1}{c}{ } & \multicolumn{2}{c}{Wave 1 ($N$ = 1{,}983)} & \multicolumn{2}{c}{Wave 1 + 2 ($N$ = 1{,}627)} \\
\cmidrule(l{3pt}r{3pt}){2-3} \cmidrule(l{3pt}r{3pt}){4-5}
Variable & Mean (SD) / \% & $N$ & Mean (SD) / \% & $N$\\
\midrule
Democrat & 52.3\% & 1983 & 54.2\% & 1627\\
Non-White & 22.7\% & 1983 & 22.9\% & 1627\\
Age & 40.38 (12.94) & 1983 & 41.28 (13.13) & 1627\\
College Graduate & 59.0\% & 1983 & 58.6\% & 1627\\
Male & 48.4\% & 1983 & 47.4\% & 1627\\
Ideology (7-pt) & 3.95 (2.24) & 1982 & 3.85 (2.24) & 1626\\
Trust in AI & 69.6\% & 1983 & 69.2\% & 1627\\
Ingroup Warmth & 74.10 (19.37) & 1983 & 73.95 (19.41) & 1627\\
Outgroup Warmth & 26.61 (23.76) & 1983 & 25.59 (23.60) & 1627\\
Affective Polarization & 47.48 (29.72) & 1983 & 48.37 (29.68) & 1627\\
Perceived Issue Polarization & 49.54 (32.41) & 1983 & 50.37 (32.51) & 1627\\
Ingroup Issue Distance & 12.96 (16.83) & 1983 & 12.67 (16.85) & 1627\\
Outgroup Issue Distance & 54.90 (33.87) & 1983 & 55.95 (33.82) & 1627\\

\bottomrule
\end{tabular}
\begin{tablenotes}
\item \textit{Note:} Binary variables (Democrat, Non-White, College Graduate, Male, Trust in AI) are reported as percentages. Continuous variables are reported as Mean (SD). Wave 1 comprises the full eligible sample. Wave 1 + 2 comprises participants who completed both survey waves (used for RQ4 analyses).
\end{tablenotes}
\end{threeparttable}}
\end{table}

\begin{table} 
\centering
\caption{\label{tab:tab:si-anes-comparison}Sample Comparison: Study Sample vs.\ ANES 2024 Time Series}
\centering
\fontsize{9}{11}\selectfont
\begin{threeparttable}
\begin{tabular}[t]{lcc}
\toprule
Variable & Study Sample (Wave 1) & ANES 2024\\
\midrule
N & 1,983 & 5,103\\
Democrat & 52.3\% & 51.8\%\\
Male & 48.4\% & 46.6\%\\
Non-White & 22.7\% & 26.6\%\\
College Graduate & 59.0\% & 46.6\%\\
Age & 40.38 (12.94) & 53.58 (17.32)\\
Ideology (7-pt) & 3.95 (2.24) & 4.14 (1.72)\\
\bottomrule
\end{tabular}
\begin{tablenotes}
\item \textit{Note: } 
\item ANES 2024 statistics are computed among self-identified Democrats and Republicans, including partisan leaners ($N$ = 5,103). Binary variables are reported as percentages; continuous variables are reported as Mean (SD). The ANES age variable is top-coded at 80.
\end{tablenotes}
\end{threeparttable}
\end{table}

\subsection{Additional Political Outcomes: Anti-Democratic Attitudes}
\label{sec:si-antidem}

Support for partisan violence is measured on a $0$ (Never) to $100$ (Always) scale and reflects support for violence by ingroup members against the outgroup, calculated as the average of two items (for these and other items, see Section~\ref{sec:si-replication}). Support for anti-democratic practices is measured on a $0$ (Strongly disagree) to $100$ (Strongly agree) scale and reflects support for the ingroup's undemocratic practices against the outgroup, also calculated as the average of two items. These outcomes were measured in both Wave 1 and Wave 2. For these additional outcomes, long-term treatment effects measured in Wave 2 are discussed below in Section~\ref{sec:si-long-term}.

\begin{figure}
    \centering
    \includegraphics[width=\linewidth]{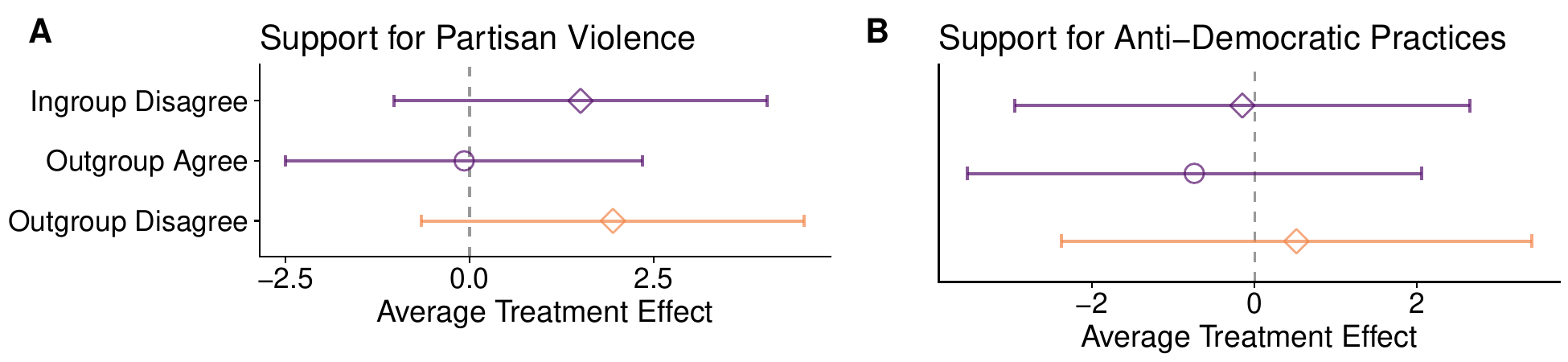}
    \caption{\textbf{Effects of the AI conversations on anti-democratic attitudes.} Each plot shows average treatment effects for the Outgroup Agree (purple circle), Ingroup Disagree (purple diamond) and Outgroup Disgree (orange diamond), relative to the Ingroup Agree condition (gray dashed line), with $95\% $ confidence intervals. Non-significant results ($q > .05$) are indicated by lighter colors and unfilled markers. Panel~(A) displays treatment effects on support for partisan violence and panel~(B) on support for anti-democratic practices, each measured on a $0-100$ scale.}
    \label{fig:si-anti-dem-outcomes}
\end{figure}

Previous research demonstrates that interventions successfully reducing political polarization often fail to alter support for undemocratic practices~\cite{voelkel2024strengthening,Voelkel2023}. Consistent with this, we find null effects, see Figure~\ref{fig:si-anti-dem-outcomes}. Neither ingroup disagreement nor outgroup agreement meaningfully altered participants' willingness to justify political violence or undemocratic actions on behalf of their party. These null results should be interpreted in light of a potential floor effect: most responses clustered near the lower end of the scale (rarely exceeding $20$ out of $100$), leaving little variance for the treatment to act upon. In other words, we observe a broad rejection of anti-democratic practices in our sample, consistent with evidence that public support for political violence is often low across the general population~\cite{westwood2022current}. 

\subsection{Long-term Treatment Effects on Main and Additional Outcomes}
\label{sec:si-long-term}

\begin{figure}
    \centering
    \includegraphics[width=1\linewidth]{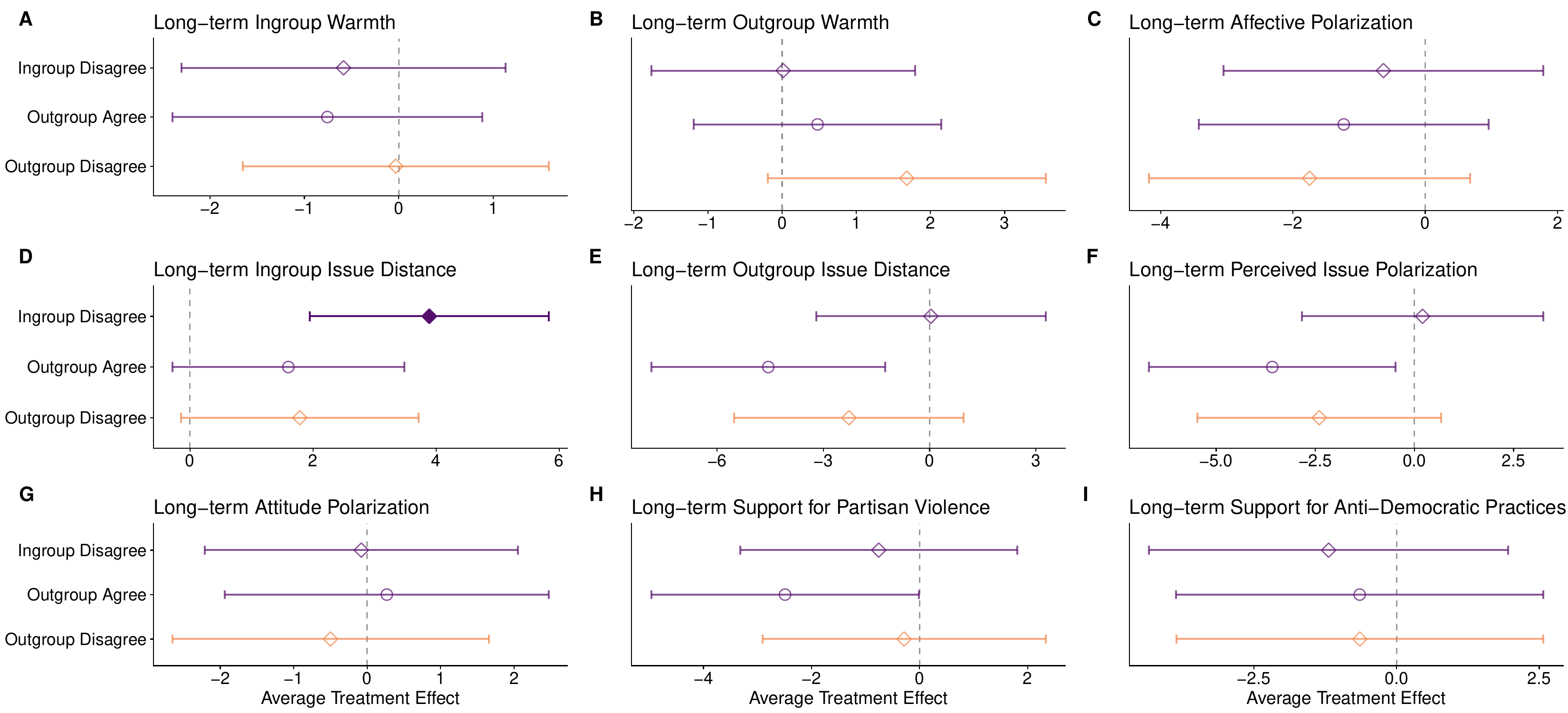}
    \caption{\textbf{Long-term effects of the AI conversations on political attitudes.} Each plot shows average treatment effects for the Outgroup Agree (purple circle), Ingroup Disagree (purple diamond) and Outgroup Disagree (orange diamond), relative to the Ingroup Agree condition (gray dashed line), with 95\% confidence intervals. Non-significant results ($q > .05$) are indicated by lighter colors and unfilled markers. The top panels display long-term treatment effects on feelings toward (A) ingroup, (B) outgroup members, and (C) their difference (affective polarization). The middle panels show long-term treatment effects on the perceived distance between participants' own issue stance and that of their (D) ingroup, (E) outgroup, as well as (F) the perceived polarization between ingroup and outgroup positions. The bottom panels illustrate long-term treatment effects on (G) support for partisan violence and (H) support for anti-democratic practices.}
    \label{fig:si-long-term-outcomes}
\end{figure}

As mentioned in the main text, most effects did not persist over time. As shown in Figure~\ref{fig:si-long-term-outcomes}, we find null effects for both affective and issue polarization (Fig.~\ref{fig:si-long-term-outcomes}~(C) and (F)). Of the two expectation-challenging conditions, only in the Outgroup Agree condition we see durable effects for long-term issue polarization. That said, when we decompose the issue polarization measure in terms of its two individual terms, we observed that expectation-challenging conversations had durable effects for issue proximity (Fig.~\ref{fig:si-long-term-outcomes}(D)) and for issue distance (Fig.~\ref{fig:si-long-term-outcomes}(E), albeit in the Outgroup Agree condition only. The results for panel~(D) are consistent with the theorized mechanism for the Ingroup Disagree condition, namely that encounters that challenge partisan expectation of that type reduce the perception that co-partisans form an homogeneous group. 

\begin{table}
\centering
\caption{Long-term Treatment Effects on Political Polarization (with FDR-adjusted q-values)}
\label{tab:si-long-term-polarization-fdr}
\centering
\resizebox{\ifdim\width>\linewidth\linewidth\else\width\fi}{!}{
\fontsize{9}{11}\selectfont
\begin{tabular}[t]{llrrrr}
\toprule
DV & Treatment & Estimate & SE & p-value & q-value\\
\midrule
 & Ingroup-Disagree & -0.585 & 0.876 & 0.504 & 1.000\\

 & Outgroup-Agree & -0.756 & 0.837 & 0.367 & 0.667\\

\multirow[t]{-3}{*}{\raggedright\arraybackslash Long-term Ingroup Love} & Outgroup-Disagree & -0.033 & 0.826 & 0.969 & 1.000\\
\cmidrule{1-6}
 & Ingroup-Disagree & 0.014 & 0.907 & 0.987 & 1.000\\

 & Outgroup-Agree & 0.478 & 0.852 & 0.575 & 1.000\\

\multirow[t]{-3}{*}{\raggedright\arraybackslash Long-term Outgroup Love} & Outgroup-Disagree & 1.681 & 0.955 & 0.078 & 0.394\\
\cmidrule{1-6}
 & Ingroup-Disagree & -0.632 & 1.234 & 0.609 & 1.000\\

 & Outgroup-Agree & -1.231 & 1.119 & 0.271 & 0.613\\

\multirow[t]{-3}{*}{\raggedright\arraybackslash Long-term Affective Polarization} & Outgroup-Disagree & -1.749 & 1.237 & 0.158 & 0.492\\
\cmidrule{1-6}
 & Ingroup-Disagree & 3.890 & 0.990 & 0.000 & 0.002\\

 & Outgroup-Agree & 1.599 & 0.962 & 0.097 & 0.409\\

\multirow[t]{-3}{*}{\raggedright\arraybackslash Long-term Ingroup Distance} & Outgroup-Disagree & 1.784 & 0.984 & 0.070 & 0.394\\
\cmidrule{1-6}
 & Ingroup-Disagree & 0.035 & 1.652 & 0.983 & 1.000\\

 & Outgroup-Agree & -4.556 & 1.684 & 0.007 & 0.074\\

\multirow[t]{-3}{*}{\raggedright\arraybackslash Long-term Outgroup Distance} & Outgroup-Disagree & -2.278 & 1.653 & 0.168 & 0.492\\
\cmidrule{1-6}
 & Ingroup-Disagree & 0.205 & 1.553 & 0.895 & 1.000\\

 & Outgroup-Agree & -3.584 & 1.584 & 0.024 & 0.178\\

\multirow[t]{-3}{*}{\raggedright\arraybackslash Long-term Issue Polarization} & Outgroup-Disagree & -2.399 & 1.565 & 0.125 & 0.476\\
\cmidrule{1-6}
 & Ingroup-Disagree & 0.080 & 1.087 & 0.941 & 1.000\\

 & Outgroup-Agree & -0.268 & 1.124 & 0.812 & 1.000\\

\multirow[t]{-3}{*}{\raggedright\arraybackslash Long-term Persuasion} & Outgroup-Disagree & 0.499 & 1.099 & 0.650 & 1.000\\
\bottomrule
\end{tabular}}
\end{table}

\subsection{Discussion-related Outcomes}
\label{sec:si-discussion-outcomes}

\begin{figure}
    \centering
    \includegraphics[width=0.75\linewidth]{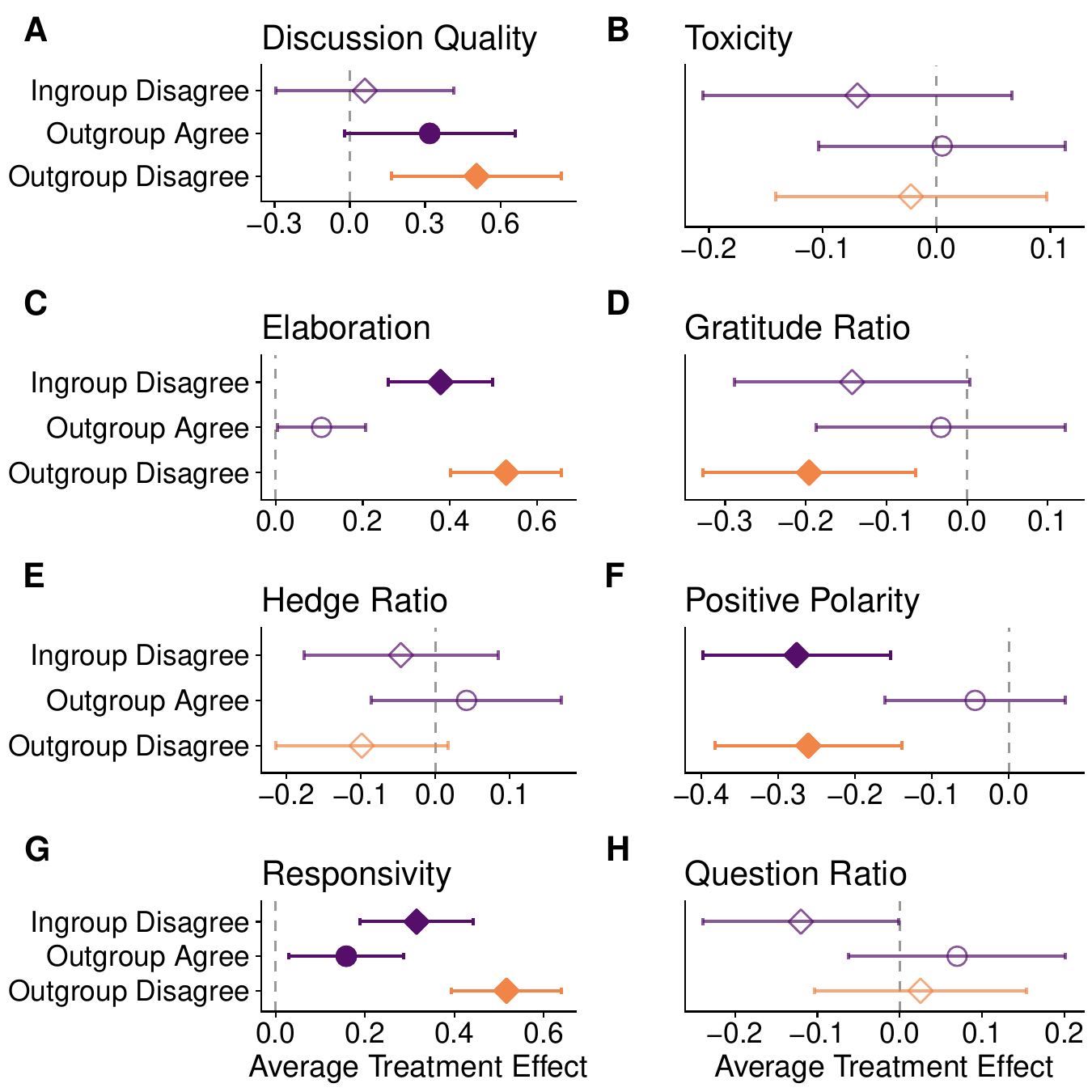}
    \caption{\textbf{Effects of the AI conversations on discussion quality index and its linguistic features.} Each plot shows average treatment effects for the Outgroup Agree (purple circle), Ingroup Disagree (purple diamond) and Outgroup Agree (orange diamond), relative to the Ingroup Agree condition (gray dashed line), with 95\% confidence intervals.  Non-significant results ($q > .05$) are indicated by lighter colors and unfilled markers. (A) main discussion quality index shown in main text; (B)--(H) the individual linguistic features.}
    \label{fig:si-discussion_quality}
\end{figure}

Figure~\ref{fig:si-discussion_quality} shows the effects on both the composite index of discussion quality shown in the main text (Fig.~\ref{fig:si-discussion_quality}(A)) and the seven linguistic features (Fig.~\ref{fig:si-discussion_quality}(B)--(H)) that went into the making of this index. Table~\ref{tab:si-discussion-quality-individual-fdr} shows the FDR-adjusted q-values for each individual linguistic features. For details about the measurement of these features, see Subsection~\ref{sec:si-discussion-quality}.

Regarding the toxicity feature (Fig.~\ref{fig:si-discussion_quality}(B)), we used the Perspective API to identify toxic messages. Raw toxicity rates returned by the API remained consistently low (below 5\%) across all conditions, with no statistically significant differences found (Ingroup Agree: 2.6\%, Ingroup Disagree: 3.9\%, Outgroup Agree: 2.8\%, Outgroup Disagree: 4.6\%). A qualitative review indicated that most flagged messages were related to sensitive topics, such as immigration or abortion.

\begin{table}
\centering
\caption{Treatment Effects on Individual Deliberative Metrics (FDR-adjusted q-values)}
\label{tab:si-discussion-quality-individual-fdr}
\centering
\resizebox{\ifdim\width>\linewidth\linewidth\else\width\fi}{!}{
\fontsize{9}{11}\selectfont
\begin{tabular}[t]{llrrrr}
\toprule
DV & Treatment & Estimate & SE & p-value & q-value\\
\midrule
 & Ingroup-Disagree & 0.378 & 0.061 & $<.001$ & $<.001$\\

 & Outgroup-Agree & 0.105 & 0.051 & 0.040 & 0.062\\

\multirow[t]{-3}{*}{\raggedright\arraybackslash Elaboration} & Outgroup-Disagree & 0.529 & 0.065 & $<.001$ & $<.001$\\
\cmidrule{1-6}
 & Ingroup-Disagree & -0.143 & 0.075 & 0.056 & 0.071\\

 & Outgroup-Agree & -0.033 & 0.079 & 0.680 & 0.471\\

\multirow[t]{-3}{*}{\raggedright\arraybackslash Gratitude Ratio} & Outgroup-Disagree & -0.196 & 0.067 & 0.004 & 0.008\\
\cmidrule{1-6}
 & Ingroup-Disagree & -0.046 & 0.067 & 0.486 & 0.377\\

 & Outgroup-Agree & 0.042 & 0.065 & 0.520 & 0.380\\

\multirow[t]{-3}{*}{\raggedright\arraybackslash Hedge Ratio} & Outgroup-Disagree & -0.099 & 0.059 & 0.094 & 0.104\\
\cmidrule{1-6}
 & Ingroup-Disagree & -0.276 & 0.062 & $<.001$ & $<.001$\\

 & Outgroup-Agree & -0.044 & 0.060 & 0.465 & 0.377\\

\multirow[t]{-3}{*}{\raggedright\arraybackslash Positive Polarity} & Outgroup-Disagree & -0.261 & 0.062 & $<.001$ & $<.001$\\
\cmidrule{1-6}
 & Ingroup-Disagree & -0.120 & 0.061 & 0.048 & 0.067\\

 & Outgroup-Agree & 0.070 & 0.067 & 0.299 & 0.255\\

\multirow[t]{-3}{*}{\raggedright\arraybackslash Question Ratio} & Outgroup-Disagree & 0.025 & 0.065 & 0.700 & 0.471\\
\cmidrule{1-6}
 & Ingroup-Disagree & 0.316 & 0.065 & $<.001$ & $<.001$\\

 & Outgroup-Agree & 0.158 & 0.066 & 0.016 & 0.029\\

\multirow[t]{-3}{*}{\raggedright\arraybackslash Responsivity} & Outgroup-Disagree & 0.517 & 0.063 & $<.001$ & $<.001$\\
\cmidrule{1-6}
 & Ingroup-Disagree & -0.070 & 0.069 & 0.316 & 0.255\\

 & Outgroup-Agree & 0.005 & 0.055 & 0.928 & 0.661\\

\multirow[t]{-3}{*}{\raggedright\arraybackslash Toxicity} & Outgroup-Disagree & -0.023 & 0.061 & 0.711 & 0.471\\
\bottomrule
\end{tabular}}
\end{table}

\begin{figure}
    \centering
    \includegraphics[width=\linewidth]{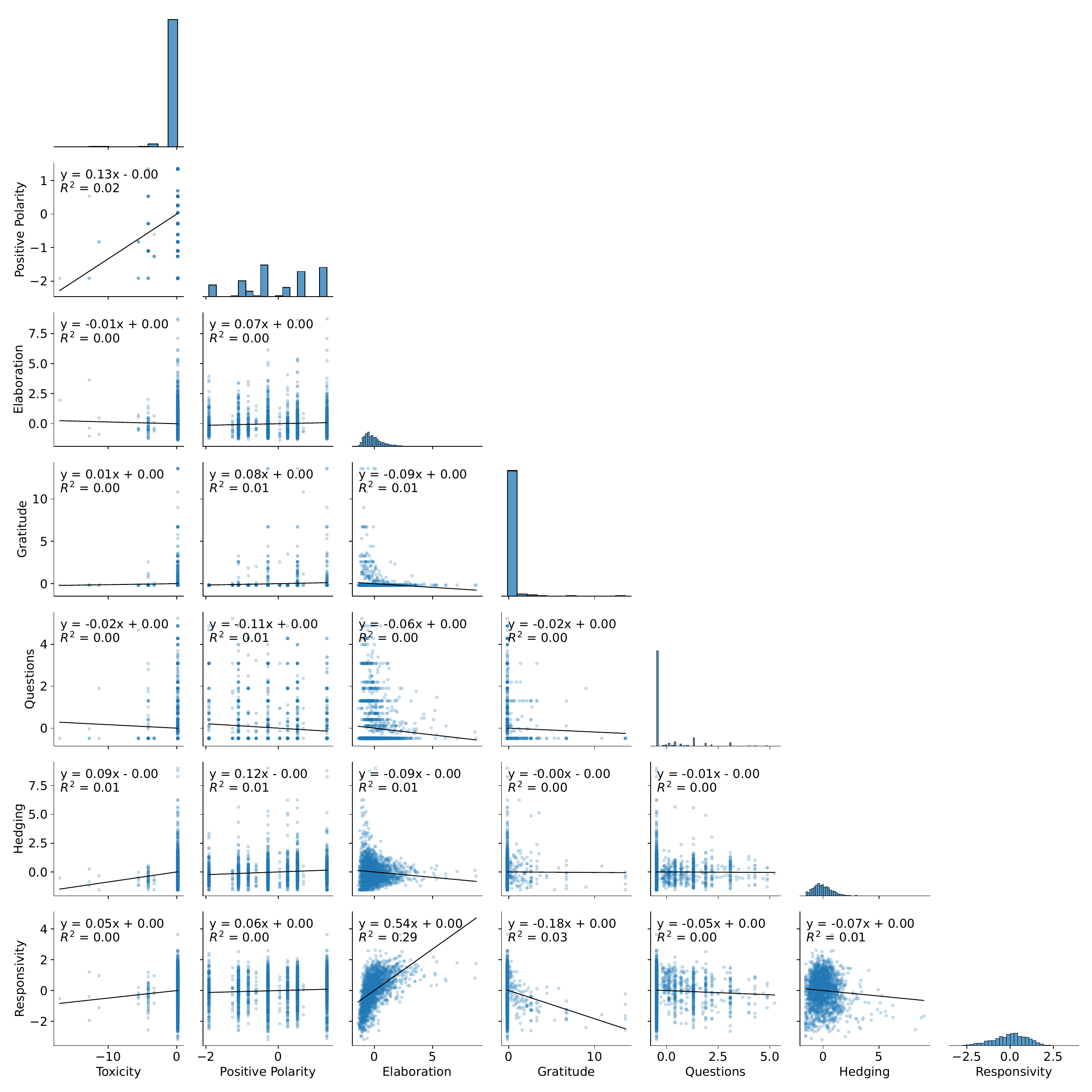}
    \caption{\textbf{Scatter plot of the individual discussion quality features.} The matrix displays scatter plots, regression lines, and $R^2$ values for the seven linguistic features used in the index. The diagonal shows histograms for each component.}
    \label{fig:si-discussion_quality_corr}
\end{figure}

To get a better sense of the multi-dimensional nature of the index, Fig.~\ref{fig:si-discussion_quality_corr} shows the pairwise correlations between individual features.
The data are generally weakly correlated ($R^2\approx0$), indicating that the various features capture distinct dimensions of discussion quality. One notable exception is the moderate positive correlation between Elaboration and Responsivity ($R^2=0.29$), suggesting that longer messages tend to be more responsive.

\subsection{Exploratory Outcome: Post-Treatment AI Trust}
\label{sec:si-ai-trust}
We also report an exploratory outcome on the potential downstream implications of interactions with the chatbot---AI trust. This outcome was not preregistered. If interactions in expectation-challenging conditions systematically reduce trust in AI, users may become less willing to engage with such systems, potentially raising concerns about the feasibility of deploying such conversational agents in politically polarized environments.

\begin{figure}
    \centering
    \includegraphics[width=0.6\linewidth]{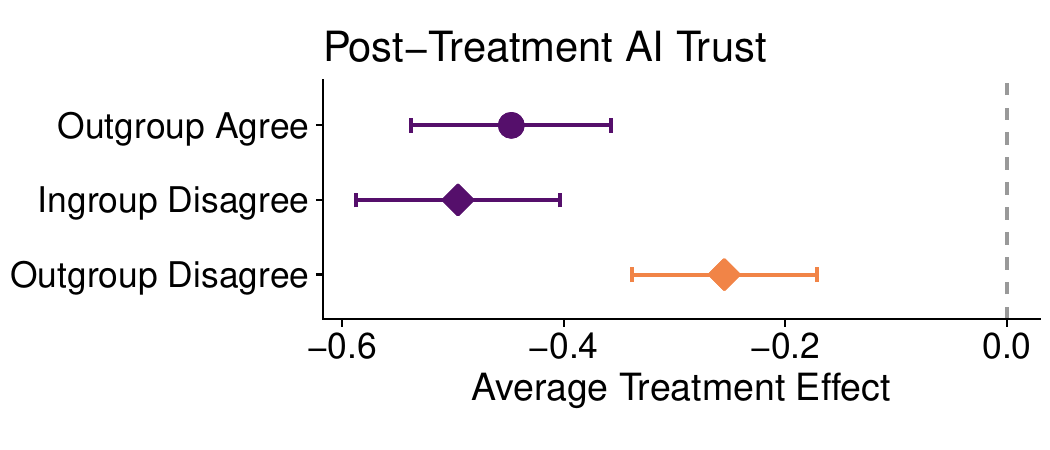}
    \caption{\textbf{Exploratory Analysis: Treatment effects on Post-Treatment AI Trust.} Each plot shows average treatment effects for the Outgroup Agree (purple circle), Ingroup Disagree (purple diamond) and Outgroup Disagree (orange diamond), relative to the Ingroup Agree condition (gray dashed line), with 95\% confidence intervals. Non-significant results ($q > .05$) are indicated by lighter colors and unfilled markers.}
    \label{fig:SI-exploratory-AI-trust}
\end{figure}

\begin{table}
\centering
\caption{Treatment Effects on Post-Treatment AI Trust}
\label{tab:SI-exploratory-AI-trust}
\centering
\resizebox{\ifdim\width>\linewidth\linewidth\else\width\fi}{!}{
\fontsize{9}{11}\selectfont
\begin{tabular}[t]{llrrrr}
\toprule
DV & Treatment & Estimate & SE & p-value & q-value\\
\midrule
 & Ingroup-Disagree & -0.495 & 0.047 & $<.001$ & $<.001$\\

 & Outgroup-Agree & -0.447 & 0.046 & $<.001$ & $<.001$\\

\multirow[t]{-3}{*}{\raggedright\arraybackslash AI Trust} & Outgroup-Disagree & -0.255 & 0.043 & $<.001$ & $<.001$ \\
\bottomrule
\end{tabular}}
\end{table}

As shown in Figure~\ref{fig:SI-exploratory-AI-trust} and Table~\ref{tab:SI-exploratory-AI-trust}, post-treatment trust in AI is significantly lower in the expectation-challenging conditions than in the Ingroup Agree condition. This finding points to a practical challenge for deploying AI systems in political contexts: although expectation-challenging interactions may reduce polarization, misalignment between a chatbot's group identity and expressed stance can undermine user trust. How such challenging encounters can be effectively promoted, encouraged, or sustained in real-world settings remains an important question for future research.

\subsection{Expectation-Challenging vs. Expectation-Confirming Contrast}
\label{sec:si-contrast}

As a complementary test, we grouped the four conditions according to whether the AI interlocutor's stance \emph{confirmed} or \emph{challenged} partisan expectations. Expectation-confirming conditions are those in which the AI behaved as partisanship would predict: an ingroup member agreeing or an outgroup member disagreeing (Ingroup Agree, Outgroup Disagree). In contrast, expectation-challenging conditions are those that violate this prediction: an ingroup member disagreeing or an outgroup member agreeing (Ingroup Disagree, Outgroup Agree). Because our central comparison concerns whether expectation-\emph{challenging} conditions reduce political polarization from expectation-\emph{confirming} conditions, we report this difference directly. For each outcome we estimated the linear contrast \text{mean(Challenging)} - \text{mean(Confirming)} from the same OLS specification used in the main analyses, and adjusted $q$-values across the family of polarization outcomes using the BKY sharpened two-stage FDR procedure.

\begin{figure}
    \centering
    \includegraphics[width=1\linewidth]{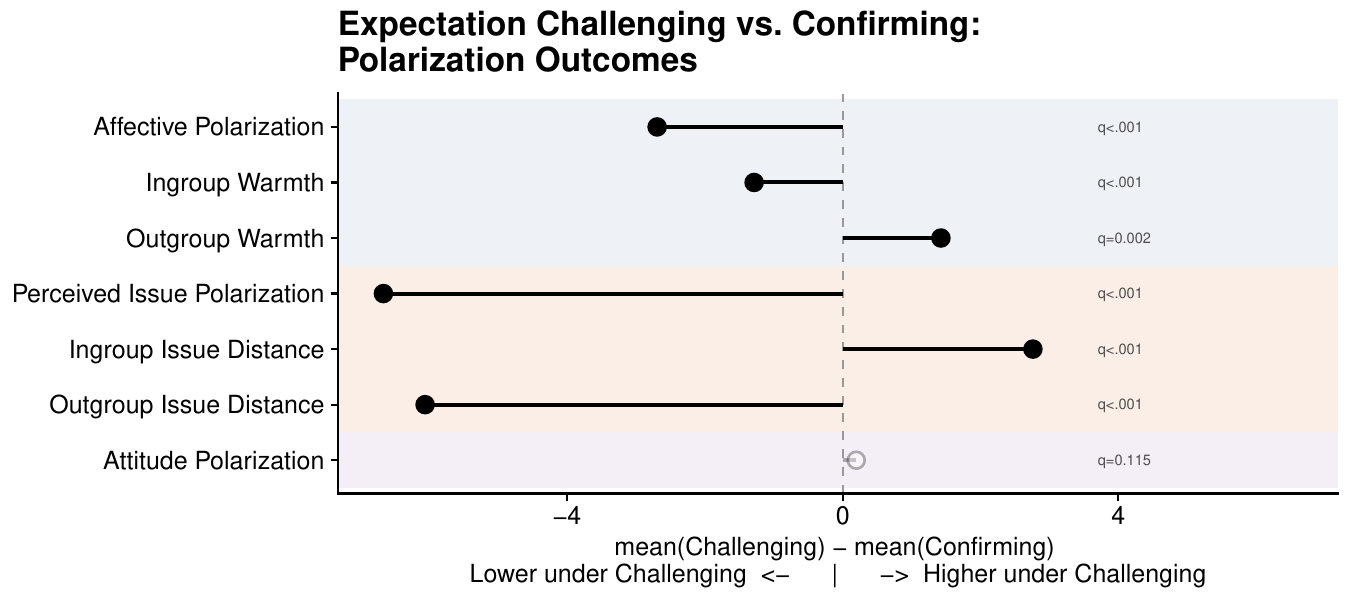}
    \caption{\textbf{Exploratory contrast of expectation-challenging versus expectation-confirming conditions on polarization outcomes.} The figure shows linear contrasts comparing expectation-challenging conditions (Ingroup Disagree and Outgroup Agree) with expectation-confirming conditions (Ingroup Agree and Outgroup Disagree). Each point represents $\text{mean(Challenging)} - \text{mean(Confirming)}$ for a given outcome. Negative estimates, shown to the left of zero, indicate lower values under expectation-challenging conditions; positive estimates, shown to the right of zero, indicate higher values under expectation-challenging conditions. Estimates are linear contrasts of treatment coefficients from OLS models with block fixed effects and LASSO-selected pre-treatment covariates, estimated with HC2 robust standard errors. Error bars indicate 95\% confidence intervals. $q$-values are BKY sharpened two-stage FDR-adjusted within this family of outcomes; filled markers denote $q \leq .05$ and hollow, faded markers denote $q > .05$. Point estimates, standard errors, sample sizes, and $p$- and $q$-values are reported in Table~\ref{tab:si-contrast}. This analysis is exploratory and was not preregistered.}
    \label{fig:si-contrast}
\end{figure}

Figure~\ref{fig:si-contrast} and Table~\ref{tab:si-contrast} summarize the results. 
Expectation-challenging conditions significantly reduces affective polarization ($C = -2.69$, $\text{SE} = 0.61$, 95\% CI $[-3.89, -1.49]$, $p < .001$) and perceived issue polarization ($C = -6.67$, $\text{SE} = 0.82$, 95\% CI $[-8.27, -5.06]$, $p < .001$), relative to expectation-confirming conditions. These reductions are driven by movement in both subcomponents: lower ingroup and higher outgroup warmth for affective polarization, and greater perceived ingroup and smaller perceived outgroup distance for issue polarization. 
Attitude polarization is the only polarization outcome for which the contrast is not distinguishable from zero ($q > .05$).

\begin{table} 
\centering
\caption{\textbf{Exploratory contrast of expectation-challenging versus expectation-confirming conditions on polarization outcomes.} Estimated linear contrast ($\text{mean(Challenging)} - \text{mean(Confirming)}$) per outcome; positive values indicate higher outcomes under challenging exposure. $q$-values are BKY two-stage FDR-adjusted within the family. Exploratory, not preregistered.}
\label{tab:si-contrast}
\centering
\resizebox{\ifdim\width>\linewidth\linewidth\else\width\fi}{!}{
\fontsize{9}{11}\selectfont
\begin{tabular}[t]{lrrcrrr}
\toprule
DV & Contrast & SE & 95\% CI & N & p-value & q-value\\
\midrule
Ingroup Warmth & -1.286 & 0.351 & {}[-1.97, -0.60] & 1,982 & <.001 & <.001\\
Outgroup Warmth & 1.425 & 0.504 & {}[0.44, 2.41] & 1,982 & 0.005 & 0.002\\
Affective Polarization & -2.693 & 0.612 & {}[-3.89, -1.49] & 1,982 & <.001 & <.001\\
Ingroup Issue Distance & 2.761 & 0.563 & {}[1.66, 3.87] & 1,983 & <.001 & <.001\\
Outgroup Issue Distance & -6.061 & 0.846 & {}[-7.72, -4.40] & 1,982 & <.001 & <.001\\
Perceived Issue Polarization & -6.666 & 0.820 & {}[-8.27, -5.06] & 1,982 & <.001 & <.001\\
Attitude Polarization & 0.197 & 0.552 & {}[-0.89, 1.28] & 1,983 & 0.722 & 0.115\\
\bottomrule
\end{tabular}}
\end{table}

\section{Experimental Validity and Data Quality}

This section assesses the validity of the experiment and the quality of the collected data. We first report whether pre-treatment covariates are balanced across experimental conditions and evaluate attrition between survey waves. We then consider several indicators of data quality as proxies for the success of the experimental manipulation. These indicators include measurements of whether participants correctly perceived the chatbot's stance and its partisan identity. They also included additional measurements meant to ensure that participants were meaningfully engaging with the conversation task. These include the response times of participants to chatbot messages and the text length of their responses. 

\subsection{Covariate Balance Across Conditions}
\label{sec:si-covar-balance}

We tested for pair-wise differences between the baseline condition (Ingroup Agree) and every other condition using Welch's $t$-tests for continuous and binary covariates (see Table~\ref{tab:balance}). None of these comparisons yielded statistically significant differences (all $p>.05$), indicating that the random assignment produced comparable groups across conditions. 

In addition, $F$-tests were computed by regressing each treatment indicator on the set of pre-treatment covariates. Missing values were replaced with a constant, and an indicator for missingness was included in the regression models to preserve the full sample. Under successful randomization, no covariate should significantly explain treatment assignment; thus, a non-significant joint $F$-test indicates that the covariates are balanced across conditions. In our analysis, none of the covariates significantly predicted treatment indicators, and the joint $F$-tests were non-significant, confirming that random assignment produced well-balanced groups where pre-treatment covariates were evenly distributed.

\begin{table}
\centering
\caption{Balance Table}
\label{tab:balance}
\resizebox{\ifdim\width>\linewidth\linewidth\else\width\fi}{!}{
\begin{tabular}[t]{l|c|ccc|ccc|cccl|c|ccc|ccc|cccl|c|ccc|ccc|cccl|c|ccc|ccc|cccl|c|ccc|ccc|cccl|c|ccc|ccc|cccl|c|ccc|ccc|cccl|c|ccc|ccc|cccl|c|ccc|ccc|cccl|c|ccc|ccc|cccl|c|ccc|ccc|ccc}
\toprule
Covariate & \makecell{Ingroup\\Agree} & \makecell{Ingroup\\Disagree} & \emph{Diff} & $p$-val. & 
\makecell{Outgroup\\Agree} & \emph{Diff} & $p$-val. & 
\makecell{Outgroup\\Disagree} & \emph{Diff} & $p$-val. \\
\midrule
Democrat & 0.52 & 0.52 & 0.00 & 0.96 & 0.53 & 0.01 & 0.76 & 0.52 & 0.00 & 0.98\\
Nonwhite & 0.25 & 0.21 & -0.04 & 0.09 & 0.21 & -0.04 & 0.10 & 0.24 & -0.01 & 0.70\\
Age & 39.49 & 40.95 & 1.46 & 0.07 & 40.13 & 0.64 & 0.43 & 40.95 & 1.46 & 0.08\\
College grad & 0.57 & 0.58 & 0.00 & 0.90 & 0.62 & 0.05 & 0.12 & 0.59 & 0.02 & 0.57\\
Male & 0.48 & 0.49 & 0.01 & 0.69 & 0.50 & 0.02 & 0.52 & 0.47 & -0.01 & 0.79\\
Ideology & 3.98 & 3.93 & -0.04 & 0.76 & 3.92 & -0.06 & 0.68 & 3.96 & -0.02 & 0.91\\
AI trust & 0.70 & 0.69 & -0.01 & 0.74 & 0.70 & -0.01 & 0.80 & 0.69 & -0.01 & 0.77\\
Ingroup love & 74.39 & 74.00 & -0.38 & 0.76 & 73.82 & -0.57 & 0.64 & 74.18 & -0.20 & 0.87\\
Outgroup hate & 26.67 & 26.35 & -0.32 & 0.83 & 27.55 & 0.88 & 0.56 & 25.90 & -0.77 & 0.62\\
Affective polarization & 47.72 & 47.65 & -0.06 & 0.97 & 46.27 & -1.45 & 0.45 & 48.29 & 0.57 & 0.77\\
Ingroup distance & 12.75 & 12.85 & 0.11 & 0.92 & 12.85 & 0.11 & 0.92 & 13.39 & 0.65 & 0.54\\
Outgroup distance & 53.14 & 54.87 & 1.73 & 0.42 & 54.79 & 1.65 & 0.45 & 56.76 & 3.62 & 0.10\\
Issue polarization & 47.66 & 49.88 & 2.22 & 0.29 & 49.35 & 1.69 & 0.41 & 51.23 & 3.57 & 0.09\\
\midrule
\# obs. & 488 & 502 &  &  & 496 &  &  & 497 &  & \\
$F$-stat & 0.73 & 0.51 &  &  & 0.59 &  &  & 0.52 &  & \\
$p$-value & 0.73 & 0.92 &  &  & 0.86 &  &  & 0.91 &  & \\
\bottomrule
\end{tabular}}
\begin{flushleft}
\emph{Note.} 
Columns labeled Ingroup Agree, Ingroup Disagree, Outgroup Agree, and Outgroup Disagree report the mean value of each pre-treatment covariate for the corresponding experimental condition. 
The \emph{Diff} column shows mean differences between the baseline condition (Ingroup Agree) and each other condition; the \emph{p-val.} column displays results from Welch's $t$-tests comparing those means. 
The bottom rows indicate: the sample size for each condition (\# obs.) and the results of $F$-test, regressing each treatment indicator on the pre-treatment covariates. 
\end{flushleft}
\end{table}

\subsection{Attrition Between Survey Waves}
\label{sec:si-attrition}

The retention rate from Wave 1 to Wave 2 was 82.05\%. All participants were invited to the second survey at least one month after completing Wave 1. Invitations to Wave 2 were sent only after all Wave 1 responses had been collected. Consequently, participants who completed Wave 1 earlier experienced a gap of nearly two months between waves. For example, those in the pilot phase who completed Wave 1 on September 17, 2025 were not invited to Wave 2 until November 11, 2025. The observed retention rate is notably high given this relatively long delay.

To address concerns about differential attrition, we conducted a $\chi$-squared test across all conditions to test the null hypothesis of no difference in attrition rates. The test failed to reject the null hypothesis (Table~\ref{tab:differential_attrition}), indicating no evidence of differential attrition across experimental conditions.

\begin{table}
\centering
\caption{Differential Attrition}
\label{tab:differential_attrition}
\centering
\begin{threeparttable}
\begin{tabular}[t]{lcc}
\toprule
Randomized Group & Attrited (\%) & Stayed (\%)\\
\midrule
Ingroup Agree & 17.8\% (87) & 82.2\% (401)\\
Ingroup Disagree & 16.5\% (83) & 83.5\% (419)\\
Outgroup Agree & 21.2\% (105) & 78.8\% (391)\\
Outgroup Disagree & 16.3\% (81) & 83.7\% (416)\\
\bottomrule
\end{tabular}
\begin{tablenotes}
{\footnotesize\item \emph{$\chi$-square = 5.099, df = 3, $p$-value = 0.165}}
\end{tablenotes}
\end{threeparttable}
\end{table}

Furthermore, we examined differences in observable characteristics between participants who attrited and those who remained in the study throughout both waves. The pre-registered characteristics included political affiliation (Democrat vs. Republican), race (white vs. nonwhite), education level (college graduate vs non-college graduate), and sex (male self-identification). As shown in Table~\ref{tab:selective_attrition}, we found no evidence of selective attrition on any of these characteristics across waves.

\begin{table}
\centering
\caption{Selective Attrition}\label{tab:selective_attrition}
\centering
\begin{tabular}[t]{llllr}
\toprule
 & Covariate & Wave~1 & Wave~2 & $p$-value\\
\midrule
\multicolumn{5}{l}{{Party ID}}\\
\hspace{1em} & Democrat & 52.3\% & 54.2\% & 0.278\\

\hspace{1em} & Republican & 47.7\% & 45.8\% & \\
\cmidrule{1-5}
\multicolumn{5}{l}{{Non-White}}\\
\hspace{1em} & White & 77.3\% & 77.1\% & 0.934\\

\hspace{1em} & Non-White & 22.7\% & 22.9\% & \\
\cmidrule{1-5}
\multicolumn{5}{l}{{College Grad}}\\
\hspace{1em} & Not College Grad & 41\% & 41.4\% & 0.875\\

\hspace{1em} & College Grad & 59\% & 58.6\% & \\
\cmidrule{1-5}
\multicolumn{5}{l}{{Male}}\\
\hspace{1em} & Not Male & 51.6\% & 52.6\% & 0.583\\

\hspace{1em} & Male & 48.4\% & 47.4\% & \\
\bottomrule
\end{tabular}
\end{table}

\subsection{Manipulation Checks}
\label{sec:si-mani-check}

\subsubsection{Participant's perception of chatbot's stance}

\begin{figure}
    \centering
    \begin{subfigure}{\linewidth} 
        \centering
        \includegraphics[width=0.75\linewidth]{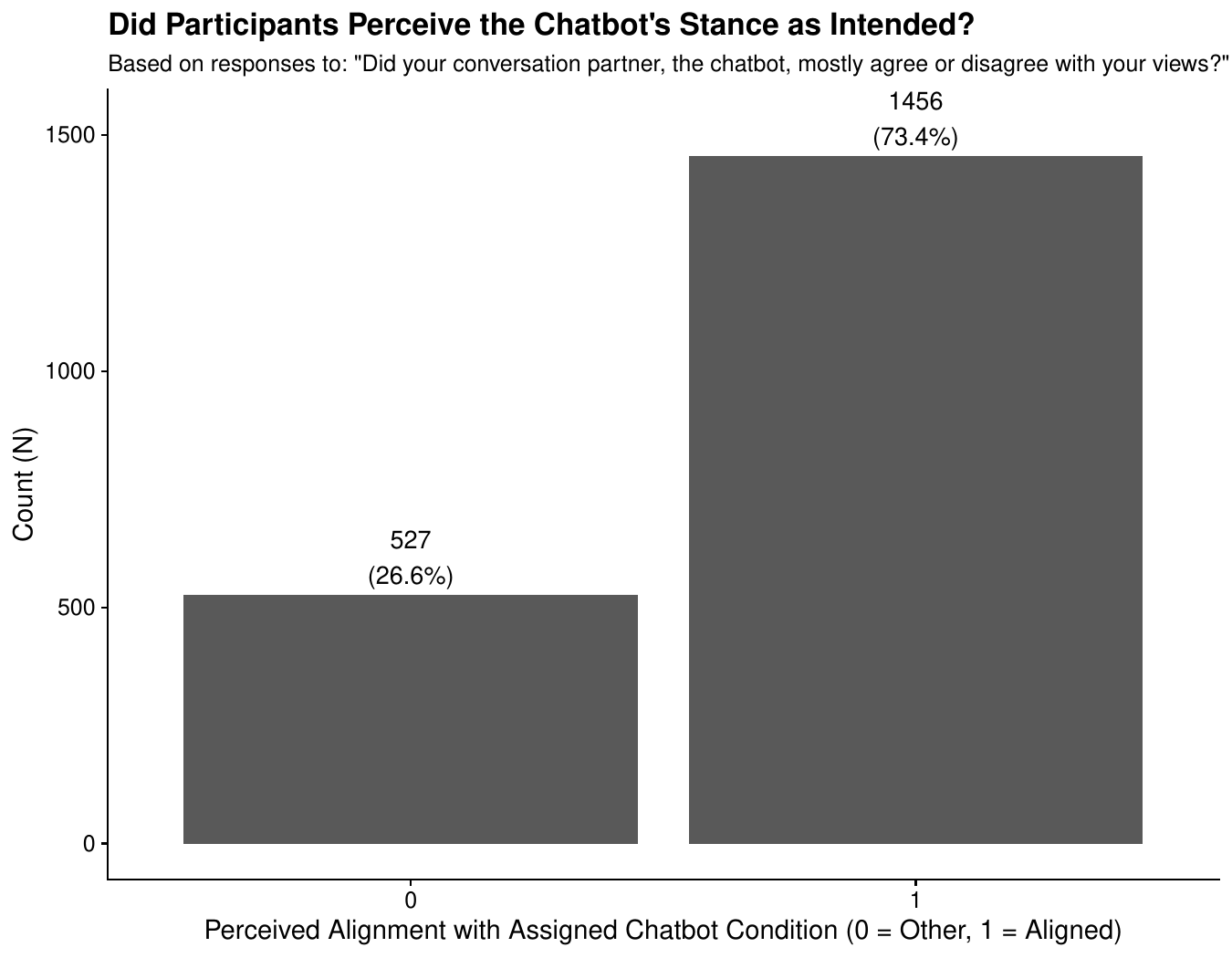}
        \label{fig:mani_check1_a}
    \end{subfigure}
    \begin{subfigure}{\linewidth}
        \centering
        \includegraphics[width=0.85\linewidth]{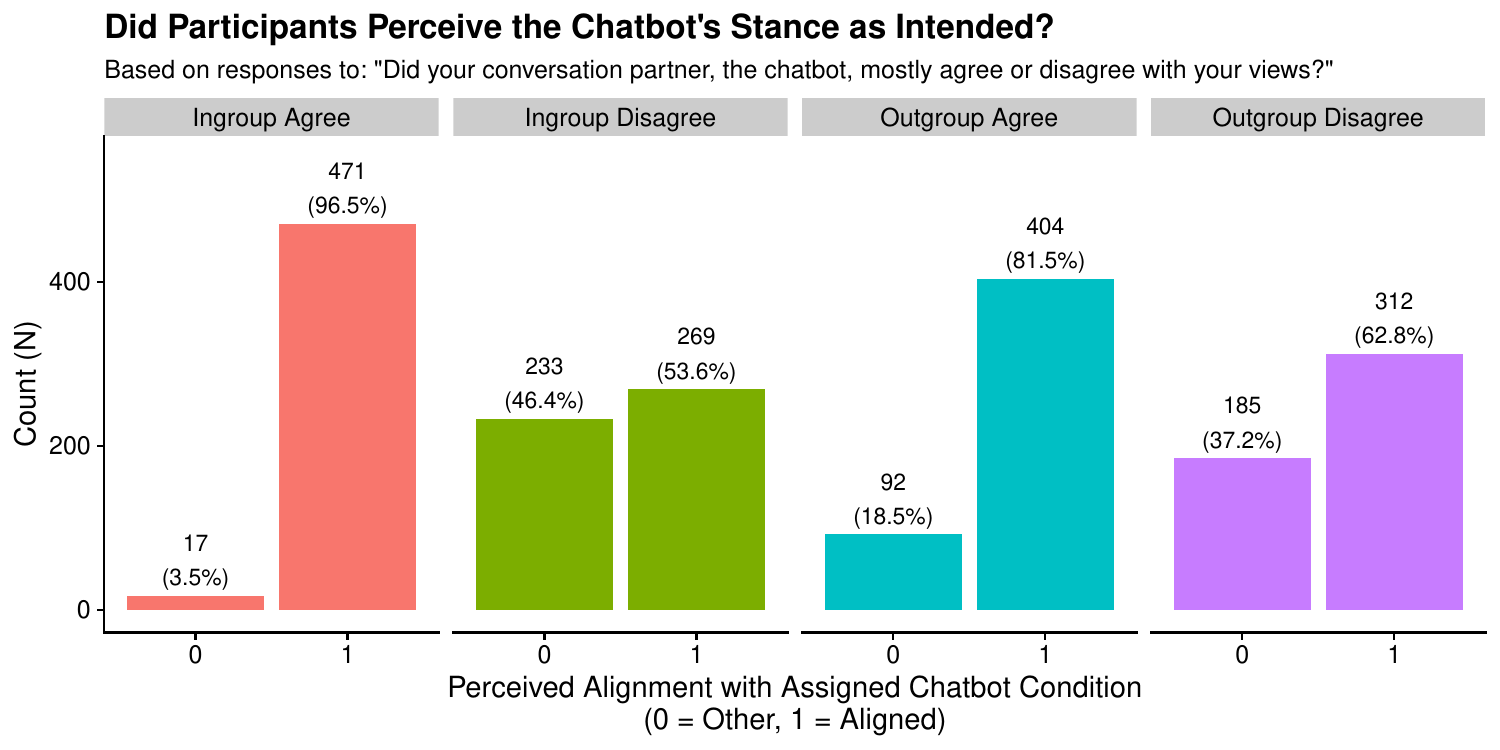}
        \label{fig:mani_check1_b}
    \end{subfigure}
    \caption{Perceived stance. Top: Proportions of participants who correctly perceived the chatbot's intended stance (1) versus those who did not (0). Alignment (1) indicates that participants in the agree condition responded ``mostly agree'' to the manipulation check question, and participants in the disagree condition responded ``mostly disagree.'' Bottom: Proportions broken down by assigned conditions.}
    \label{fig:mani_check1_combined}
\end{figure}

We examine whether participants correctly perceived the chatbot's stance on the issue. Immediately after the conversation, participants answered a manipulation check question: \emph{``Did your conversation partner, the chatbot, mostly agree or disagree with your views?''}. The majority of participants perceived the chatbot's issue stance as intended (see Figure~\ref{fig:mani_check1_combined}, top). That is, participants in the agree condition tended to report that the chatbot ``mostly agreed,'' whereas those in the disagree condition tended to report that it ``mostly disagreed.''

The break-down of these proportions by experimental condition shows that participants assigned to chatbots that shared similar stances were more likely to perceive the chatbot as aligned (i.e., Ingroup Agree and Outgroup Agree participants perceiving chatbots as mostly agreeing). In contrast, perceived alignment was lower when the chatbots had opposing stances, particularly in the Ingroup Disagree (Figure~\ref{fig:mani_check1_combined}, bottom). One possible explanation is that the respectful tone of the chatbots may have led participants to perceive their disagreeing responses as less explicitly oppositional. Despite this potential ambiguity in perceived disagreement, our results are robust when the estimates are restricted to a subset of the sample of participants who clearly perceived the chatbot's intended issue stance (see Section~\ref{sec:si-robustness-manip} below).

\subsubsection{Participant's perception of chatbot's partisanship}

Although the chatbot's partisanship (Democrat/Republican) was stated in the instructions, we further examine the extent to which participants felt the chatbot spoke authentically for that group by asking the question: \emph{``How accurately do you think the chatbot represented \texttt{\{group\}}'s views?''}

\begin{figure}
    \centering
    \begin{subfigure}{\linewidth} 
        \centering
        \includegraphics[width=0.8\linewidth]{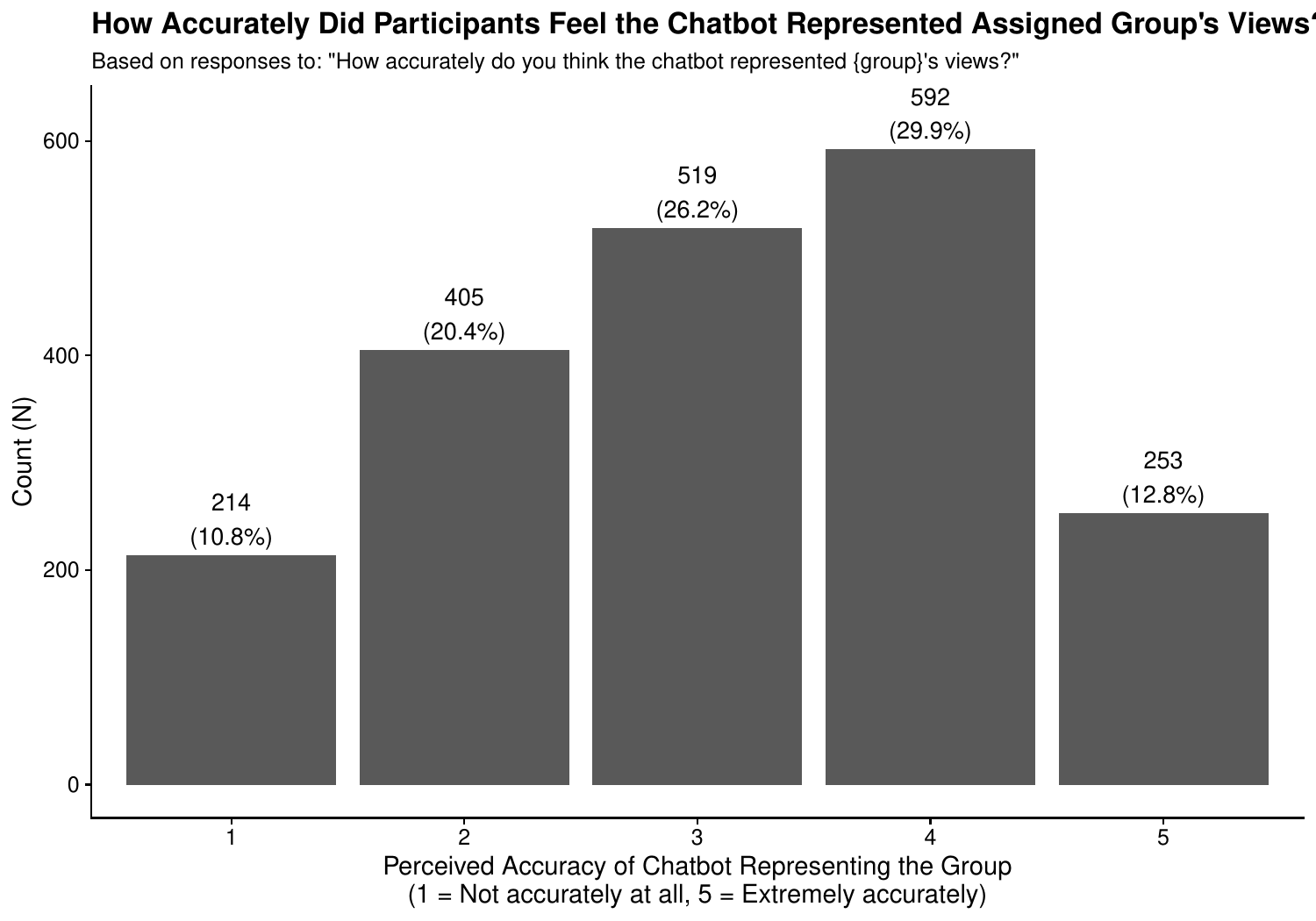}
        \label{fig:mani_check2_a}
    \end{subfigure}
    \begin{subfigure}{\linewidth}
        \centering
        \includegraphics[width=0.9\linewidth]{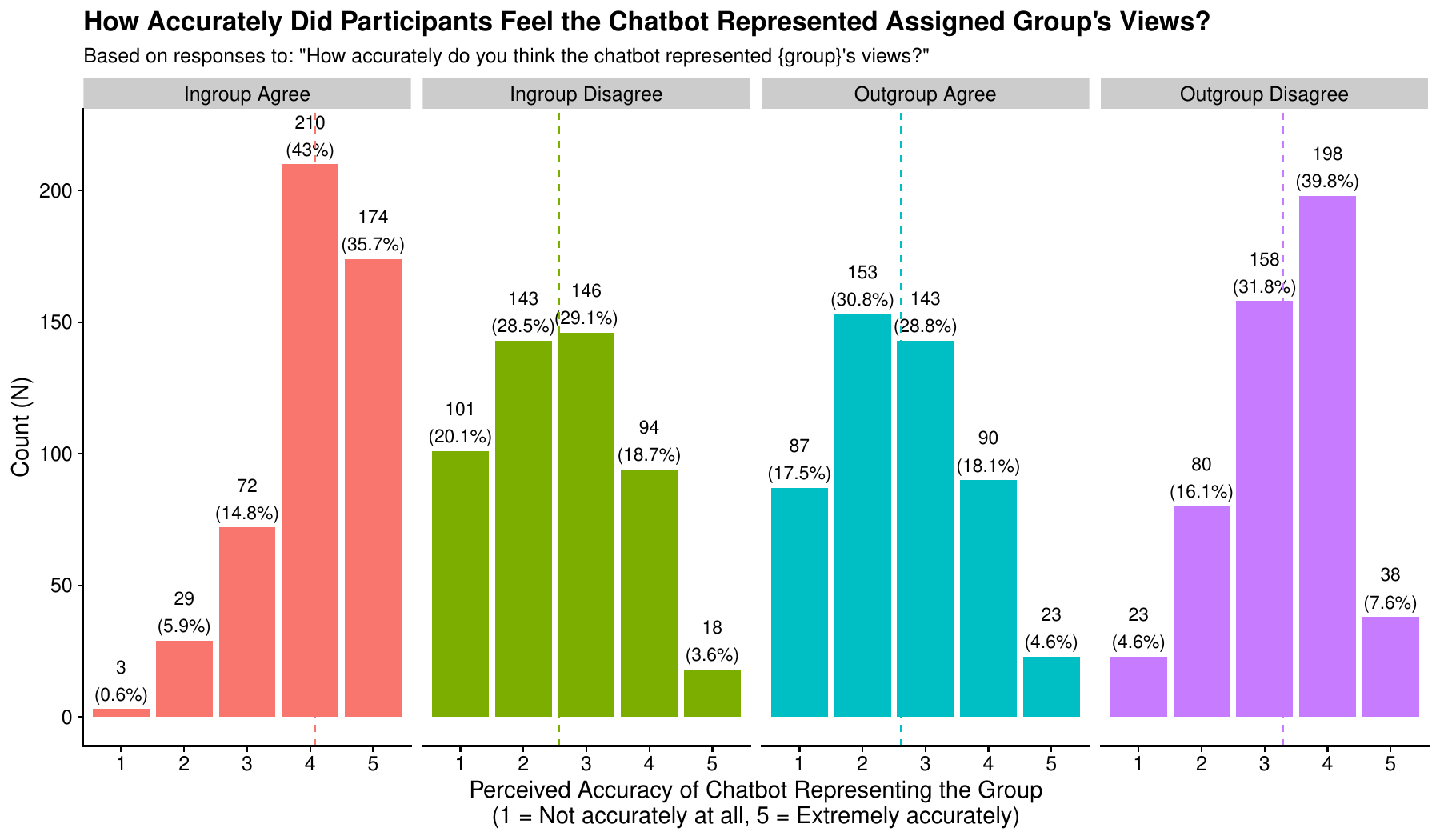}
        \label{fig:mani_check2_b}
    \end{subfigure}
    \caption{
    Perceived partisanship. Top: Distribution of scores given by participants about their perception that the chatbot accurately represented the assigned group status on a 5-point scale. Bottom: Proportions broken down by assigned conditions. Dotted lines indicate the mean value within each condition.} 
    \label{fig:mani_check2_combined}
\end{figure}

Most participants perceived the chatbot's accuracy in representing the assigned partisanship as at or above the midpoint (Figure~\ref{fig:mani_check2_combined}, top panel). The perceptions vary across conditions. Participants in the expectation-confirming conditions (Ingroup Agree and Outgroup Disagree) reported perceived accuracy scores above the midpoint (see the bottom panel of Figure~\ref{fig:mani_check2_combined}), indicating that they viewed the chatbot as appropriately representing the assigned group opinions. In contrast, participants in the expectation-challenging conditions (Ingroup Disagree and Outgroup Agree) rated perceived accuracy below the midpoint, suggesting poorer perceived representation.

In other words, participants viewed the chatbot as accurate when its stance aligned with their expectations and less accurate when it did not. This pattern suggests that participants were responsive to the intended manipulations. The lower accuracy ratings in expectation-challenging conditions may also reflect cognitive dissonance, as participants discounted positions that conflicted with their group-based expectations as inaccurate representation.

\subsection{Attention and Engagement Checks}
\label{sec:si-attention-check}

To assess whether participants meaningfully paid attention to the conversation, we examine participant response times to chatbot messages across conversational turns. Using interaction log data, we operationalize response times in two ways: reading time, defined as the elapsed time between receiving a chatbot message and initiating typing within the same turn, and writing time, defined as the interval between initiating typing and submitting the corresponding message. As participants could engage in up to three exchanges, writing time is observed for turns 1--3 only. All timestamps were recorded as Unix epoch milliseconds and converted to seconds for analysis. Empty or missing entries are treated as missing values.

\begin{figure}
    \centering
    \begin{subfigure}{\linewidth} 
        \centering
        \includegraphics[width=\linewidth]{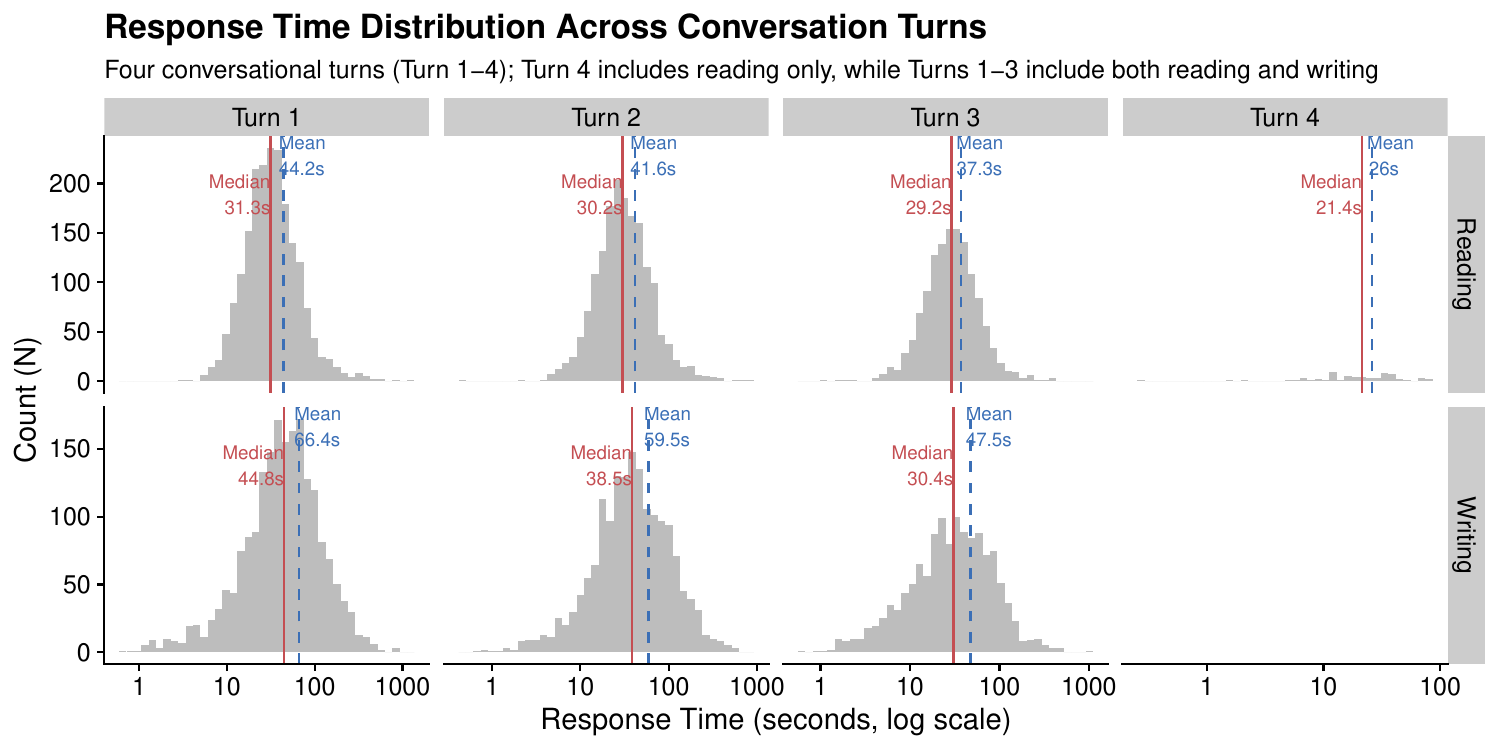}
    \end{subfigure}
    \begin{subfigure}{\linewidth} 
        \centering
        \includegraphics[width=\linewidth]{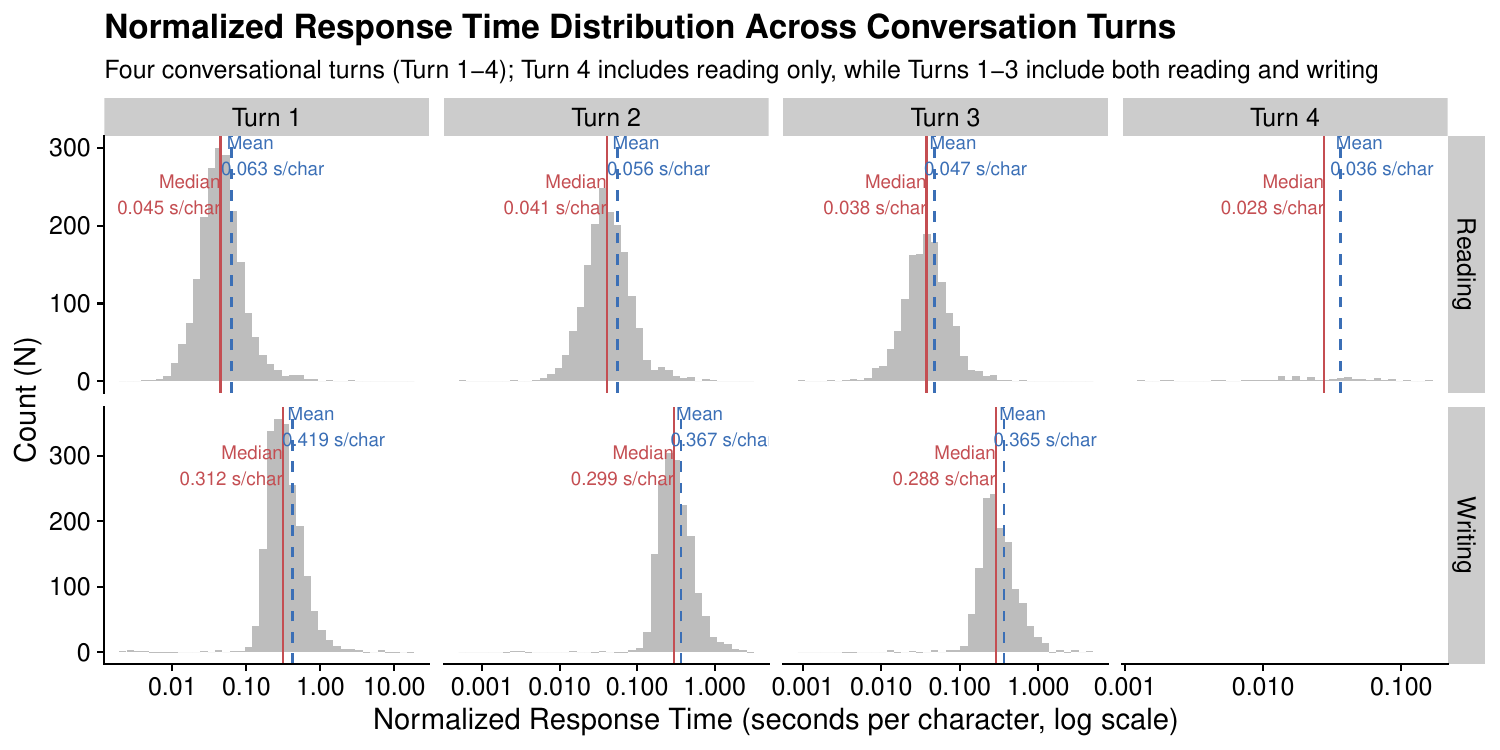}
    \end{subfigure} 
    \caption{Response time distributions across conversational turns: reading (top rows) and writing (bottom rows). The top panel shows raw response times; the bottom panel shows normalized response times (seconds per character), adjusted for message length. Turns 1--3 include both reading and writing, while only reading time is available for turn 4. Vertical lines indicate the mean (dashed) and median (solid) times.}
    \label{fig:si-response-time}
\end{figure}

Both reading and writing times remained relatively long and consistent throughout the interactions, suggesting that participants consistently took time to read and formulate their responses (Figure~\ref{fig:si-response-time}, top). The median is approximately 30 seconds for reading time and 30--45 seconds for writing time across turns. Response times decrease across turns, indicating that participants became faster as the conversation progressed. 

The length-normalized response times exhibit similar patterns, indicating that the observed patterns in response times are not driven solely by variation in message length (Figure~\ref{fig:si-response-time}, top). Notably, writing times remained substantially longer than reading times after adjusting for message length, indicating that participants spent more time per character when producing responses than when processing chatbot messages.  

\begin{figure}
    \centering
    \includegraphics[width=0.95\linewidth]{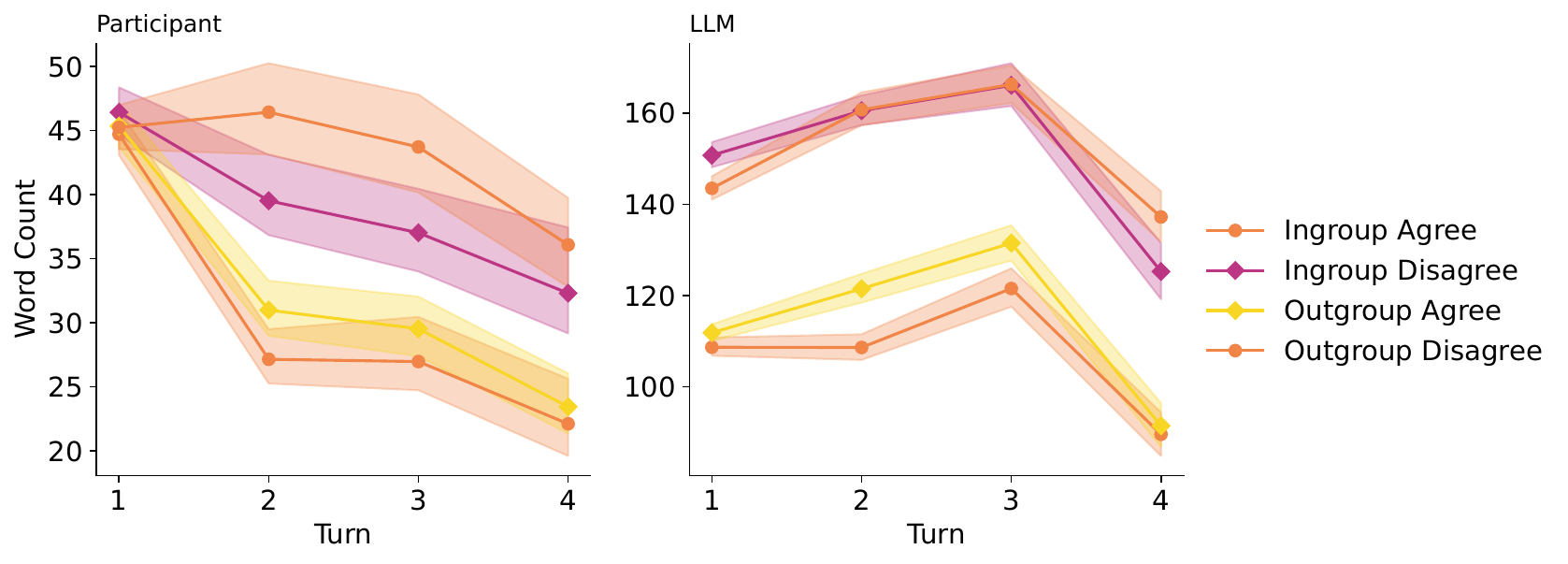}
    \caption{\textbf{Response lengths by turn and treatment}. Lines represent bootstrapped mean values and shaded areas represent 95\% confidence intervals.}
    \label{fig:response_len}
\end{figure}

Figure~\ref{fig:response_len} further illustrates the evolution of response lengths in word counts across conversational turns. The results show that LLMs are consistently more verbose than human participants across all conditions.
Notably, both humans and LLMs tend to use significantly more words in outgroup conditions compared to ingroup ones.

\subsection{Analysis of Data Missingness due to API Technical Failures} 
\label{sec:si-technical-failure}

As described in the manuscript, 65 participants were excluded due to technical
failures during the LLM-based discussion task: 50 for whom there was a failure
to generate the initial opinion summaries and 15 who experienced technical
errors during the conversation. We conducted the following checks to verify that
these exclusions did not introduce systematic bias into the analytic sample. See
Tables~\ref{tab:technical_exclusion}
and~\ref{tab:technical_exclusion_by_condition} below. 

To assess whether technical failures were purely at random and did not follow
any systematic pattern, we compared excluded ($n = 65$) and retained
participants ($n = 1{,}983$) on key demographics and pre-treatment attitudes
(age, gender, ethnicity, race, education, ideology, and prior AI trust).
Individual $t$-tests revealed no significant differences on any variable (
$p > .05$ throughout), and a $\chi$-squared test showed no differential failure rates across
treatment conditions ($\chi^2(3) = 5.40$, $p = .145$), with failure rates
ranging from 2.0\% to 4.3\% across cells. These results support the assumption
that LLM API failures were random and unrelated to participant characteristics
or treatment assignment.

\begin{table}
\centering
\caption{Balance Between Retained and Excluded Participants}
\label{tab:technical_exclusion}
\centering
\begin{threeparttable}
\begin{tabular}[t]{lllll}
\toprule
\multicolumn{1}{c}{ } & \multicolumn{2}{c}{Mean} & \multicolumn{2}{c}{ } \\
\cmidrule(l{3pt}r{3pt}){2-3}
Variable & Retained & Excluded & Difference & $p$-value\\
\midrule
Age & 40.38 & 39.60 & 0.78 & 0.67\\
Male & 0.48 & 0.55 & -0.07 & 0.27\\
Hispanic & 0.09 & 0.14 & -0.05 & 0.30\\
Non-White & 0.23 & 0.19 & 0.04 & 0.43\\
College Graduate & 0.59 & 0.49 & 0.10 & 0.13\\
Ideology & 3.95 & 3.91 & 0.04 & 0.89\\
Prior AI Trust & 0.70 & 0.75 & -0.06 & 0.30\\
\bottomrule
\end{tabular}
\begin{tablenotes}
\item \emph{Note: } Individual $t$-tests comparing retained ($n = 1{,}983$) and excluded ($n = 65$) participants.  
\end{tablenotes}
\end{threeparttable}
\end{table}

\begin{table}
\centering
\caption{Failure Rate by Treatment Condition}\label{tab:technical_exclusion_by_condition}
\centering
\begin{threeparttable}
\begin{tabular}[t]{lrrl}
\toprule
Condition & $N$ & $N$ Excluded & \% Excluded\\
\midrule
Ingroup--Agree & 510 & 22 & 4.3\%\\
Ingroup--Disagree & 512 & 10 & 2\%\\
Outgroup--Agree & 515 & 19 & 3.7\%\\
Outgroup--Disagree & 511 & 14 & 2.7\%\\
\bottomrule
\end{tabular}
\begin{tablenotes}
\item \emph{Note: } Chi-square test across treatment conditions: $\chi^2(3) = 5.40$, $p = .145$.
\end{tablenotes}
\end{threeparttable}
\end{table}

\section{Robustness and Sensitivity Analyses}
\label{sec:si-robustness-check}

We conducted three robustness checks. First, we re-estimated treatment effects after restricting the sample to participants who passed the manipulation check. Second, we excluded participants recruited during the pilot phase. Third, for political polarization measures collected at both pre- and post-treatment time points, we re-estimated models using mixed effects specifications as an alternative modeling approach that explicitly accounts for within-person change over time. Across three robustness checks, the main findings remained stable.

\subsection{Analyses Restricted to Manipulation-Check Passers}
\label{sec:si-robustness-manip}

\begin{figure}
    \centering
    \includegraphics[width=\linewidth]{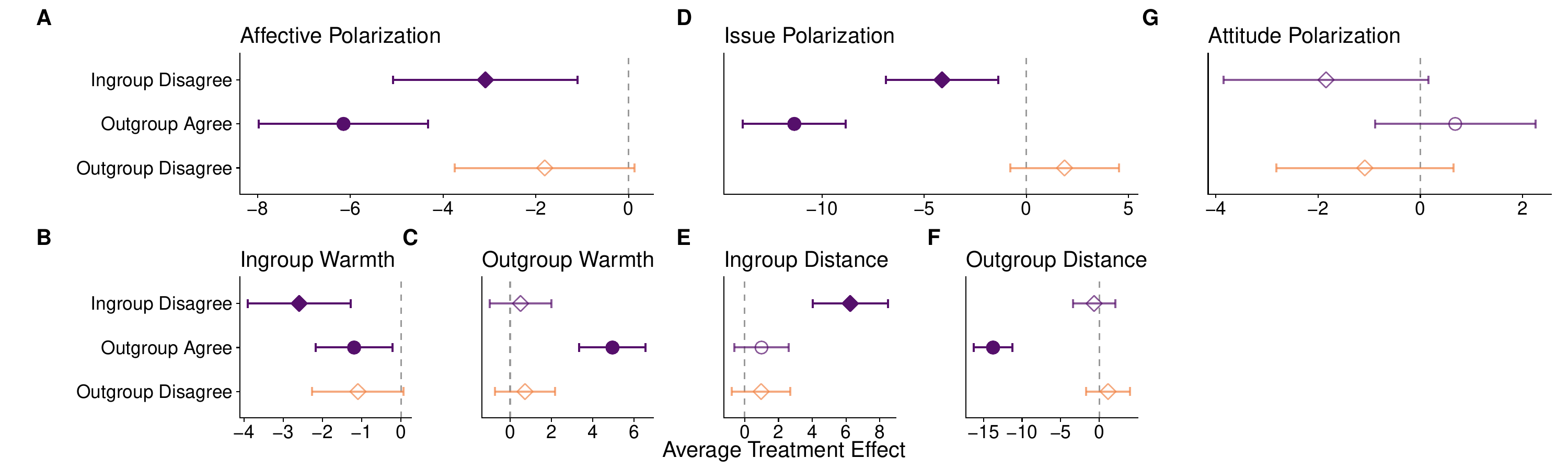}
    \caption{\textbf{Treatment effects on political polarization, among those who passed the manipulation check.}
    Each plot shows average treatment effects for the Outgroup Agree (purple circle), Ingroup Disagree (purple diamond) and Outgroup Agree (orange diamond), relative to the Ingroup Agree condition (gray dashed line), with 95\% confidence intervals. Non-significant results ($q > .05$) are indicated by lighter colors and unfilled markers.
    (A)~Expectation-challenging conversations (Outgroup Agree and Ingroup Disagree) and conversations in the Outgroup Disagree condition reduce affective polarization.
    (B,C)~The reduction in affective polarization in the Outgroup Agree condition is driven by both warmer feelings toward the outgroup and colder feelings toward the ingroup. In the Ingroup Disagree condition, the decrease is driven by colder feelings toward the ingroup.
    (D)~Expectation-challenging conversations also reduce perceived issue polarization.
    (E,F)~The reduction in perceived issue polarization in the Outgroup Agree condition is driven by a decrease in perceived outgroup distance, while in the Ingroup Disagree condition it is largely driven by a increase in perceived ingroup distance.
    (G)~We find no evidence that challenging partisan expectations persuade participants either by moderating or strengthening their previous attitudes on the issue.}
    \label{fig:si-robustness-check1-polarization}
\end{figure}
\begin{figure}
    \centering
    \includegraphics[width=1\linewidth]{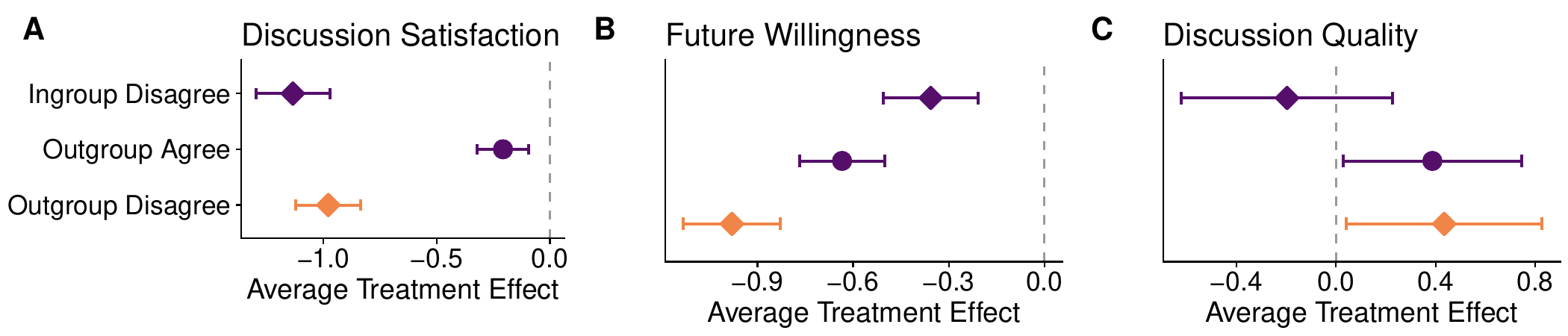}
    \caption{\textbf{Treatment effects on discussion-related outcomes, among those who passed the manipulation check.} Each plot shows estimated treatment effects with 95\% confidence intervals. Non-significant results ($q > .05$) are indicated by lighter colors and unfilled markers. The baseline is `Ingroup Agree' condition.}
    \label{fig:si-robustness-check1-discussion}
\end{figure}

As shown in Section~\ref{sec:si-mani-check}, some participants did not perceive the chatbot's issue stance as intended by the assigned treatment. To assess the robustness of our results to this ambiguity, we re-run models restricting the analysis to participants who passed the manipulation check ($N=1,456$). This analysis replicates our main results, with even larger estimated effect sizes for political polarization (See Figures~\ref{fig:si-robustness-check1-polarization}, and~\ref{fig:si-robustness-check1-discussion}). This suggests that our reported effects are conservative estimations of the treatment effects. Clearer perception of the intended treatment would likely have produced even stronger effects.

\begin{table} 
\centering
\caption{Treatment effects on political polarization, among those who passed the manipulation check (FDR-adjusted q-values)}
\label{tab:si-robustness-check1-polarization-FDR}
\centering
\resizebox{\ifdim\width>\linewidth\linewidth\else\width\fi}{!}{
\fontsize{9}{11}\selectfont
\begin{tabular}[t]{llrrrr}
\toprule
DV & Treatment & Estimate & SE & p-value & q-value\\
\midrule
 & Ingroup-Disagree & -2.593 & 0.668 & <.001 & <.001\\

 & Outgroup-Agree & -1.193 & 0.500 & 0.017 & 0.026\\

\multirow[t]{-3}{*}{\raggedright\arraybackslash Ingroup Warmth} & Outgroup-Disagree & -1.096 & 0.594 & 0.065 & 0.077\\
\cmidrule{1-6}
 & Ingroup-Disagree & 0.504 & 0.761 & 0.507 & 0.296\\

 & Outgroup-Agree & 4.953 & 0.820 & <.001 & <.001\\

\multirow[t]{-3}{*}{\raggedright\arraybackslash Outgroup Warmth} & Outgroup-Disagree & 0.718 & 0.743 & 0.334 & 0.215\\
\cmidrule{1-6}
 & Ingroup-Disagree & -3.088 & 1.014 & 0.002 & 0.006\\

 & Outgroup-Agree & -6.147 & 0.932 & <.001 & <.001\\

\multirow[t]{-3}{*}{\raggedright\arraybackslash Affective Polarization} & Outgroup-Disagree & -1.809 & 0.989 & 0.068 & 0.077\\
\cmidrule{1-6}
 & Ingroup-Disagree & 6.248 & 1.135 & <.001 & <.001\\

 & Outgroup-Agree & 0.992 & 0.824 & 0.229 & 0.160\\

\multirow[t]{-3}{*}{\raggedright\arraybackslash Ingroup Issue Distance} & Outgroup-Disagree & 0.971 & 0.886 & 0.274 & 0.182\\
\cmidrule{1-6}
 & Ingroup-Disagree & -0.672 & 1.389 & 0.629 & 0.369\\

 & Outgroup-Agree & -13.756 & 1.271 & <.001 & <.001\\

\multirow[t]{-3}{*}{\raggedright\arraybackslash Outgroup Issue Distance} & Outgroup-Disagree & 1.139 & 1.437 & 0.428 & 0.255\\
\cmidrule{1-6}
 & Ingroup-Disagree & -4.126 & 1.400 & 0.003 & 0.007\\

 & Outgroup-Agree & -11.356 & 1.286 & <.001 & <.001\\

\multirow[t]{-3}{*}{\raggedright\arraybackslash Perceived Issue Polarization} & Outgroup-Disagree & 1.869 & 1.356 & 0.168 & 0.143\\
\cmidrule{1-6}
 & Ingroup-Disagree & -1.844 & 1.021 & 0.071 & 0.077\\

 & Outgroup-Agree & 0.682 & 0.802 & 0.395 & 0.247\\

\multirow[t]{-3}{*}{\raggedright\arraybackslash Attitude Polarization} & Outgroup-Disagree & -1.086 & 0.884 & 0.219 & 0.160\\
\bottomrule
\end{tabular}}
\end{table}

\begin{table} 
\centering
\caption{Treatment effects on discussion-related outcomes, among those who passed the manipulation check (FDR-adjusted q-values)}
\label{tab:si-robustness-check1-discussion-FDR}
\centering
\resizebox{\ifdim\width>\linewidth\linewidth\else\width\fi}{!}{
\fontsize{9}{11}\selectfont
\begin{tabular}[t]{llrrrr}
\toprule
DV & Treatment & Estimate & SE & p-value & q-value\\
\midrule
 & Ingroup-Disagree & -0.196 & 0.217 & 0.365 & 0.043\\

 & Outgroup-Agree & 0.387 & 0.183 & 0.034 & 0.014\\

\multirow[t]{-3}{*}{\raggedright\arraybackslash Discussion Quality} & Outgroup-Disagree & 0.435 & 0.201 & 0.031 & 0.014\\
\cmidrule{1-6}
 & Ingroup-Disagree & -0.357 & 0.076 & <.001 & <.001\\

 & Outgroup-Agree & -0.635 & 0.068 & <.001 & <.001\\

\multirow[t]{-3}{*}{\raggedright\arraybackslash Future Willingness} & Outgroup-Disagree & -0.980 & 0.077 & <.001 & <.001\\
\cmidrule{1-6}
 & Ingroup-Disagree & -1.135 & 0.083 & <.001 & <.001\\

 & Outgroup-Agree & -0.208 & 0.057 & <.001 & <.001\\

\multirow[t]{-3}{*}{\raggedright\arraybackslash Discussion Satisfaction} & Outgroup-Disagree & -0.978 & 0.073 & <.001 & <.001\\
\bottomrule
\end{tabular}}
\end{table}

\subsection{Exclusion of Pilot-Phase Participants}

We also examined whether the results hold when excluding participants recruited during the pilot phase ($N = 97$). Although the pilot and main study shared the same sampling frame, survey instruments, and experimental procedures, we conducted this robustness check to rule out the possibility that pilot participants differed systematically in unobserved ways. 

\begin{figure}
    \centering
    \includegraphics[width=\linewidth]{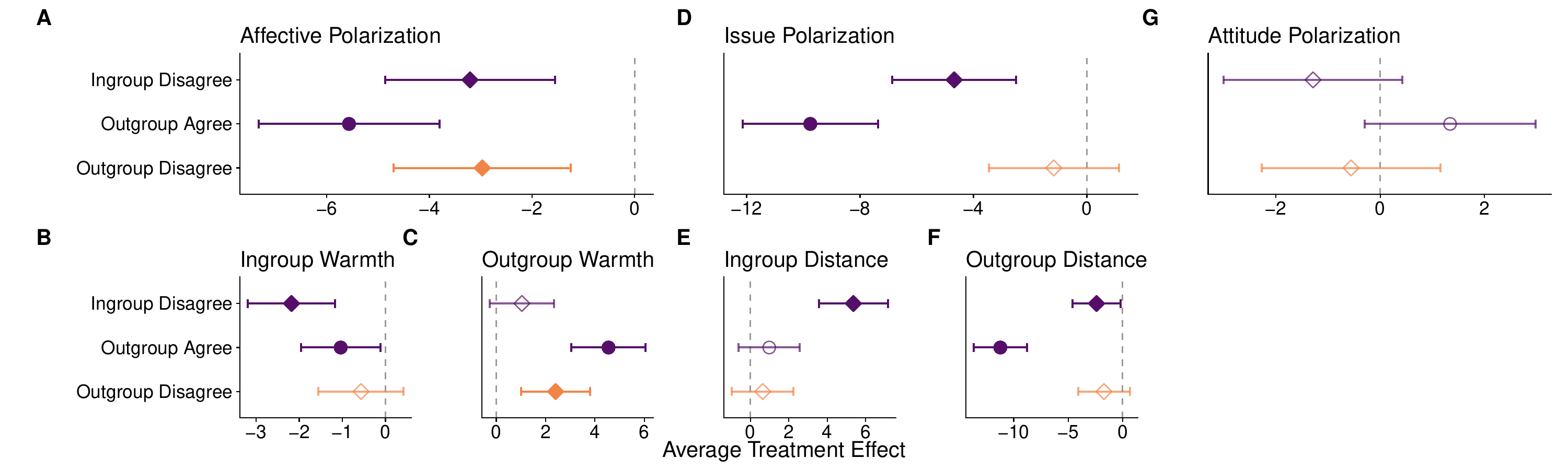}
    \caption{\textbf{Treatment effects on political polarization, excluding pilot study participants.} Each plot shows estimated treatment effects with 95\% confidence intervals. Non-significant results ($q > .05$) are indicated by lighter colors and unfilled markers. The baseline is `Ingroup Agree' condition. The top panels display predicted mean feelings toward (A) ingroup, (B) outgroup members, and (C) their difference (affective polarization). The bottom panels show the perceived distance between participants' own issue stance and that of their (D) ingroup, (E) outgroup, as well as (F) the perceived polarization between ingroup and outgroup positions. }
    \label{fig:si-robustness-check2-polarization}
\end{figure}

\begin{figure}
    \centering
    \includegraphics[width=\linewidth]{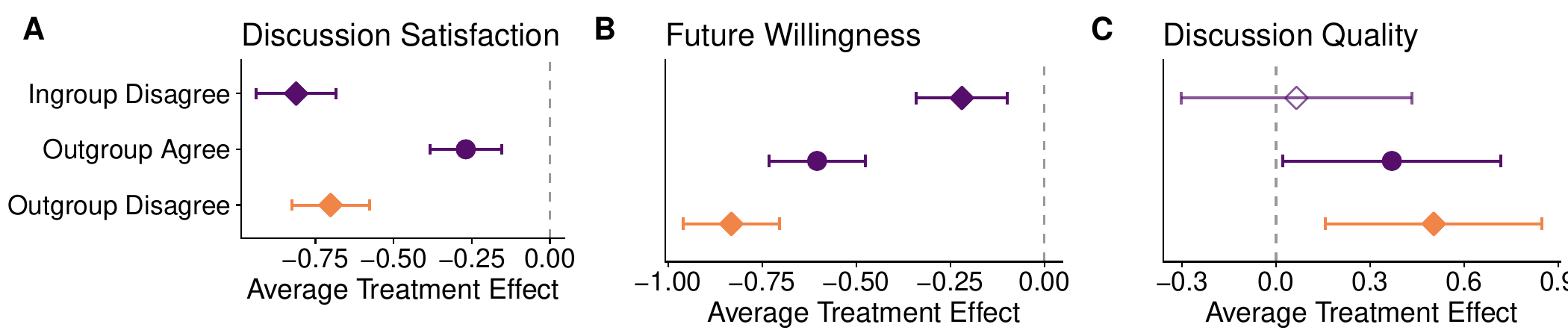}
    \caption{\textbf{Treatment effects on discussion-related outcomes, excluding pilot study participants.} Each plot shows estimated treatment effects with 95\% confidence intervals. Non-significant results ($q > .05$) are indicated by lighter colors and unfilled markers. The baseline is `Ingroup Agree' condition.}
    \label{fig:si-robustness-check2-discussion}
\end{figure}

As shown in Figures~\ref{fig:si-robustness-check2-polarization} and~\ref{fig:si-robustness-check2-discussion}, excluding the pilot sample does not change the substantive conclusions. These results indicate that pooling the pilot and main-study data is appropriate and that the inclusion of pilot participants does not influence the main findings.

\begin{table} 
\centering
\caption{Treatment effects on political polarization, excluding pilot study participants (FDR-adjusted q-values)}
\label{tab:si-robustness-check2-polarization-FDR}
\centering
\resizebox{\ifdim\width>\linewidth\linewidth\else\width\fi}{!}{
\fontsize{9}{11}\selectfont
\begin{tabular}[t]{llrrrr}
\toprule
DV & Treatment & Estimate & SE & p-value & q-value\\
\midrule
 & Ingroup-Disagree & -2.181 & 0.518 & <.001 & <.001\\

 & Outgroup-Agree & -1.039 & 0.471 & 0.028 & 0.029\\

\multirow[t]{-3}{*}{\raggedright\arraybackslash Ingroup Warmth} & Outgroup-Disagree & -0.567 & 0.505 & 0.261 & 0.151\\
\cmidrule{1-6}
 & Ingroup-Disagree & 1.039 & 0.665 & 0.119 & 0.083\\

 & Outgroup-Agree & 4.551 & 0.769 & <.001 & <.001\\

\multirow[t]{-3}{*}{\raggedright\arraybackslash Outgroup Warmth} & Outgroup-Disagree & 2.404 & 0.715 & <.001 & 0.002\\
\cmidrule{1-6}
 & Ingroup-Disagree & -3.209 & 0.845 & <.001 & <.001\\

 & Outgroup-Agree & -5.566 & 0.898 & <.001 & <.001\\

\multirow[t]{-3}{*}{\raggedright\arraybackslash Affective Polarization} & Outgroup-Disagree & -2.970 & 0.880 & <.001 & 0.002\\
\cmidrule{1-6}
 & Ingroup-Disagree & 5.362 & 0.913 & <.001 & <.001\\

 & Outgroup-Agree & 0.982 & 0.810 & 0.226 & 0.136\\

\multirow[t]{-3}{*}{\raggedright\arraybackslash Ingroup Issue Distance} & Outgroup-Disagree & 0.641 & 0.820 & 0.434 & 0.216\\
\cmidrule{1-6}
 & Ingroup-Disagree & -2.393 & 1.137 & 0.035 & 0.034\\

 & Outgroup-Agree & -11.221 & 1.240 & <.001 & <.001\\

\multirow[t]{-3}{*}{\raggedright\arraybackslash Outgroup Issue Distance} & Outgroup-Disagree & -1.715 & 1.210 & 0.157 & 0.097\\
\cmidrule{1-6}
 & Ingroup-Disagree & -4.679 & 1.113 & <.001 & <.001\\

 & Outgroup-Agree & -9.753 & 1.215 & <.001 & <.001\\

\multirow[t]{-3}{*}{\raggedright\arraybackslash Perceived Issue Polarization} & Outgroup-Disagree & -1.168 & 1.171 & 0.319 & 0.178\\
\cmidrule{1-6}
 & Ingroup-Disagree & -1.287 & 0.873 & 0.141 & 0.093\\

 & Outgroup-Agree & 1.339 & 0.836 & 0.110 & 0.083\\

\multirow[t]{-3}{*}{\raggedright\arraybackslash Attitude Polarization} & Outgroup-Disagree & -0.559 & 0.875 & 0.523 & 0.217\\
\bottomrule
\end{tabular}}
\end{table}

\begin{table} 
\centering
\caption{Treatment effects on effects on discussion-related outcomes, excluding pilot study participants (FDR-adjusted q-values)}
\label{tab:si-robustness-check2-discussion-FDR}
\centering
\resizebox{\ifdim\width>\linewidth\linewidth\else\width\fi}{!}{
\fontsize{9}{11}\selectfont
\begin{tabular}[t]{llrrrr}
\toprule
DV & Treatment & Estimate & SE & p-value & q-value\\
\midrule
 & Ingroup-Disagree & 0.065 & 0.187 & 0.729 & 0.089\\

 & Outgroup-Agree & 0.369 & 0.177 & 0.037 & 0.010\\

\multirow[t]{-3}{*}{\raggedright\arraybackslash Discussion Quality} & Outgroup-Disagree & 0.503 & 0.176 & 0.004 & 0.002\\
\cmidrule{1-6}
 & Ingroup-Disagree & -0.219 & 0.062 & <.001 & <.001\\

 & Outgroup-Agree & -0.604 & 0.065 & <.001 & <.001\\

\multirow[t]{-3}{*}{\raggedright\arraybackslash Future Willingness} & Outgroup-Disagree & -0.832 & 0.065 & <.001 & <.001\\
\cmidrule{1-6}
 & Ingroup-Disagree & -0.812 & 0.066 & <.001 & <.001\\

 & Outgroup-Agree & -0.269 & 0.059 & <.001 & <.001\\

\multirow[t]{-3}{*}{\raggedright\arraybackslash Discussion Satisfaction} & Outgroup-Disagree & -0.702 & 0.063 & <.001 & <.001\\
\bottomrule
\end{tabular}}
\end{table}

\subsection{Mixed Effects Model}

\begin{figure}
    \centering
    \begin{subfigure}{\linewidth}
        \centering
        \includegraphics[width=\linewidth]{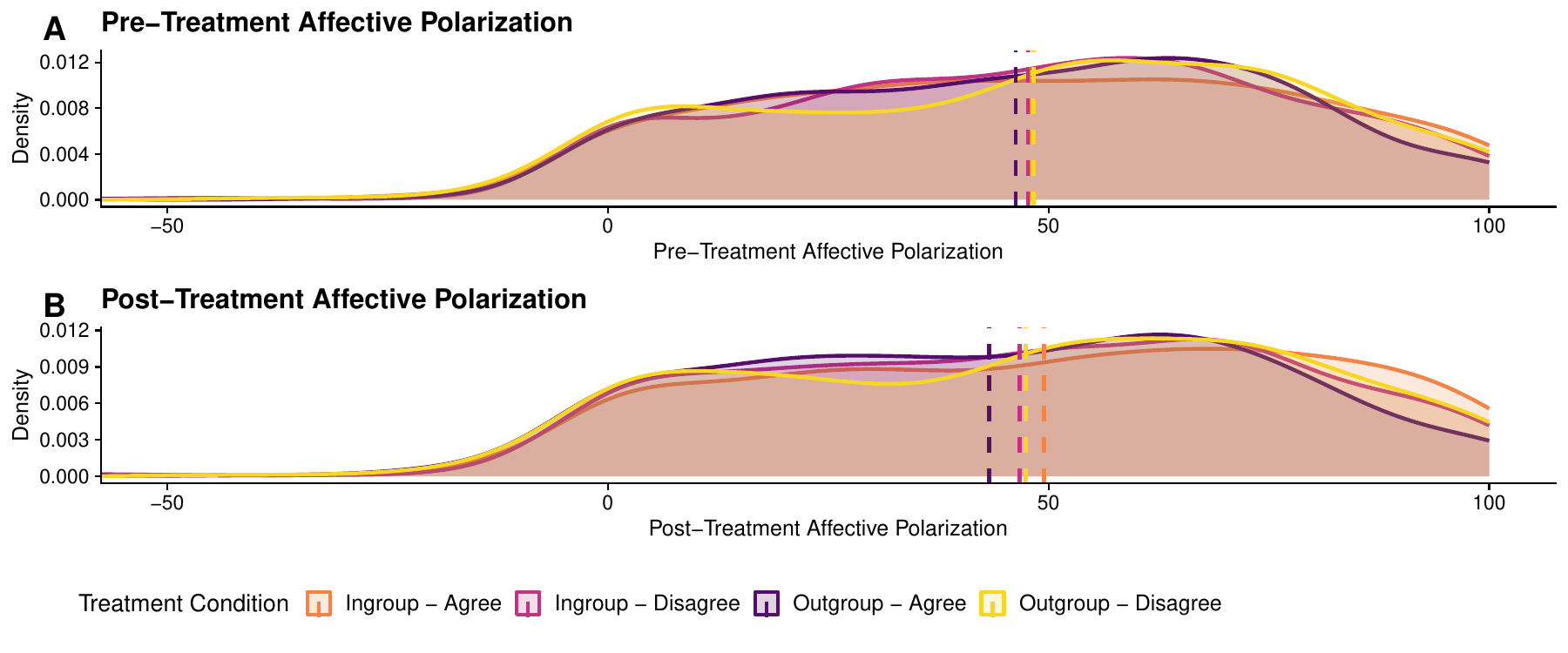}
    \end{subfigure}
    \begin{subfigure}{\linewidth} 
        \centering
        \includegraphics[width=\linewidth]{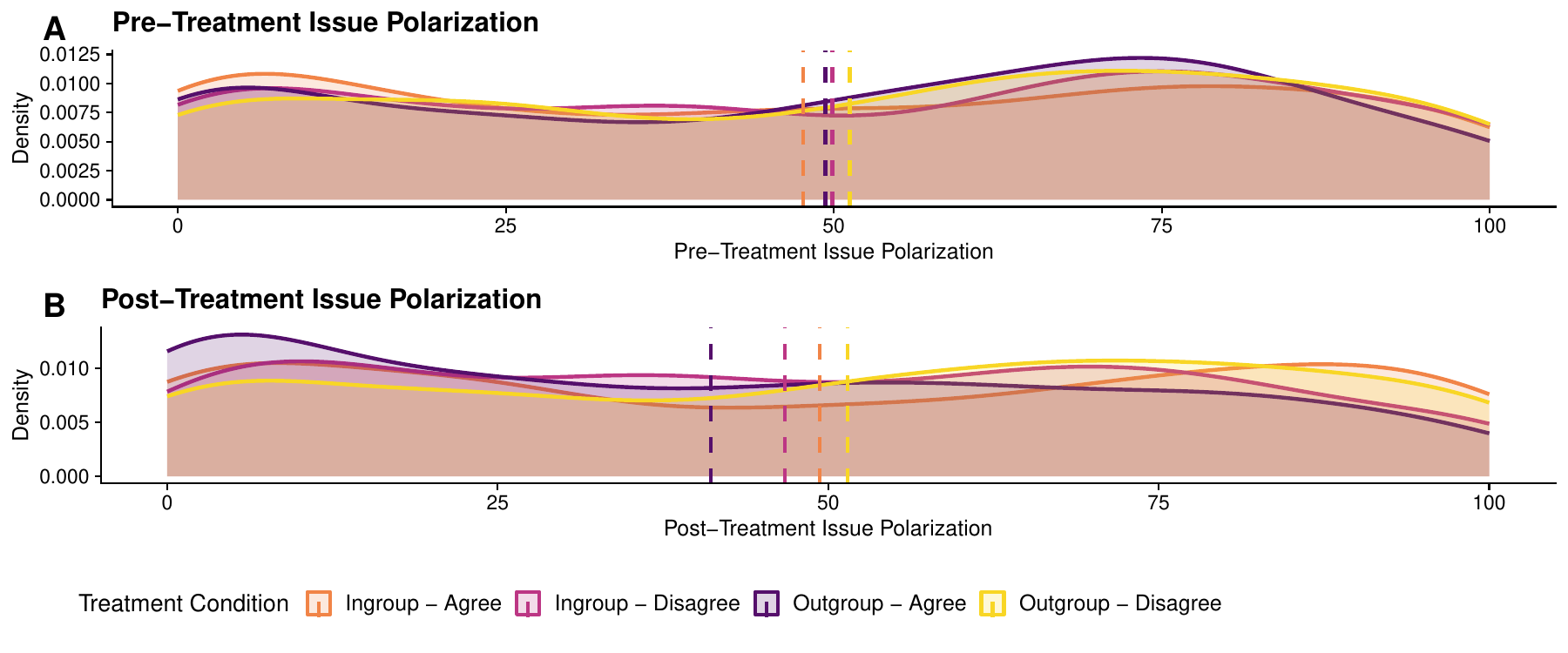}
    \end{subfigure}
    \caption{Pre- and post-treatment changes in polarization outcomes. Dashed lines indicate sample means.}
    \label{fig:prepost}
\end{figure}

Figure~\ref{fig:prepost} shows raw sample means to provide an overview of polarization outcomes across treatment groups and time points. Baseline levels of polarization outcomes at pre-treatment period are broadly comparable across conditions, while post-treatment means diverge. (For the full table, see Table~\ref{tab:si-mixed-effects-model}.) 

We conducted a robustness check using linear mixed-effects models for the outcomes that were measured by means a pre--post comparison (i.e., affective polarization and perceived issue polarization only, but excluding attitude polarization). These models include participant-level random intercepts and adjust for block indicators and covariates selected through LASSO.

\begin{figure}
    \centering
    \includegraphics[width=\linewidth]{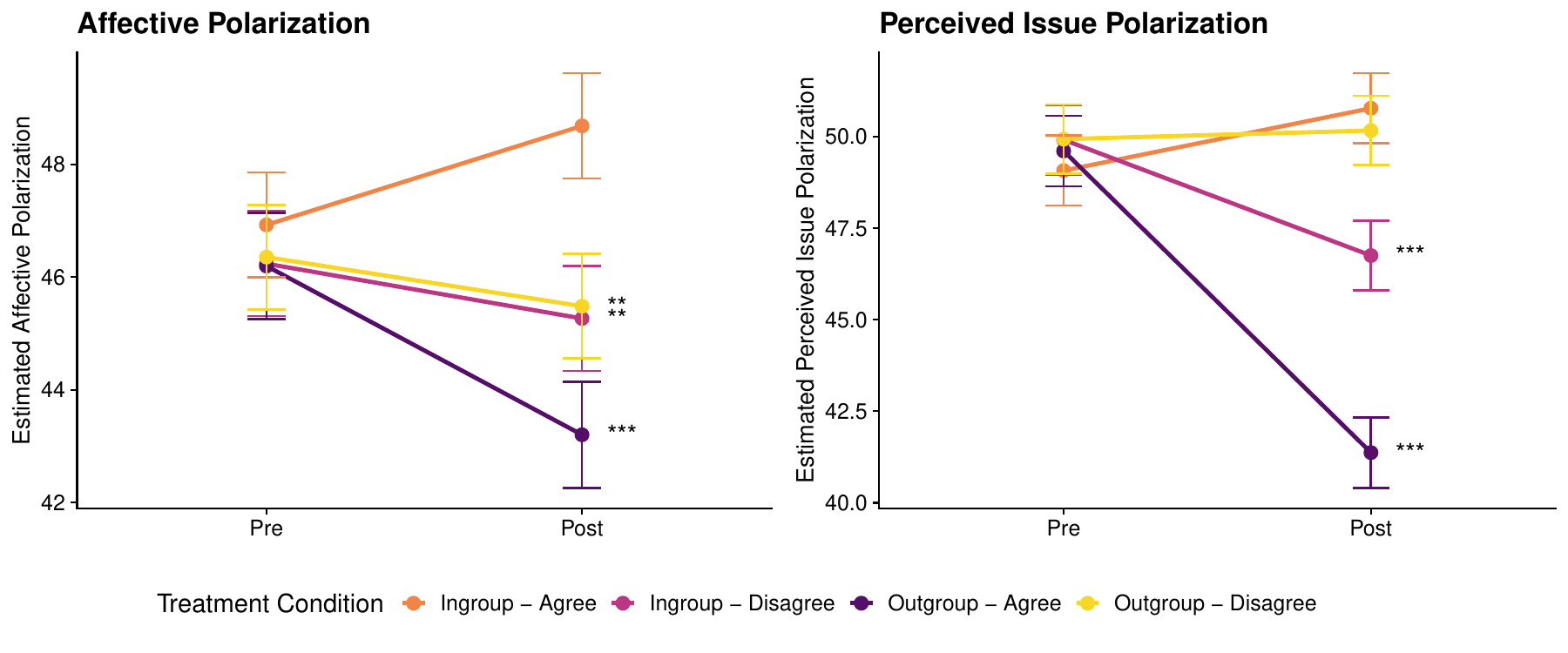}
    \caption{\textbf{Pre–Post Changes in Affective Polarization and Perceived Issue Polarization by Treatment Condition}. Estimated marginal means (EMMs) from linear mixed-effects models are shown for affective polarization (left panel) and perceived issue polarization (right panel) across pre- and post-treatment time points. Models include fixed effects for treatment condition, time (pre vs. post), their interaction, block indicators, and LASSO-selected covariates, with random intercepts for participants. Error bars represent $\pm$1 standard error. Asterisks indicate treatment groups whose polarization increased or decreased more from Pre to Post than in the baseline `ingroup-agree' group.}
    \label{fig:mixed_effects}
\end{figure}

The results of the mixed-effects models are fully consistent with our main analyses. Across both affective polarization and perceived issue polarization, the direction and statistical significance of treatment effects closely mirror those obtained from the primary OLS models. Specifically, the expectation-challenging conditions
(Outgroup Agree and Ingroup Disagree) show greater reductions in polarization relative to the Ingroup Agree baseline. This robustness check demonstrates that the substantive conclusions are not sensitive to modeling choices, and hold when
accounting for within-person correlation across pre- and post-treatment measurements.

\section{Heterogeneous Treatment Effects}
\label{sec:si-hte}

This section explores whether treatment effects on primary outcomes vary across subgroups. Specifically, we examine heterogeneity by discussion topic and by participant partisan identity. 

\subsection{By Discussion Topic}
\label{si:treatment-by-topic}

\begin{figure}
    \centering
    \includegraphics[width=0.8\linewidth]{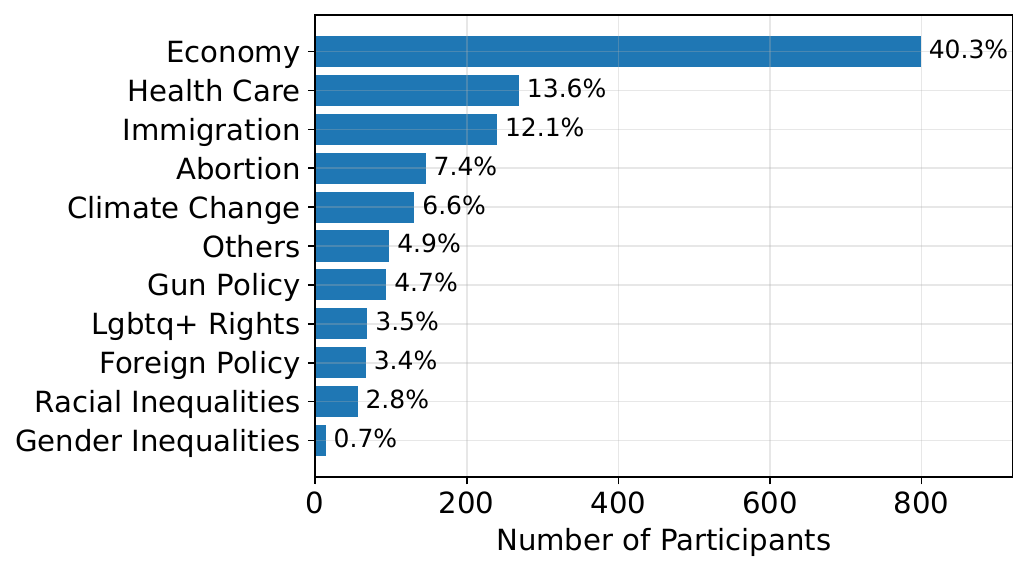}
    \caption{Distribution of chosen topics across participants}
    \label{fig:topic_dist}
\end{figure}

We examine whether treatment effects vary by discussion topics. Morally charged issues, such as gun control and immigration, are more likely to
trigger defensive identity responses~\cite{jung2025varieties}, and are also more polarizing than others~\cite{bor2023quantifying}. On the other hand, economic issues are typically easier to talk about, as they are often considered as broadly-shared concerns~\cite{pew2019topic}. In our sample, `Economy' was the most frequently chosen topic among participants (see Figure~\ref{fig:topic_dist}). Thus, to test whether the effects of expectation-challenging conversations are moderated by topic selection (economy ($n=799$) vs. non-economy ($n=1{,}183$)), we choose the economy as a comparatively less contentious topic (i.e., compared to all other topics that were available for selection), and estimate separate models for economic and non-economic conversations.  

To assess topic-level heterogeneity, we estimate OLS models that include interaction terms between treatment condition and a binary topic indicator (Economy = 1 if the participant selected an economic topic, 0 otherwise), along with block fixed effects and LASSO-selected covariates. This specification allows treatment effects to vary across topic types, with the Ingroup Agree condition serving as the reference category. Subgroup-specific treatment effects are computed as linear combinations of the corresponding main and interaction terms. Statistical significance is evaluated based on these subgroup-specific estimates, with false discovery rate adjustment using the Benjamini–Krieger–Yekutieli (BKY) procedure.

\begin{figure}
    \centering
    \includegraphics[width=0.8\linewidth]{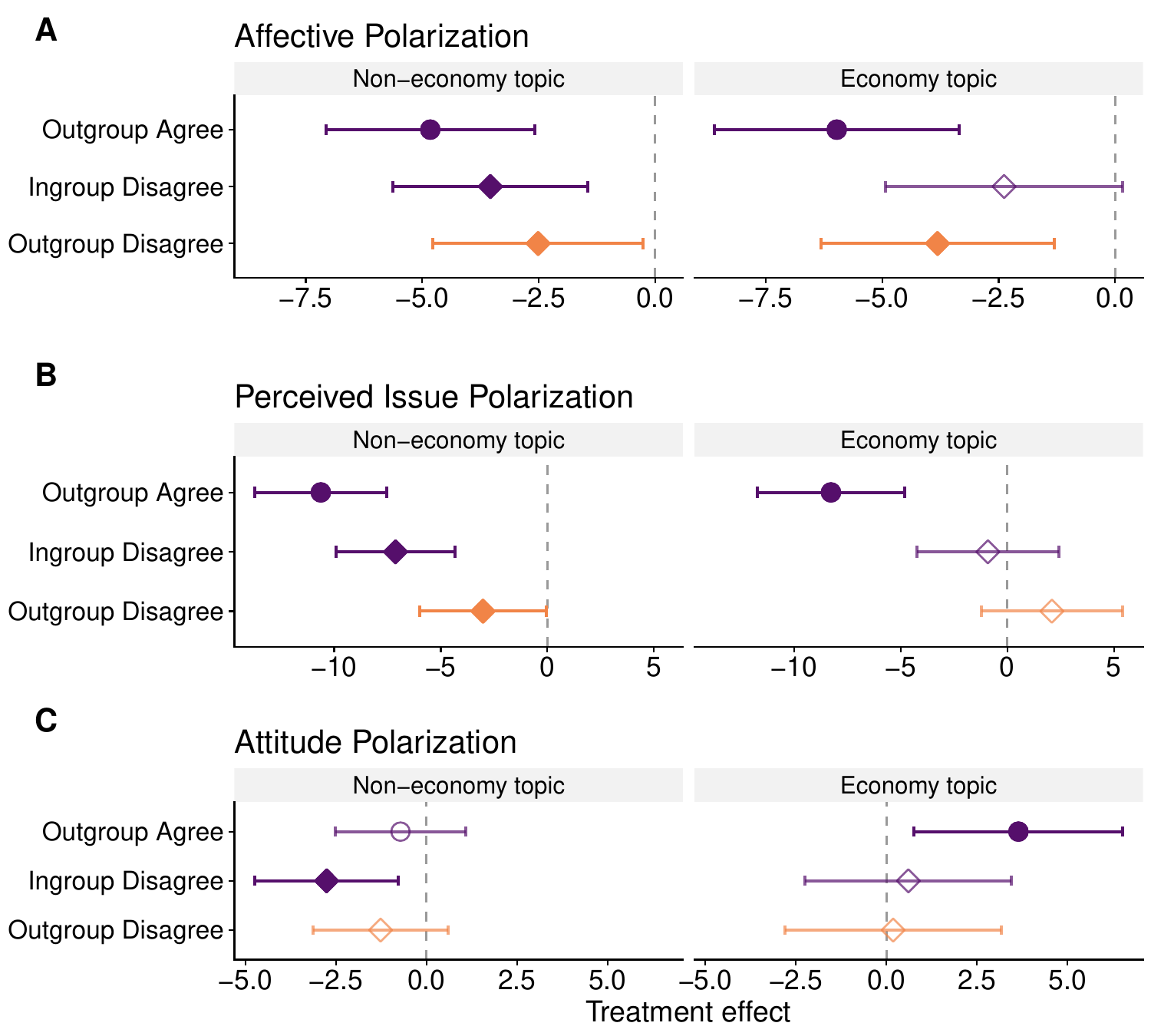}
    \caption{Heterogeneous treatment effects by topic on political polarization. Each panel displays estimated treatment effects (relative to Ingroup Agree) with 95\% confidence intervals, separately for non-economy (left; $n=1,183$) and economy (right; $n=799$) topics. Estimates are obtained from OLS models with block fixed effects, LASSO-selected covariate adjustment, and Treatment $\times$ Topic interactions. For economic topics, treatment effects are computed as linear combinations of the main and interaction terms. Statistical significance is assessed based on these subgroup-specific effects, with false discovery rate (FDR) adjustment (BKY). Non-significant results ($q > .05$) are indicated by lighter colors and unfilled markers.
    Cell sizes were as follows: for non-economy topics, Ingroup-Agree ($n = 287$), Ingroup-Disagree ($n = 308$), Outgroup-Agree ($n = 298$), and Outgroup-Disagree ($n = 290$); for economy topics, the corresponding counts were $n = 201$, $193$, $198$, and $207$, respectively.}
\label{fig:HTE_by_topic}
\end{figure}

For more contentious non-economy topics, the expectation-challenging conditions reduce both affective polarization (Figure~\ref{fig:HTE_by_topic}(A); Ingroup-Disagree: $b = -3.54$ [$-5.63$, $-1.44$], $p < .001$, $q = .003$; Outgroup-Agree: $b = -4.83$ [$-7.07$, $-2.58$], $p < .001$, $q < .001$) and perceived issue polarization (Figure~\ref{fig:HTE_by_topic}(B); Ingroup-Disagree: $b = -7.12$ [$-9.90$, $-4.35$], $p < .001$, $q < .001$; Outgroup-Agree: $b = -10.63$ [$-13.73$, $-7.53$], $p < .001$, $q < .001$). 
By contrast, when the discussion centers on the economy, these effects weaken substantially, especially for the Ingroup-Disagree condition, which is no longer statistically distinguishable from zero for affective polarization ($b = -2.38$ [$-4.93$, $0.16$], $p = .067$, $q = .063$) or perceived issue polarization ($b = -0.91$ [$-4.25$, $2.43$], $p = .593$, $q = .300$).

This pattern suggests that different types of expectation-challenging encounters may operate through distinct psychological mechanism. The effects of Outgroup Agree condition consistently strong effect across topics, suggesting that surprises stemming from cross-partisan agreement are less dependent on the issue level of politicization. By comparison, ingroup disagreement is more sensitive to topic context. Its depolarizing effect emerges primarily for contentious, identity-relevant issues, but becomes attenuated and statistically indistinguishable from zero in the economic domain for both outcomes. One possible explanation is that disagreement from a co-partisan on morally charged topics creates a stronger violation of expectations, prompting greater affective updating. By contrast, when the topic is less identity-threatening, such as the economy, this mechanism is weaker.

Consistent with the main results, expectation-challenging conversations show inconsistent or null effects on attitude polarization, even when decomposed by topic. In the economic domain, exposure to outgroup agreement increases issue polarization, whereas ingroup disagreement leads individuals to adopt more moderate positions (Figure~\ref{fig:HTE_by_topic}(C)), possibly due to the different mechanisms hypothesized above.

\begin{figure}
    \centering
    \includegraphics[width=1\linewidth]{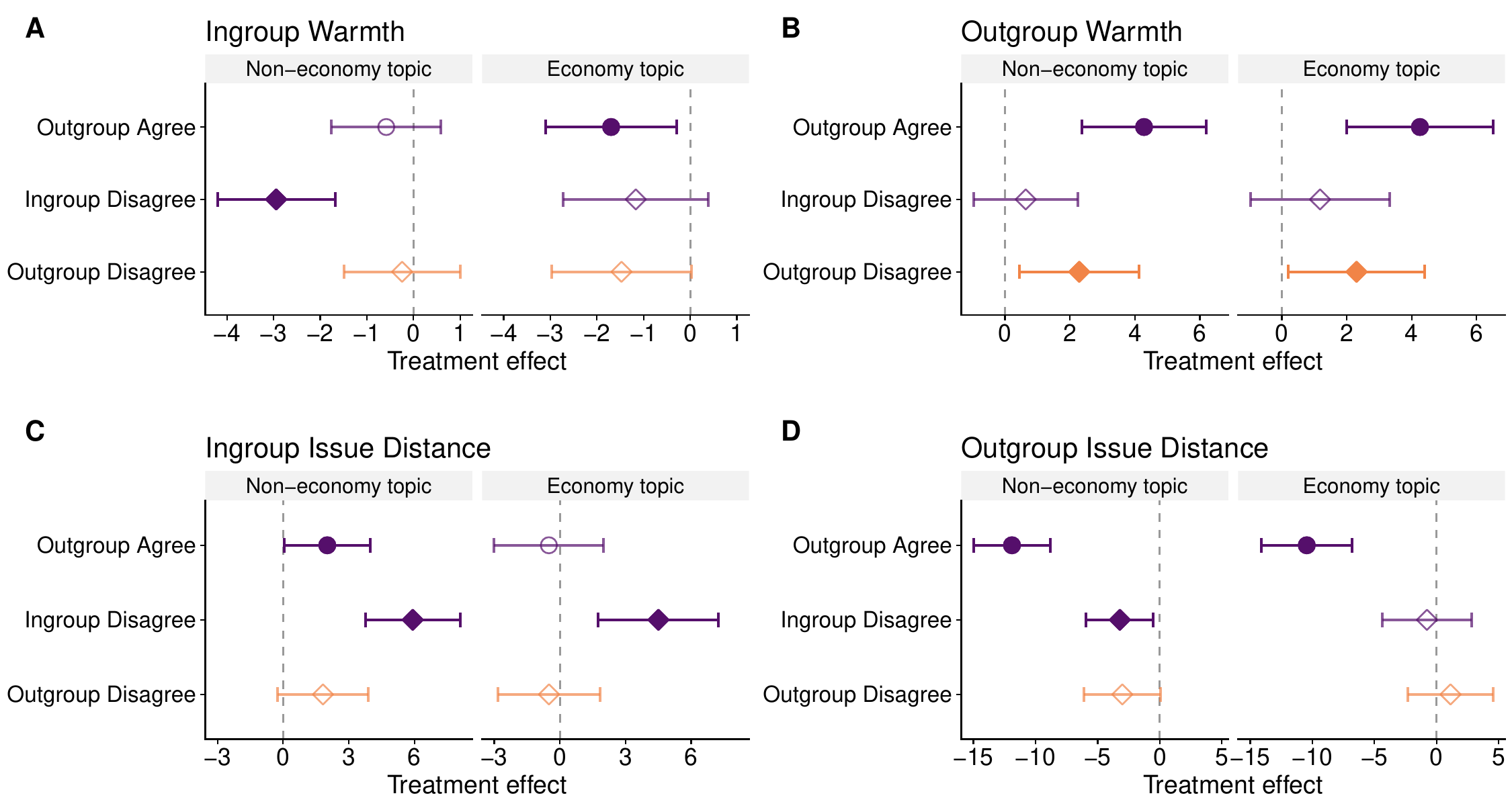}
    \caption{Heterogeneous treatment effects by topic on political polarization (decomposed). Each panel displays estimated treatment effects (relative to Ingroup Agree) with 95\% confidence intervals, separately for non-economy (left) and economy (right) topics. Estimates are obtained from OLS models with block fixed effects, LASSO-selected covariate adjustment, and Treatment $\times$ Topic interactions. For economic topics, treatment effects are computed as linear combinations of the main and interaction terms. Statistical significance is assessed based on these subgroup-specific effects, with false discovery rate (FDR) adjustment (BKY). Non-significant results ($q > .05$) are indicated by lighter colors and unfilled markers.}
\label{fig:HTE_by_topic2}
\end{figure}

\begin{table}
\centering
\caption{Heterogeneous Treatment Effect by Topic (FDR-adjusted q-values)}
\label{tab:si-hte-topic-polarization}
\centering
\resizebox{\ifdim\width>\linewidth\linewidth\else\width\fi}{!}{
\fontsize{9}{11}\selectfont
\begin{tabular}[t]{lllrrrr}
\toprule
DV & Treatment & Group & Estimate & SE & p-value & q-value\\
\midrule
 & Ingroup-Disagree & Non-economy topic & -3.538 & 1.069 & $<.001$ & 0.003\\

 & Outgroup-Agree & Non-economy topic & -4.828 & 1.145 & $<.001$ & $<.001$\\

 & Outgroup-Disagree & Non-economy topic & -2.512 & 1.150 & 0.029 & 0.039\\

 & Ingroup-Disagree & Economy topic & -2.383 & 1.299 & 0.067 & 0.063\\

 & Outgroup-Agree & Economy topic & -5.975 & 1.337 & $<.001$ & $<.001$\\

\multirow[t]{-6}{*}{\raggedright\arraybackslash Affective Polarization} & Outgroup-Disagree & Economy topic & -3.812 & 1.278 & 0.003 & 0.007\\
\cmidrule{1-7}
 & Ingroup-Disagree & Non-economy topic & -7.123 & 1.417 & $<.001$ & $<.001$\\

 & Outgroup-Agree & Non-economy topic & -10.628 & 1.580 & $<.001$ & $<.001$\\

 & Outgroup-Disagree & Non-economy topic & -3.021 & 1.515 & 0.046 & 0.049\\

 & Ingroup-Disagree & Economy topic & -0.910 & 1.704 & 0.593 & 0.300\\

 & Outgroup-Agree & Economy topic & -8.270 & 1.769 & $<.001$ & $<.001$\\

\multirow[t]{-6}{*}{\raggedright\arraybackslash Perceived Issue Polarization} & Outgroup-Disagree & Economy topic & 2.093 & 1.693 & 0.217 & 0.131\\
\cmidrule{1-7}
 & Ingroup-Disagree & Non-economy topic & -2.759 & 1.012 & 0.006 & 0.013\\

 & Outgroup-Agree & Non-economy topic & -0.718 & 0.917 & 0.433 & 0.240\\

 & Outgroup-Disagree & Non-economy topic & -1.268 & 0.948 & 0.181 & 0.119\\

 & Ingroup-Disagree & Economy topic & 0.607 & 1.453 & 0.676 & 0.315\\

 & Outgroup-Agree & Economy topic & 3.646 & 1.469 & 0.013 & 0.023\\

\multirow[t]{-6}{*}{\raggedright\arraybackslash Attitude Polarization} & Outgroup-Disagree & Economy topic & 0.187 & 1.526 & 0.902 & 0.388\\
\cmidrule{1-7}
 & Ingroup-Disagree & Non-economy topic & -2.944 & 0.643 & $<.001$ & $<.001$\\

 & Outgroup-Agree & Non-economy topic & -0.584 & 0.600 & 0.331 & 0.184\\

 & Outgroup-Disagree & Non-economy topic & -0.244 & 0.635 & 0.701 & 0.315\\

 & Ingroup-Disagree & Economy topic & -1.167 & 0.794 & 0.142 & 0.103\\

 & Outgroup-Agree & Economy topic & -1.696 & 0.714 & 0.018 & 0.028\\

\multirow[t]{-6}{*}{\raggedright\arraybackslash Ingroup Warmth} & Outgroup-Disagree & Economy topic & -1.474 & 0.766 & 0.055 & 0.054\\
\cmidrule{1-7}
 & Ingroup-Disagree & Non-economy topic & 0.638 & 0.822 & 0.438 & 0.240\\

 & Outgroup-Agree & Non-economy topic & 4.281 & 0.974 & $<.001$ & $<.001$\\

 & Outgroup-Disagree & Non-economy topic & 2.286 & 0.938 & 0.015 & 0.025\\

 & Ingroup-Disagree & Economy topic & 1.177 & 1.093 & 0.282 & 0.159\\

 & Outgroup-Agree & Economy topic & 4.254 & 1.146 & $<.001$ & $<.001$\\

\multirow[t]{-6}{*}{\raggedright\arraybackslash Outgroup Warmth} & Outgroup-Disagree & Economy topic & 2.302 & 1.071 & 0.032 & 0.040\\
\cmidrule{1-7}
 & Ingroup-Disagree & Non-economy topic & 5.921 & 1.105 & $<.001$ & $<.001$\\

 & Outgroup-Agree & Non-economy topic & 2.016 & 1.001 & 0.044 & 0.049\\

 & Outgroup-Disagree & Non-economy topic & 1.819 & 1.054 & 0.084 & 0.071\\

 & Ingroup-Disagree & Economy topic & 4.501 & 1.402 & 0.001 & 0.004\\

 & Outgroup-Agree & Economy topic & -0.505 & 1.275 & 0.692 & 0.315\\

\multirow[t]{-6}{*}{\raggedright\arraybackslash Ingroup Issue Distance} & Outgroup-Disagree & Economy topic & -0.499 & 1.190 & 0.675 & 0.315\\
\cmidrule{1-7}
 & Ingroup-Disagree & Non-economy topic & -3.217 & 1.378 & 0.020 & 0.029\\

 & Outgroup-Agree & Non-economy topic & -11.896 & 1.584 & $<.001$ & $<.001$\\

 & Outgroup-Disagree & Non-economy topic & -3.016 & 1.572 & 0.055 & 0.054\\

 & Ingroup-Disagree & Economy topic & -0.770 & 1.840 & 0.675 & 0.315\\

 & Outgroup-Agree & Economy topic & -10.447 & 1.865 & $<.001$ & $<.001$\\

\multirow[t]{-6}{*}{\raggedright\arraybackslash Outgroup Issue Distance} & Outgroup-Disagree & Economy topic & 1.125 & 1.740 & 0.518 & 0.271\\
\bottomrule
\end{tabular}}
\begin{threeparttable}
\begin{tablenotes}[flushleft]
\footnotesize
\item \textit{Notes:} Entries report estimated treatment effects (relative to Ingroup Agree) from OLS models with block fixed effects, LASSO-selected covariates, and Treatment $\times$ Topic interactions. For economic topics, effects are computed as linear combinations of the main and interaction terms. $q$-values are adjusted for multiple testing using the Benjamini--Krieger--Yekutieli (BKY) procedure.
\end{tablenotes}
\end{threeparttable}
\end{table}

Figure~\ref{fig:HTE_by_topic2} further illustrates the moderating effects of topic selection by decomposing affective polarization into ingroup- and outgroup warmth, and perceived issue polarization into ingroup- and outgroup issue distance. 

\subsection{By Partisanship}
\label{sec:si-robustness-partisanship}

\begin{figure}
    \centering
    \includegraphics[width=0.8\linewidth]{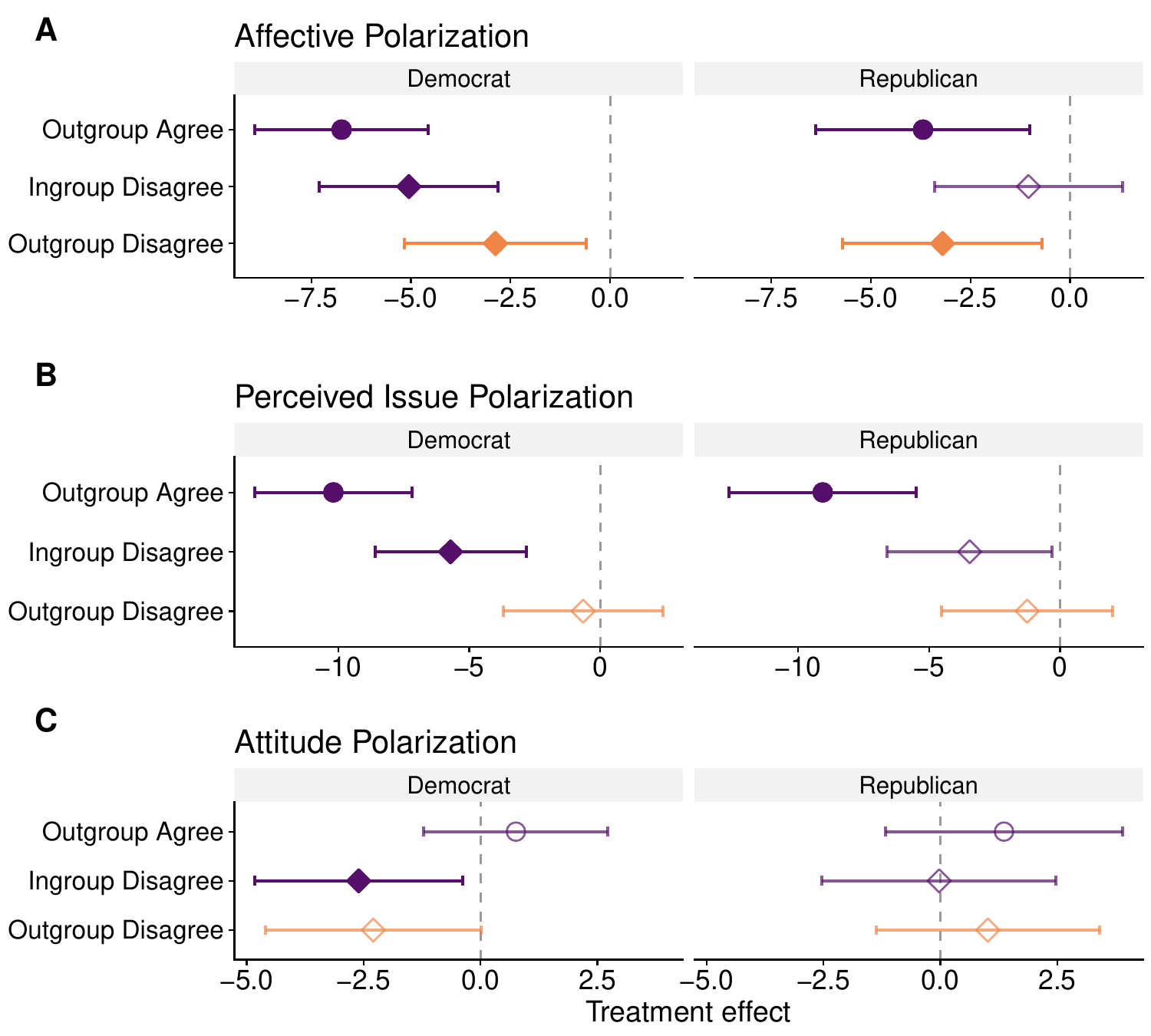}
    \caption{Heterogeneous treatment effects by partisanship on political polarization. Each panel displays estimated treatment effects (relative to Ingroup Agree) with 95\% confidence intervals, separately for self-identified Democrats (left) and Republicans (right). Estimates are obtained from OLS models with block fixed effects, LASSO-selected covariate adjustment, and Treatment $\times$ Partisanship interactions. For economic topics, treatment effects are computed as linear combinations of the main and interaction terms. Statistical significance is assessed based on these subgroup-specific effects, with false discovery rate (FDR) adjustment (BKY). Non-significant results ($q > .05$) are indicated by lighter colors and unfilled markers.
    Cell sizes were as follows: for Democrats, Ingroup-Agree ($n = 254$), Ingroup-Disagree ($n = 261$), Outgroup-Agree ($n = 263$), and Outgroup-Disagree ($n = 259$); for Republicans, Ingroup-Agree ($n = 234$), Ingroup-Disagree ($n = 240$), Outgroup-Agree ($n = 233$), and Outgroup-Disagree ($n = 238$).}
\label{fig:HTE_by_pid}
\end{figure}

We further examine whether treatment effects vary by partisanship, revealing a pronounced partisan asymmetry. Among Democrats, all expectation-challenging conditions produce significant reductions in affective polarization (Ingroup Disagree and Outgroup Agree) yield meaningful decreases in affective polarization (Figure~\ref{fig:HTE_by_pid}(A)) and perceived issue polarization (Figure~\ref{fig:HTE_by_pid}(B)). Among Republicans, by contrast, only Outgroup Agree condition produces statistically significant effects. This asymmetry suggests that the Democrats in our sample are more willing to revise their affective evaluations, while Republicans show greater resistance to updating feelings toward partisan groups following a single AI-mediated conversation. 

\begin{figure}
    \centering
    \includegraphics[width=\linewidth]{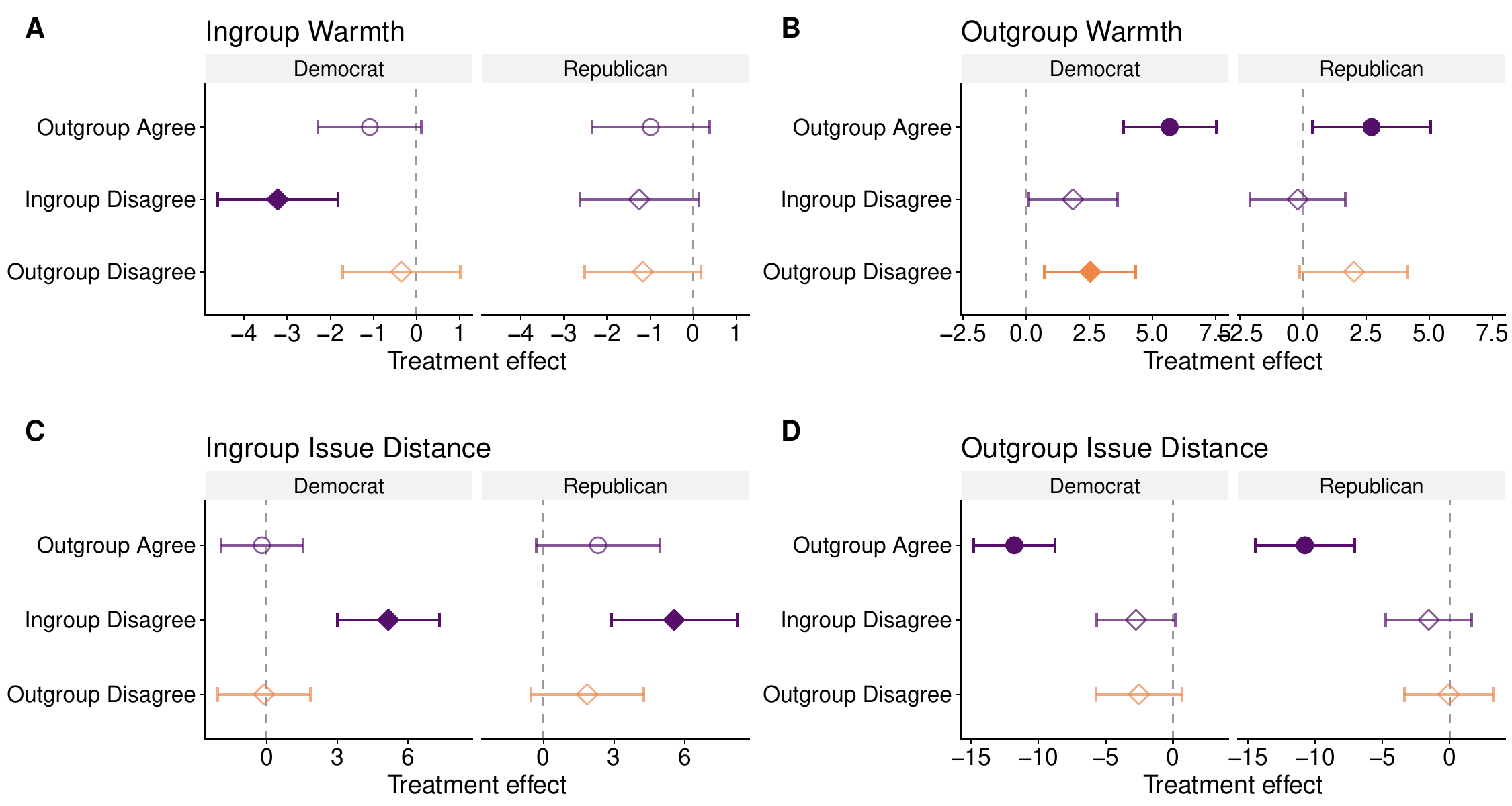}
    \caption{Heterogeneous treatment effects by partisanship on political polarization (decomposed). Each panel displays estimated treatment effects (relative to Ingroup Agree) with 95\% confidence intervals, separately for self-identified Democrats (left) and Republicans (right). Estimates are obtained from OLS models with block fixed effects, LASSO-selected covariate adjustment, and Treatment $\times$ Partisanship interactions. For economic topics, treatment effects are computed as linear combinations of the main and interaction terms. Statistical significance is assessed based on these subgroup-specific effects, with false discovery rate (FDR) adjustment (BKY). Non-significant results ($q > .05$) are indicated by lighter colors and unfilled markers.}
\label{fig:HTE_by_pid2}
\end{figure}

\begin{table} 
\centering
\caption{Heterogeneous Treatment Effect by Partisanship (FDR-adjusted q-values)}
\label{tab:si-hte-pid-polarization}
\centering
\resizebox{\ifdim\width>\linewidth\linewidth\else\width\fi}{!}{
\fontsize{9}{11}\selectfont
\begin{tabular}[t]{lllrrrr}
\toprule
DV & Treatment & Group & Estimate & SE & p-value & q-value\\
\midrule
 & Ingroup-Disagree & Democrat & -5.050 & 1.144 & $<.001$ & $<.001$\\

 & Outgroup-Agree & Democrat & -6.740 & 1.111 & $<.001$ & $<.001$\\

 & Outgroup-Disagree & Democrat & -2.880 & 1.165 & 0.013 & 0.027\\

 & Ingroup-Disagree & Republican & -1.044 & 1.206 & 0.386 & 0.235\\

 & Outgroup-Agree & Republican & -3.689 & 1.371 & 0.007 & 0.018\\

\multirow[t]{-6}{*}{\raggedright\arraybackslash Affective Polarization} & Outgroup-Disagree & Republican & -3.198 & 1.277 & 0.012 & 0.027\\
\cmidrule{1-7}
 & Ingroup-Disagree & Democrat & -5.711 & 1.473 & $<.001$ & $<.001$\\

 & Outgroup-Agree & Democrat & -10.181 & 1.530 & $<.001$ & $<.001$\\

 & Outgroup-Disagree & Democrat & -0.648 & 1.550 & 0.676 & 0.312\\

 & Ingroup-Disagree & Republican & -3.447 & 1.605 & 0.032 & 0.051\\

 & Outgroup-Agree & Republican & -9.057 & 1.827 & $<.001$ & $<.001$\\

\multirow[t]{-6}{*}{\raggedright\arraybackslash Perceived Issue Polarization} & Outgroup-Disagree & Republican & -1.254 & 1.662 & 0.450 & 0.240\\
\cmidrule{1-7}
 & Ingroup-Disagree & Democrat & -2.605 & 1.133 & 0.022 & 0.040\\

 & Outgroup-Agree & Democrat & 0.757 & 1.004 & 0.451 & 0.240\\

 & Outgroup-Disagree & Democrat & -2.292 & 1.177 & 0.052 & 0.070\\

 & Ingroup-Disagree & Republican & -0.026 & 1.276 & 0.984 & 0.438\\

 & Outgroup-Agree & Republican & 1.359 & 1.294 & 0.294 & 0.186\\

\multirow[t]{-6}{*}{\raggedright\arraybackslash Attitude Polarization} & Outgroup-Disagree & Republican & 1.014 & 1.221 & 0.406 & 0.235\\
\cmidrule{1-7}
 & Ingroup-Disagree & Democrat & -3.225 & 0.714 & $<.001$ & $<.001$\\

 & Outgroup-Agree & Democrat & -1.085 & 0.613 & 0.077 & 0.084\\

 & Outgroup-Disagree & Democrat & -0.354 & 0.697 & 0.611 & 0.294\\

 & Ingroup-Disagree & Republican & -1.257 & 0.704 & 0.074 & 0.084\\

 & Outgroup-Agree & Republican & -0.988 & 0.697 & 0.157 & 0.135\\

\multirow[t]{-6}{*}{\raggedright\arraybackslash Ingroup Warmth} & Outgroup-Disagree & Republican & -1.172 & 0.687 & 0.088 & 0.089\\
\cmidrule{1-7}
 & Ingroup-Disagree & Democrat & 1.847 & 0.904 & 0.041 & 0.061\\

 & Outgroup-Agree & Democrat & 5.682 & 0.938 & $<.001$ & $<.001$\\

 & Outgroup-Disagree & Democrat & 2.534 & 0.922 & 0.006 & 0.016\\

 & Ingroup-Disagree & Republican & -0.210 & 0.960 & 0.827 & 0.381\\

 & Outgroup-Agree & Republican & 2.715 & 1.193 & 0.023 & 0.040\\

\multirow[t]{-6}{*}{\raggedright\arraybackslash Outgroup Warmth} & Outgroup-Disagree & Republican & 2.015 & 1.091 & 0.065 & 0.080\\
\cmidrule{1-7}
 & Ingroup-Disagree & Democrat & 5.171 & 1.109 & $<.001$ & $<.001$\\

 & Outgroup-Agree & Democrat & -0.205 & 0.892 & 0.818 & 0.381\\

 & Outgroup-Disagree & Democrat & -0.117 & 1.010 & 0.908 & 0.419\\

 & Ingroup-Disagree & Republican & 5.553 & 1.360 & $<.001$ & $<.001$\\

 & Outgroup-Agree & Republican & 2.322 & 1.337 & 0.083 & 0.087\\

\multirow[t]{-6}{*}{\raggedright\arraybackslash Ingroup Issue Distance} & Outgroup-Disagree & Republican & 1.856 & 1.222 & 0.129 & 0.119\\
\cmidrule{1-7}
 & Ingroup-Disagree & Democrat & -2.749 & 1.489 & 0.065 & 0.080\\

 & Outgroup-Agree & Democrat & -11.791 & 1.552 & $<.001$ & $<.001$\\

 & Outgroup-Disagree & Democrat & -2.529 & 1.635 & 0.122 & 0.117\\

 & Ingroup-Disagree & Republican & -1.561 & 1.627 & 0.338 & 0.212\\

 & Outgroup-Agree & Republican & -10.759 & 1.879 & $<.001$ & $<.001$\\

\multirow[t]{-6}{*}{\raggedright\arraybackslash Outgroup Issue Distance} & Outgroup-Disagree & Republican & -0.081 & 1.678 & 0.962 & 0.438\\
\bottomrule
\end{tabular}}
\begin{threeparttable}
\begin{tablenotes}[flushleft]
\footnotesize
\item \textit{Notes:} Entries report estimated treatment effects (relative to Ingroup Agree) from OLS models with block fixed effects, LASSO-selected covariates, and Treatment $\times$ Topic interactions. For economic topics, effects are computed as linear combinations of the main and interaction terms. $q$-values are adjusted for multiple testing using the Benjamini--Krieger--Yekutieli (BKY) procedure.
\end{tablenotes}
\end{threeparttable}
\end{table}

Why this partisan difference emerges remains an open question and beyond the
scope of this paper. Several possibilities merit exploration for future studies:
Republicans may exhibit stronger prior commitment to partisan identities, making
affective evaluations less susceptible to a single interaction with an AI
chatbot; they may be more skeptical of AI-mediated information that contradicts
their expectations; or they may process expectation-violating information
differently due to distinct motivated reasoning patterns.

In terms of perceived issue polarization (Figure~\ref{fig:HTE_by_pid}(B)), among
Democrats, both Ingroup Disagree and Outgroup Agree conditions produce
significant reductions in perceived issue polarization relative to the Ingroup
Agree baseline. Among Republicans, only the Outgroup Agree condition yields a
statistically significant reduction. This pattern mirrors the asymmetry observed
for affective polarization: Republicans appear less
responsive to within-group disagreement when recalibrating their estimates of
partisan policy distance, whereas cross-partisan agreement proves sufficient to
shift perceived polarization for both groups. 

These analyses of heterogeneous treatment effects reconfirm that the two
expectation-challenging conditions operate through fundamentally distinct
psychological pathways: Ingroup Disagree corrects the belief that one's own
party is uniformly aligned and unequivocally right -- reducing both in-group
love and perceived in-group distance -- while Outgroup Agree corrects the
stereotype that the out-group is extreme and hostile, increasing out-group love
and reducing perceived out-group Distance. These condition-specific mechanisms,
however, are not equally accessible across all contexts. Both pathways are most
pronounced among Democrats and for non-economic, highly politicized topics.
Among Republicans and for economic topics, the effects are more muted, pointing
to boundary conditions under which AI-mediated intergroup contact may be less
sufficient to shift affective and cognitive assessments of political
polarization.

\section{Measurements}
\label{sec:si-measurement}

This section describes how we operationalized participant's open-ended texts
into outcome measures, including initial issue stance and discussion quality. 

\subsection{Measurement of Issue Topic and Initial Stances}
\label{sec:si-issue-measurement}

Participants were asked to select a salient political issue and indicate their
stance toward it. This personalized approach allows us to adapt to each
participant, as individuals choose topics they personally find important and may
emphasize different aspects or dimensions of those issues. As a result, it
provides a conservative test of whether our treatment has depolarizing effects,
even for issues of high personal importance that may be especially resistant to
attitude change.

However, even though this approach allows to capture the heterogeneity in
responses across participants, it comes at the risk of interpretability of the
responses, since the free-form text provided by participants may not be as
concise as a standard issue stance survey item. To address this, we followed
prior work~\cite{costello_durably_2024,velez_confronting_2024} and transformed
participants' original responses into clear and concise statements on the
selected issue and asked them to validate these statements. These validated
statements were then used to measure participants' own issue attitudes and
perceived partisan polarization in both pre- and post-treatment stages. 

The procedure consists of three steps. First, participants were elicited an
issue and asked to provide their response in an open-ended text box with a
minimum text length of 100 characters, to elicit sufficient details for the
summarization step to work as intended.

\begin{quote}
    Reflecting on the [\emph{chosen topic; e.g., climate change}] issue you selected, what is your position on it? Please write a few sentences explaining your view and the reasons behind it. Your response will help initiate a political conversation with a chatbot on this topic.
\end{quote}

Second, the participant's open-ended response was subsequently summarized by
GPT-4.1. Each response was sent to the model together with the following
instruction as the system prompt: ``Create a one-sentence summary of the
argument. The summary should start with `I believe' and only express one concept
regarding the issue at a time; ignore the rest of the argument if needed.'' Participants were shown the AI-generated summary of their response
in the form of a concise statement:

\begin{quote}
    Here is an AI-generated summary of what you wrote earlier: Statement: [LLM-summarized opinion]. Do you think this summary accurately represents your view?
\end{quote} 

Participants selected either ``Yes'' or No.'' If they selected ``No,'' they were
given the opportunity to revise the AI-generated summary to better reflect their
views, with the option to freely edit, rephrase, or rewrite the statement. The
revised version then replaced the original AI-generated summary.

Third, participants reported their attitudes with the validated statement on a
scale from 0 to 100 (where 0 indicates strong disagreement, and 100 indicates
strong agreement). They then indicated how strongly they believe Democrats and
Republicans would agree with the same statement using the same scale. These
measures capture perceived polarization on the issue.

\subsection{Discussion Quality Measurements}
\label{sec:si-discussion-quality}

This section describes the linguistic measures used to assess the quality of
political discussions and the rationale for their selection. Each measure is
normalized to have a mean of 0 and a standard deviation of 1 across all
participants. The normalized scores are then summed to construct a Discussion
Quality Index for each participant. We briefly motivate each measure and describe the way we operationalized in practice.

\paragraph{Gratitude} Signaling gratitude during conversations strengthens
social connections~\cite{algoe2012find}, indicates prosocial
intentions~\cite{you2006gratitude}, and prevents conflict escalation during
disagreement~\cite{yeomans2020conversational}. At message level, we measure the
ratio of gratitude-related terms (e.g., ``thank you,'' ``appreciate
it'')~\cite{bao2021alrightconversations} to all words.

\paragraph{Hedge Usage} Hedges signal intellectual humility and soften
disagreement~\cite{seckin2025identifying, zhang2018awryconversations}. At
message level, we take the ratio of tentative terms (e.g., ``perhaps,'' ``it
seems'') to total words.

\paragraph{Question-asking} Asking questions in online discussions increases
perceived responsiveness~\cite{huang2017doesn} and
trustworthiness~\cite{saltz2024re}. We operationalize this feature as the ratio
of sentences identified as questions to the total number of sentences in a post.

\paragraph{Toxicity} Uncivil language can exacerbate
polarization~\cite{anderson2014nasty}, erode trust, and diminish the perceived
credibility of news media~\cite{meltzer2015journalistic,
thorson2010credibility}. Therefore, uncivil language can reduce perceived
quality of the discussion. While there are various ways to detect
incivility~\cite{rossini2022beyond}, we use toxicity as our proxy, employing the
widely used Perspective API~\footnote{\url{https://www.perspectiveapi.com/}} to
flag messages as toxic or not. We then calculate the ratio of toxic messages to
the total message count. Since toxicity is negatively related to discussion
quality, we reverse this score so that its presence contributes negatively to
the index.

\paragraph{Responsivity} A high-quality discussion is characterized by active
engagement, defined as the extent to which a participant demonstrates uptake of
the previous turn rather than talking past it~\cite{scudder2020beyond,
stromer2007measuring}. We operationalize this as semantic similarity ---
specifically, the average cosine similarity between an AI's message and the
immediately preceding participant turns --- to capture how much the
participant's response was substantively grounded in the ongoing
conversation~\cite{hughes2024pursuit}.

\paragraph{Elaboration} Details and coherence in messages improve the perceived
quality of communication~\cite{crossley2016say, crossley2011text}. Incorporating
rhetorical devices and supporting evidence enhances persuasiveness and
acceptance~\cite{feng2023effects}. Therefore, more elaborated messages can be
expected to increase the perceived quality of discussions. We operationalize
\emph{elaboration} through lexical diversity~\cite{johansson2008lexical},
measured by the average number of unique words within key lexical categories
(nouns, verbs, adjectives, and adverbs) in all messages.

\paragraph{Polarity} Positive affect increases perceived communication
quality~\cite{han2018you} and leads to more positive engagement from others in
online discussions~\cite{chen2024online}. We used VADER~\cite{hutto2014vader} to
classify each message as positive or not, then calculated the proportion of
messages that conveyed positive sentiment to the total number of messages.

\section{Technical Implementation}
\label{sec:si-technical}

This section describes the technical details at each step of the experiment.
First, prior to the IRB application and data collection, we conducted validation
and guardrail procedures to ensure that the LLM model that we used for the
chatbot, GPT-4.1, did not generate toxic responses and that it adhered to the
assigned treatment conditions. Second, we provide details to the conversation
interface and the prompt to assign chatbot condition. Lastly, we report the
procedures used to handle technical failures during the experiment. 

\subsection{Chatbot Validation Prior to the Experiment}
\label{sec:si-llm-validation}

\subsubsection{Civility and Toxicity Guardrails}

We conducted a test to assess whether the chatbot produced toxic or uncivil
responses when interacting with participant inputs. In the experimental prompts,
the chatbot was explicitly instructed to ``maintain respect throughout the
conversation.'' This validation was designed to test the robustness of this
safety constraint by examining whether the chatbot would continue to produce
civil responses across both hostile and non-hostile conversational contexts.

To simulate such scenarios, we generated conversations in which participant
messages were either intentionally toxic or non-toxic. For each political topic
used in the experiment, we first generated four initial questions representing
prominent aspects of that topic. For each question, we simulated both Democratic
and Republican participants. For each simulated participant, we generated two
types of responses to the initial question: one toxic and one non-toxic. The
toxic responses were generated using prompts that instructed the model to
produce offensive or uncivil language consistent with partisan rhetoric, whereas
the non-toxic responses were generated without such instructions.

For each simulated participant response, the same message was then reused across
all four experimental conditions of the study design (ingroup–agree,
ingroup–disagree, outgroup–agree, outgroup–disagree) when generating the chatbot
reply. Holding the participant message constant ensured that differences in
chatbot responses across simulations were driven only by the experimental
condition rather than variation in the participant input. This procedure
yielded 64 simulated conversations per topic (4 questions × 2 participant
parties × 2 toxicity conditions × 4 experimental conditions).

The chatbot responses were then generated using the same prompting framework
used in the experiment. To evaluate whether the chatbot produced toxic outputs,
we analyzed all generated responses using the Perspective
API~\footnote{\url{https://www.perspectiveapi.com/}}, which assigns a toxicity
score between 0 and 1 indicating the likelihood that a message would be
perceived as toxic. Across all simulated conversations, the maximum toxicity
score observed for any chatbot response was 0.27. According to the Perspective
API documentation, messages with toxicity scores below 0.30 are categorized as
not toxic. Therefore, all chatbot responses generated in this validation
exercise fell within the non-toxic range, indicating that the prompt
instructions requiring the chatbot to maintain respectful language were robust
across both toxic and non-toxic participant inputs.

To ensure that the actual responses from the chatbot to participants followed a
similar pattern, we measured again toxicity of the actual responses of the
chatbot from the main study and found consistent results with this pre-study
analysis.

\subsubsection{Adherence to Treatment Condition}

To verify that the chatbot adhered to the experimental treatment conditions, we
conducted a validation exercise to assess whether the model consistently
maintained the assigned partisan identity and stance (agreement vs.
disagreement) in its responses. For this purpose, we randomly sampled 100
chatbot responses from the simulations generated for the toxicity validation
described above. These responses were drawn from simulations covering a wide
range of political topics used in the experiment.

One member of the research team manually reviewed each sampled chatbot response
and coded whether the response adhered to the assigned experimental condition.
Specifically, the coder evaluated whether the chatbot maintained the partisan
identity specified in the prompt and whether the response appropriately agreed
or disagreed with the participant’s stance according to the assigned treatment
condition.

Out of the 100 responses reviewed, 97 clearly adhered to the assigned
experimental condition and were judged satisfactory. In two cases, the chatbot
maintained the correct identity and stance but the response quality was
considered somewhat unsatisfactory (e.g., weakly articulated alignment with the
assigned stance). In one case, the chatbot appeared to deviate from the assigned
condition and partially broke character. Overall, these results indicate that
the prompting framework reliably generated responses consistent with the
intended treatment conditions.

\subsection{Conversation Interface}
\label{sec:si-interface}

\begin{figure}
  \centering
  \begin{subfigure}{0.49\textwidth}
\centering
    \includegraphics[width=\textwidth]{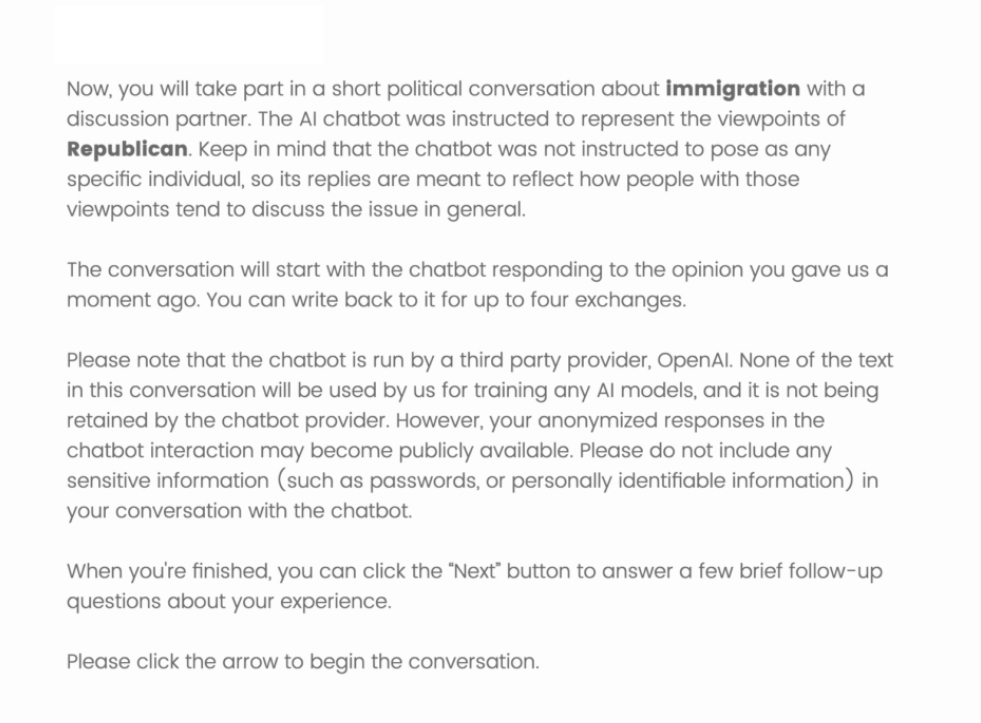}%
\caption{Instruction page}
\end{subfigure}\hfill
  \begin{subfigure}{0.49\textwidth}
\centering
    \includegraphics[width=\textwidth]{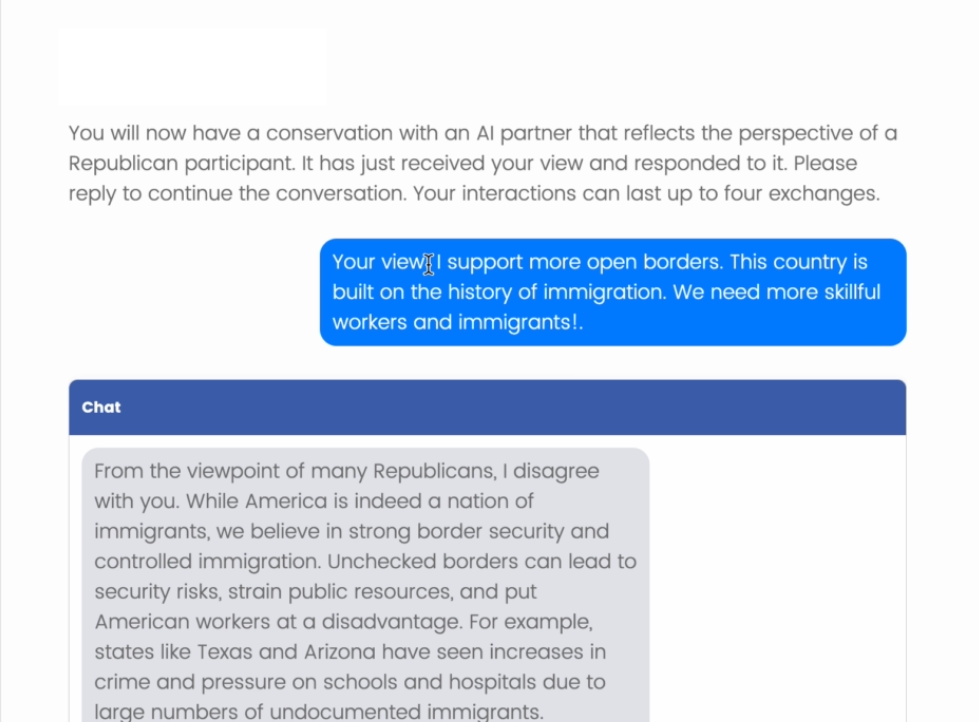}%
\caption{Initial stance from \emph{Issue Block} shown in blue bubble (``Your view: ...'')}
\end{subfigure}\\[0.1ex]

  \begin{subfigure}{0.49\textwidth}
\centering
    \includegraphics[width=\textwidth]{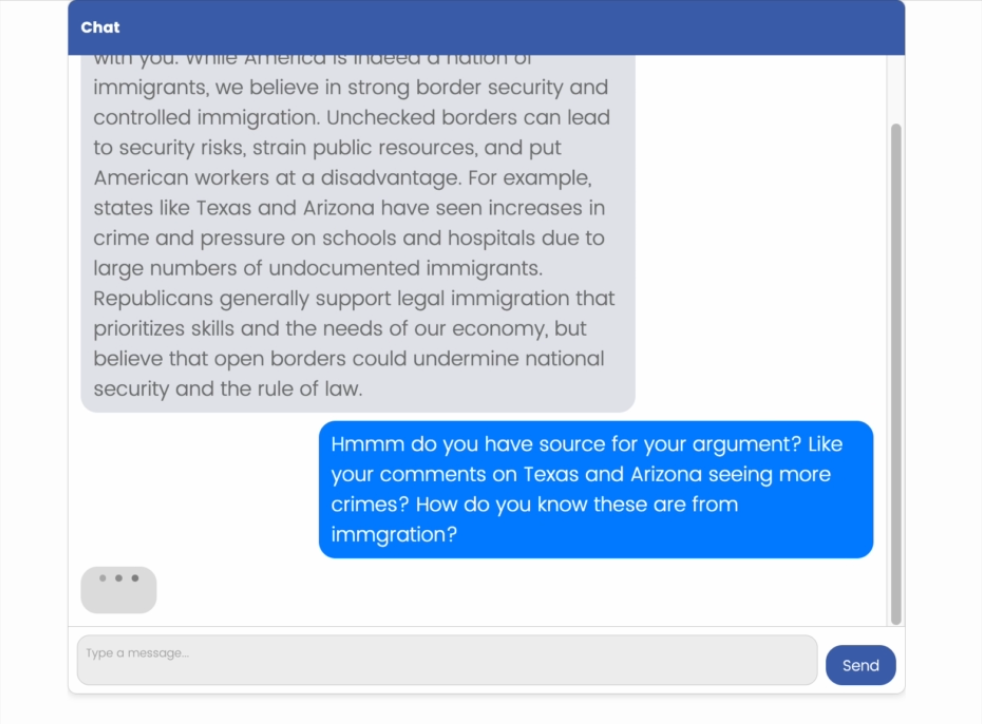}%
\caption{Chat interface showing typing indicator while waiting for a response}
\end{subfigure}\hfill
  \begin{subfigure}{0.49\textwidth}
\centering
    \includegraphics[width=\textwidth]{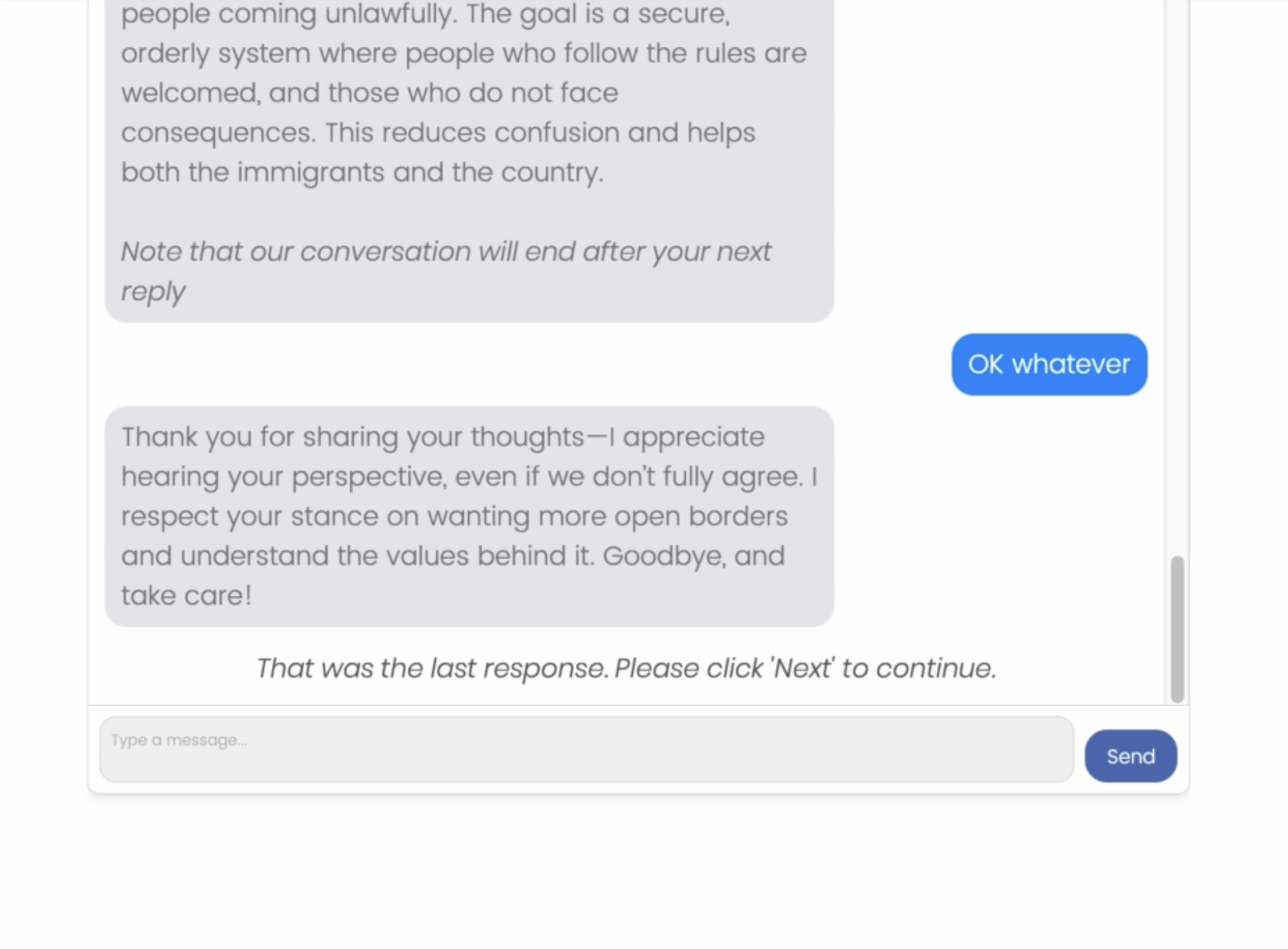}%
\caption{Final interaction and the instruction to move to the next page}
\end{subfigure}

  \caption{Example of the chat interface}
  \label{fig:survey_chat_interface}
\end{figure}

To deliver the treatment, participants were asked to engage in a structured
conversation with our chatbot through a text-based chat interface embedded
directly in Qualtrics (see Figure~\ref{fig:survey_chat_interface}). The
conversation was initiated by the chatbot with an opening message that responds
to the original (i.e. non-summarized) participant response on their stance on
the selected opinion issue (see Sec.~\ref{sec:si-issue-measurement} above).
Participants were prevented from leaving the interface until the opening message
from the chatbot had been displayed and they had sent a reply to it. From that
point on, participants could leave the chat at any time. They were allowed to
continue the conversation for up to three turns, for a maximum of up to four
back-and-forth exchanges with the chatbot. 

To improve the realism of the user experience, after participants sent their
messages the interface would briefly display an animated ellipsis indicator to
indicate that the chatbot was taking time to process a response.

LLM responses were generated by the OpenAI API (GPT-4.1) via OpenRouter. After
each participant reply, the full accumulated chat history (including prior user
and chatbot turns) was sent to the API as context to maintain coherence across
turns. The final participant reply triggered the display of an end-of-chat
notice, after which participants were prompted to click ``Next'' to continue.

\subsection{LLM Prompt}
\label{sec:si-llm-prompts}

For the main interaction, we used the following prompt template as a system
prompt to instruct GPT-4.1 to adopt a specific partisan identity and stance:

\begin{quote}
You are a discussion partner who holds the values and viewpoints typically associated with \{GROUP\}.\\
You will be having a conversation with a person who is \{PID\}.\\
\ [AGREEMENT\_INSTRUCTION]  Respond to the participant's stance by affirming it and sharing your relevant point of view.\\
\ [DISAGREEMENT\_INSTRUCTION] People often assume that in-group members share similar views, while out-group members hold more extreme and opposing positions. In reality, people tend to overestimate how much they agree with in-partisans and underestimate the extent of agreement with out-partisans. For instance, some Democrats support stricter immigration policies, while some Republicans acknowledge human-driven climate change and advocate for responsible measures. Change this misperception by engaging with the partner, even if you don't agree with their viewpoint.\\
You represent the stance that \{STANCE\}s with the participant's opinion on \{TOPIC\}.\\
Keep your responses short and concise. Present well-reasoned supporting arguments; use concrete examples when appropriate.
Maintain respect throughout the conversation and use simple language that an average person can understand.
Have a natural discussion; do not say goodbye unless explicitly instructed.
Only say goodbye and acknowledge the user's stance if the message is marked with $<$user-last-message:$>$.\\
The first (and only the first) response should start with:
`From the viewpoint of many \{GROUP\}s , I \{STANCE\} with you.'
\end{quote}

The placeholders \texttt{GROUP}, \texttt{PID}, \texttt{STANCE}, and
\texttt{TOPIC} were dynamically populated using embedded survey data.
\texttt{PID} denotes the participant’s self-reported partisan identity.
\texttt{TOPIC} corresponds to the salient political issue chosen by the
participant from a list of political topics in the survey instrument. The
combination of \texttt{GROUP} (ingroup vs. outgroup) and \texttt{STANCE} (agree
vs. disagree) was determined by the experimental condition.

For example, in the Ingroup Agree condition, \texttt{GROUP} matched the
participant’s party (\texttt{PID}) and \texttt{STANCE} was set to ``agree.'' In
contrast, in the Outgroup Agree condition, \texttt{GROUP} referred to the
opposing party (e.g., if the participant’s \texttt{PID} = Democrat, then
\texttt{GROUP} = Republican), while \texttt{STANCE} remained ``agree.'' The same
logic applied for the disagreement conditions (Ingroup Disagree,
\emph{outgroup–disagree}) with \texttt{STANCE} = ``disagree.''

We tested several versions of this prompt and qualitatively evaluated the
chatbot's responses, optimizing for outputs that conveyed partisanship and
stance in a clear and engaging fashion, while remaining concise and civil. The
above prompt, which is the result of this iterative process, includes explicit
instructions for how to agree or disagree with participants (marked by
[AGREEMENT\_INSTRUCTION] and [DISAGREEMENT\_INSTRUCTION] in the text). The disagreement instruction is longer than the agreement instruction because we added additional contextual information into the system prompt to steer the model toward disagreement\footnote{See \url{https://disagreebetter.us/how-to/}.}, rather than producing overly accommodating responses, especially when asked to disagree with an in-party member.

As mentioned above, for each API call during the discussion phase, the full
accumulated conversation history was sent to the model. This included (1)~the
system prompt, (2)~all prior participant messages as user-role entries, and (3)~all previous chatbot responses as assistant-role entries. 

\subsection{Handling Technical Failures during the Experiment}

Our Qualtrics survey required external API access for text generation and summarization. Any technical failures due to network glitches or other API errors (e.g. downtime, internal errors, etc.) could have prevented participants from completing the survey. To prevent disruptions to participants in these cases we included in our code a number of mechanisms to: detect these errors, attempt to retry the external API request that was causing the issue, and if none of these attempts would succeed, inform participants of the error and let them complete the survey, potentially with incomplete data.

If an error occurred at the stage of generating the summary of participants' initial issue stance (i.e., before the chat stage), participants were shown the
following message:

\begin{quote}
We are facing some trouble with the chatbot right now. Please click Next and
retry again in a few minutes. If the problem persists, please contact us at
$<$EMAIL$>$. Thank you for the patience!
\end{quote}

In these cases, since no summary could be displayed, participants were allowed
to complete the survey without a pre-treatment measurement of their stance on
the issue. As mentioned earlier (see Sec.~\ref{sec:si-technical-failure}), we
manually removed these cases from our analysis.

If a failure occurred during the chat, participants were instead shown:

\begin{quote}
We are facing some trouble with the chatbot right now. Please retry in a few
moments by typing your message and clicking the Send button. If the problem
persists, please contact us at $<$EMAIL$>$. Thank you for the patience!
\end{quote}

If the issue persisted after two retry attempts, participants were permitted to
complete the survey. We also manually removed these cases from our analysis.

\section{Deviations from Preregistration and Clarifications}
\label{sec:si-pre-reg}

We pre-registered the study on August 26, 2025~\cite{kim2025effects}. In this section we describe any deviations from the preregistered analysis plan and provide clarifications for any aspect of the pre-registration that had not been fully specified.

\subsection{Balance Check}

In the preregistered analysis plan, we initially specified that covariate
balance would be assessed across all four experimental conditions using ANOVA
for continuous variables, $\chi$-squared tests for categorical variables, and
$t$-tests for binary variables. In the final analysis, we slightly deviated from
this plan for greater interpretability. Specifically, we compared the baseline
condition (Ingroup Agree) with each of the other treatment conditions using
$t$-tests for continuous and binary variables, and computed regression-based
$F$-tests by regressing treatment indicators on the set of pre-treatment
covariates. Missing values were replaced with a constant, and missingness
indicators were included in the models. These deviations do not affect the
substantive interpretation of the results but provide a more transparent and
statistically consistent assessment of covariate balance.

\subsection{Statistical Model}
The preregistration specified that we would estimate Conditional Average Treatment Effects (CATEs) and differences-in-CATEs within a $2\times2$ factorial design. In the final analysis, we revised the estimands to focus on Average Treatment Effects (ATEs) defined as pairwise contrasts relative to the baseline condition (Ingroup Agree). Accordingly, we moved from an interaction-centered factorial parameterization to a specification that directly estimates simple contrasts between each treatment condition and the baseline category. 

This change was made to improve interpretability and better reflect the theoretical comparisons central to the study. Note that this change shifts the emphasis from interaction-based CATEs to baseline-referenced ATEs, but relies on the same experimental randomization and identification strategy. The substantive conclusions remain unchanged.

\subsection{Outcome Classification}
In the preregistration, affective polarization, perceived issue polarization, persuasion, and future willingness to engage were grouped together as primary outcomes capturing different dimensions of political polarization, while anti-democratic attitudes (support for partisan violence, support for anti-democratic practices) were designated as secondary outcomes. 

In the manuscript, we deviated from this classification in two respects. First, we reverse-coded `persuasion' and relabeled it as `attitude polarization' within our primary outcome group of political polarization measures. Previously, under the persuasion measure, higher values indicated that respondents adopted more moderate positions following the conversation. Now, consistent with other political polarization outcomes, the higher values of `attitude polarization' indicate more extreme and polarized positions.

Second, we grouped `future willingness' (i.e., willingness to engage with future political conversations) as a secondary outcome in the main text, rather than a primary outcome. Specifically, we group it with other discussion-related measures, such as discussion satisfaction and the discussion quality index. Although conceptually related to political polarization, `future willingness' captures distinct aspects of the construct (namely, behavioral inclinations to engage in political discussions), and are therefore more naturally discussed separately from other political polarization measures. 

Third, we report the pre-registered anti-democratic attitude measures in the Supplementary Information rather than as secondary outcomes in the main text, as they are not the central focus of the present study.

Overall, these deviations are largely organizational and do not affect the substantive interpretation of the results. All other aspects of the analysis follow the preregistered plan.

\section{Full Regression Tables}
\label{sec:si-full-regression}

This section presents full regression tables corresponding to the analyses
summarized in the main text and earlier sections of the Supplementary Materials.
We report complete model specifications, including all covariates selected in
the LASSO procedure. These tables are provided to support replication and
detailed inspection of the reported results. HC2 robust standard errors are
reported in parentheses, and 95\% confidence intervals are shown in brackets.

\begin{table}
\centering
\caption{Treatment Effects on Political Polarization Outcomes (FDR-adjusted q-values)}
\label{tab:SI_H1_FDR}
\centering
\resizebox{\ifdim\width>\linewidth\linewidth\else\width\fi}{!}{
\fontsize{9}{11}\selectfont
\begin{tabular}[t]{llrrrr}
\toprule
DV & Treatment & Estimate & SE & p-value & q-value\\
\midrule
 & Ingroup-Disagree & -2.279 & 0.501 & <.001 & <.001\\

 & Outgroup-Agree & -1.034 & 0.461 & 0.025 & 0.026\\

\multirow[t]{-3}{*}{\raggedright\arraybackslash Ingroup Warmth} & Outgroup-Disagree & -0.742 & 0.492 & 0.132 & 0.093\\
\cmidrule{1-6}
 & Ingroup-Disagree & 0.858 & 0.660 & 0.194 & 0.123\\

 & Outgroup-Agree & 4.272 & 0.757 & <.001 & <.001\\

\multirow[t]{-3}{*}{\raggedright\arraybackslash Outgroup Warmth} & Outgroup-Disagree & 2.279 & 0.711 & 0.001 & 0.002\\
\cmidrule{1-6}
 & Ingroup-Disagree & -3.124 & 0.830 & <.001 & <.001\\

 & Outgroup-Agree & -5.285 & 0.879 & <.001 & <.001\\

\multirow[t]{-3}{*}{\raggedright\arraybackslash Affective Polarization} & Outgroup-Disagree & -3.022 & 0.861 & <.001 & <.001\\
\cmidrule{1-6}
 & Ingroup-Disagree & 5.356 & 0.870 & <.001 & <.001\\

 & Outgroup-Agree & 0.995 & 0.787 & 0.206 & 0.123\\

\multirow[t]{-3}{*}{\raggedright\arraybackslash Ingroup Issue Distance} & Outgroup-Disagree & 0.830 & 0.787 & 0.292 & 0.161\\
\cmidrule{1-6}
 & Ingroup-Disagree & -2.182 & 1.096 & 0.047 & 0.045\\

 & Outgroup-Agree & -11.298 & 1.203 & <.001 & <.001\\

\multirow[t]{-3}{*}{\raggedright\arraybackslash Outgroup Issue Distance} & Outgroup-Disagree & -1.358 & 1.173 & 0.247 & 0.141\\
\cmidrule{1-6}
 & Ingroup-Disagree & -4.621 & 1.079 & <.001 & <.001\\

 & Outgroup-Agree & -9.643 & 1.178 & <.001 & <.001\\

\multirow[t]{-3}{*}{\raggedright\arraybackslash Perceived Issue Polarization} & Outgroup-Disagree & -0.933 & 1.135 & 0.412 & 0.205\\
\cmidrule{1-6}
 & Ingroup-Disagree & -1.370 & 0.845 & 0.105 & 0.088\\

 & Outgroup-Agree & 1.054 & 0.806 & 0.191 & 0.123\\

\multirow[t]{-3}{*}{\raggedright\arraybackslash Attitude Polarization} & Outgroup-Disagree & -0.709 & 0.843 & 0.400 & 0.205\\
\bottomrule
\end{tabular}}
\end{table}

\begin{table}
\caption{Treatment Effects on Political Polarization Outcomes (Full Models)}
\label{tab:SI_H1}
\centering
\begin{adjustbox}{width=\textwidth}
\begin{tabular}{llllllll}
\toprule
& Ingroup Warmth & Outgroup Warmth & Affective Polarization & Ingroup Issue Distance & Outgroup Issue Distance & Perceived Issue Polarization & Attitude Polarization \\
\midrule
Intercept & \num{8.588}*** & \num{5.452}*** & \num{3.184} & \num{6.368}*** & \num{14.796}*** & \num{11.476}*** & \num{6.360}*** \\
& (\num{1.222}) & (\num{1.498}) & (\num{1.945}) & (\num{1.712}) & (\num{2.541}) & (\num{2.676}) & (\num{1.710}) \\
& [\num{6.192}, \num{10.984}] & [\num{2.514}, \num{8.391}] & [\num{-0.632}, \num{6.999}] & [\num{3.010}, \num{9.726}] & [\num{9.812}, \num{19.779}] & [\num{6.229}, \num{16.723}] & [\num{3.007}, \num{9.713}] \\
Ingroup-Disagree & \num{-2.279}*** & \num{0.858} & \num{-3.124}*** & \num{5.356}*** & \num{-2.182}* & \num{-4.621}*** & \num{-1.370} \\
& (\num{0.501}) & (\num{0.660}) & (\num{0.830}) & (\num{0.870}) & (\num{1.096}) & (\num{1.079}) & (\num{0.845}) \\
& [\num{-3.261}, \num{-1.297}] & [\num{-0.436}, \num{2.151}] & [\num{-4.751}, \num{-1.496}] & [\num{3.651}, \num{7.062}] & [\num{-4.331}, \num{-0.032}] & [\num{-6.738}, \num{-2.504}] & [\num{-3.027}, \num{0.287}] \\
Outgroup-Agree & \num{-1.034}* & \num{4.272}*** & \num{-5.285}*** & \num{0.995} & \num{-11.298}*** & \num{-9.643}*** & \num{1.054} \\
\end{tabular}
\end{adjustbox}
\end{table} 

\begin{table} 
\caption{Long-term Treatment Effects on Political Polarization Outcomes (FDR-adjusted q-values)}
\label{tab:SI_H1_Long-term_FDR}
\centering
\resizebox{\ifdim\width>\linewidth\linewidth\else\width\fi}{!}{
\fontsize{9}{11}\selectfont
\begin{tabular}[t]{llrrrr}
\toprule
DV & Treatment & Estimate & SE & p-value & q-value\\
\midrule
 & Ingroup-Disagree & -0.585 & 0.876 & 0.504 & 1.000\\

 & Outgroup-Agree & -0.756 & 0.837 & 0.367 & 0.667\\

\multirow[t]{-3}{*}{\raggedright\arraybackslash Long-term Ingroup Warmth} & Outgroup-Disagree & -0.033 & 0.826 & 0.969 & 1.000\\
\cmidrule{1-6}
 & Ingroup-Disagree & 0.014 & 0.907 & 0.987 & 1.000\\

 & Outgroup-Agree & 0.478 & 0.852 & 0.575 & 1.000\\

\multirow[t]{-3}{*}{\raggedright\arraybackslash Long-term Outgroup Warmth} & Outgroup-Disagree & 1.681 & 0.955 & 0.078 & 0.394\\
\cmidrule{1-6}
 & Ingroup-Disagree & -0.632 & 1.234 & 0.609 & 1.000\\

 & Outgroup-Agree & -1.231 & 1.119 & 0.271 & 0.613\\

\multirow[t]{-3}{*}{\raggedright\arraybackslash Long-term Affective Polarization} & Outgroup-Disagree & -1.749 & 1.237 & 0.158 & 0.492\\
\cmidrule{1-6}
 & Ingroup-Disagree & 3.890 & 0.990 & <.001 & 0.002\\

 & Outgroup-Agree & 1.599 & 0.962 & 0.097 & 0.409\\

\multirow[t]{-3}{*}{\raggedright\arraybackslash Long-term Ingroup Issue Distance} & Outgroup-Disagree & 1.784 & 0.984 & 0.070 & 0.394\\
\cmidrule{1-6}
 & Ingroup-Disagree & 0.035 & 1.652 & 0.983 & 1.000\\

 & Outgroup-Agree & -4.556 & 1.684 & 0.007 & 0.074\\

\multirow[t]{-3}{*}{\raggedright\arraybackslash Long-term Outgroup Issue Distance} & Outgroup-Disagree & -2.278 & 1.653 & 0.168 & 0.492\\
\cmidrule{1-6}
 & Ingroup-Disagree & 0.205 & 1.553 & 0.895 & 1.000\\

 & Outgroup-Agree & -3.584 & 1.584 & 0.024 & 0.178\\

\multirow[t]{-3}{*}{\raggedright\arraybackslash Long-term Perceived Issue Polarization} & Outgroup-Disagree & -2.399 & 1.565 & 0.125 & 0.476\\
\cmidrule{1-6}
 & Ingroup-Disagree & -0.080 & 1.087 & 0.941 & 1.000\\

 & Outgroup-Agree & 0.268 & 1.124 & 0.812 & 1.000\\

\multirow[t]{-3}{*}{\raggedright\arraybackslash Long-term Attitude Polarization} & Outgroup-Disagree & -0.499 & 1.099 & 0.650 & 1.000\\
\bottomrule
\end{tabular}}
\end{table}

\begin{table}
\caption{Long-term Treatment Effects on Political Polarization Outcomes (Full Models)}
\label{tab:SI_H1_Long-term}
\centering
\begin{adjustbox}{width=\textwidth}
\begin{tabular}{llllllll}
\toprule
& Long-term Ingroup Warmth & Long-term Outgroup Warmth & Long-term Affective Polarization & Long-term Ingroup Issue Distance & Long-term Outgroup Issue Distance & Long-term Perceived Issue Polarization & Long-term Attitude Polarization \\
\midrule
Intercept & \num{6.415}** & \num{6.242}*** & \num{-0.887} & \num{16.451}*** & \num{24.378}*** & \num{14.469}*** & \num{6.762}** \\
& (\num{2.042}) & (\num{1.444}) & (\num{2.819}) & (\num{2.498}) & (\num{3.869}) & (\num{3.547}) & (\num{2.605}) \\
& [\num{2.410}, \num{10.420}] & [\num{3.408}, \num{9.075}] & [\num{-6.415}, \num{4.642}] & [\num{11.551}, \num{21.351}] & [\num{16.789}, \num{31.968}] & [\num{7.512}, \num{21.426}] & [\num{1.651}, \num{11.872}] \\
Ingroup-Disagree & \num{-0.585} & \num{0.014} & \num{-0.632} & \num{3.890}*** & \num{0.035} & \num{0.205} & \num{-0.080} \\
& (\num{0.876}) & (\num{0.907}) & (\num{1.234}) & (\num{0.990}) & (\num{1.652}) & (\num{1.553}) & (\num{1.087}) \\
& [\num{-2.303}, \num{1.132}] & [\num{-1.765}, \num{1.793}] & [\num{-3.051}, \num{1.788}] & [\num{1.948}, \num{5.832}] & [\num{-3.204}, \num{3.275}] & [\num{-2.842}, \num{3.252}] & [\num{-2.211}, \num{2.051}] \\
Outgroup-Agree & \num{-0.756} & \num{0.478} & \num{-1.231} & \num{1.599}+ & \num{-4.556}** & \num{-3.584}* & \num{0.268} \\
\end{tabular}
\end{adjustbox}
\end{table}

\begin{table}
\centering
\caption{Treatment Effects on Discussion-related Outcomes (FDR-adjusted q-values)}
\label{si:tab:SI_H3_FDR}
\centering
\resizebox{\ifdim\width>\linewidth\linewidth\else\width\fi}{!}{
\fontsize{9}{11}\selectfont
\begin{tabular}[t]{llrrrr}
\toprule
DV & Treatment & Estimate & SE & p-value & q-value\\
\midrule
 & Ingroup-Disagree & 0.060 & 0.181 & 0.742 & 0.090\\

 & Outgroup-Agree & 0.318 & 0.173 & 0.066 & 0.017\\

\multirow[t]{-3}{*}{\raggedright\arraybackslash Discussion Quality} & Outgroup-Disagree & 0.505 & 0.172 & 0.003 & 0.002\\
\cmidrule{1-6}
 & Ingroup-Disagree & -0.225 & 0.060 & <.001 & <.001\\

 & Outgroup-Agree & -0.626 & 0.063 & <.001 & <.001\\

\multirow[t]{-3}{*}{\raggedright\arraybackslash Future Willingness} & Outgroup-Disagree & -0.815 & 0.063 & <.001 & <.001\\
\cmidrule{1-6}
 & Ingroup-Disagree & -0.803 & 0.064 & <.001 & <.001\\

 & Outgroup-Agree & -0.277 & 0.057 & <.001 & <.001\\

\multirow[t]{-3}{*}{\raggedright\arraybackslash Discussion Satisfaction} & Outgroup-Disagree & -0.705 & 0.062 & <.001 & <.001\\
\bottomrule
\end{tabular}}
\end{table}

\begin{table}
\caption{Treatment Effects on Discussion-related Outcomes (Full Models)}
\label{si:tab:SI_H3}
\centering
\begin{adjustbox}{width=0.7\textwidth}
\begin{tabular}{llll}
\toprule
& Future Willingness & Discussion Satisfaction & Discussion Quality \\
\midrule
Intercept & \num{3.841}*** & \num{3.318}*** & \num{1.279}*** \\
& (\num{0.156}) & (\num{0.153}) & (\num{0.352}) \\
& [\num{3.536}, \num{4.146}] & [\num{3.018}, \num{3.617}] & [\num{0.589}, \num{1.969}] \\
Ingroup-Disagree & \num{-0.225}*** & \num{-0.803}*** & \num{0.060} \\
& (\num{0.060}) & (\num{0.064}) & (\num{0.181}) \\
& [\num{-0.343}, \num{-0.108}] & [\num{-0.929}, \num{-0.677}] & [\num{-0.296}, \num{0.416}] \\
Outgroup-Agree & \num{-0.626}*** & \num{-0.277}*** & \num{0.318}+ \\
\end{tabular}
\end{adjustbox}
\end{table}

\begin{table} 
\centering
\caption{Treatment Effects on Each Discussion Quality (FDR-adjusted q-values)}
\label{tab:SI_H3_sep_FDR}
\centering
\resizebox{\ifdim\width>\linewidth\linewidth\else\width\fi}{!}{
\fontsize{9}{11}\selectfont
\begin{tabular}[t]{llrrrr}
\toprule
DV & Treatment & Estimate & SE & p-value & q-value\\
\midrule
 & Ingroup-Disagree & 0.378 & 0.061 & <.001 & <.001\\

 & Outgroup-Agree & 0.105 & 0.051 & 0.040 & 0.062\\

\multirow[t]{-3}{*}{\raggedright\arraybackslash Elaboration} & Outgroup-Disagree & 0.529 & 0.065 & <.001 & <.001\\
\cmidrule{1-6}
 & Ingroup-Disagree & -0.143 & 0.075 & 0.056 & 0.071\\

 & Outgroup-Agree & -0.033 & 0.079 & 0.680 & 0.471\\

\multirow[t]{-3}{*}{\raggedright\arraybackslash Gratitude Ratio} & Outgroup-Disagree & -0.196 & 0.067 & 0.004 & 0.008\\
\cmidrule{1-6}
 & Ingroup-Disagree & -0.046 & 0.067 & 0.486 & 0.377\\

 & Outgroup-Agree & 0.042 & 0.065 & 0.520 & 0.380\\

\multirow[t]{-3}{*}{\raggedright\arraybackslash Hedge Ratio} & Outgroup-Disagree & -0.099 & 0.059 & 0.094 & 0.104\\
\cmidrule{1-6}
 & Ingroup-Disagree & -0.276 & 0.062 & <.001 & <.001\\

 & Outgroup-Agree & -0.044 & 0.060 & 0.465 & 0.377\\

\multirow[t]{-3}{*}{\raggedright\arraybackslash Positive Polarity} & Outgroup-Disagree & -0.261 & 0.062 & <.001 & <.001\\
\cmidrule{1-6}
 & Ingroup-Disagree & -0.120 & 0.061 & 0.048 & 0.067\\

 & Outgroup-Agree & 0.070 & 0.067 & 0.299 & 0.255\\

\multirow[t]{-3}{*}{\raggedright\arraybackslash Question Ratio} & Outgroup-Disagree & 0.025 & 0.065 & 0.700 & 0.471\\
\cmidrule{1-6}
 & Ingroup-Disagree & 0.316 & 0.065 & <.001 & <.001\\

 & Outgroup-Agree & 0.158 & 0.066 & 0.016 & 0.029\\

\multirow[t]{-3}{*}{\raggedright\arraybackslash Responsivity} & Outgroup-Disagree & 0.517 & 0.063 & <.001 & <.001\\
\cmidrule{1-6}
 & Ingroup-Disagree & -0.070 & 0.069 & 0.316 & 0.255\\

 & Outgroup-Agree & 0.005 & 0.055 & 0.928 & 0.661\\

\multirow[t]{-3}{*}{\raggedright\arraybackslash Toxicity} & Outgroup-Disagree & -0.023 & 0.061 & 0.711 & 0.471\\
\bottomrule
\end{tabular}}
\end{table}

\begin{table}
\caption{Treatment Effects on Each Discussion Quality (Full Models)}
\label{tab:SI_H3_sep}
\centering
\begin{adjustbox}{width=\textwidth}
\begin{tabular}{llllllll}
\toprule
& Toxicity & Positive Polarity & Elaboration & Gratitude & Question Ratio & Hedge Ratio & Responsivity\\
\midrule
Intercept & \num{0.380}+ & \num{0.456}** & \num{-0.286}*** & \num{0.097} & \num{0.510}*** & \num{0.260}* & \num{-0.180} \\
& (\num{0.202}) & (\num{0.146}) & (\num{0.085}) & (\num{0.068}) & (\num{0.116}) & (\num{0.131}) & (\num{0.129}) \\
& [\num{-0.016}, \num{0.775}] & [\num{0.170}, \num{0.742}] & [\num{-0.453}, \num{-0.119}] & [\num{-0.037}, \num{0.230}] & [\num{0.282}, \num{0.739}] & [\num{0.004}, \num{0.516}] & [\num{-0.433}, \num{0.072}] \\
Ingroup-Disagree & \num{-0.070} & \num{-0.276}*** & \num{0.378}*** & \num{-0.143}+ & \num{-0.120}* & \num{-0.046} & \num{0.316}*** \\
& (\num{0.069}) & (\num{0.062}) & (\num{0.061}) & (\num{0.075}) & (\num{0.061}) & (\num{0.067}) & (\num{0.065}) \\
& [\num{-0.205}, \num{0.066}] & [\num{-0.398}, \num{-0.154}] & [\num{0.258}, \num{0.499}] & [\num{-0.289}, \num{0.004}] & [\num{-0.239}, \num{-0.001}] & [\num{-0.177}, \num{0.084}] & [\num{0.188}, \num{0.443}] \\
Outgroup-Agree & \num{0.005} & \num{-0.044} & \num{0.105}* & \num{-0.033} & \num{0.070} & \num{0.042} & \num{0.158}* \\
\end{tabular} 
\end{adjustbox}
\end{table}

\begin{table}
\caption{Treatment Effects on Anti-democratic Attitudes (Full Models)}
\label{tab:SI_H2}
\centering
\begin{adjustbox}{totalheight=0.88\textheight, keepaspectratio}
\begin{tabular}{lll}
\toprule
& Support for Partisan Violence & Support for Anti-Democratic Practices \\
\midrule
Intercept & \num{3.358} & \num{3.252} \\
& (\num{3.035}) & (\num{3.392}) \\
& [\num{-2.594}, \num{9.310}] & [\num{-3.400}, \num{9.904}] \\
Ingroup-Disagree & \num{1.507} & \num{-0.149} \\
& (\num{1.292}) & (\num{1.429}) \\
& [\num{-1.027}, \num{4.041}] & [\num{-2.951}, \num{2.653}] \\
Outgroup-Agree & \num{-0.073} & \num{-0.741} \\
\end{tabular} 
\end{adjustbox}
\end{table}

\begin{table}
\caption{Long-term Treatment Effects on Anti-democratic Attitudes (Full Models)}
\label{tab:SI_H2_Long-term}
\centering
\begin{adjustbox}{width=0.95\textwidth}
\begin{tabular}{lll}
\toprule
& Long-term Support for Partisan Violence & Long-term Support for Anti-Democratic Practices \\
\midrule
Intercept & \num{7.706}** & \num{6.063}* \\
& (\num{2.858}) & (\num{3.029}) \\
& [\num{2.099}, \num{13.312}] & [\num{0.121}, \num{12.005}] \\
Ingroup-Disagree & \num{-0.756} & \num{-1.189} \\
& (\num{1.307}) & (\num{1.605}) \\
& [\num{-3.320}, \num{1.808}] & [\num{-4.337}, \num{1.958}] \\
Outgroup-Agree & \num{-2.491}* & \num{-0.646} \\
\end{tabular} 
\end{adjustbox}
\end{table}


\begin{table}
\caption{Treatment Effects on Political Polarization Outcomes Among Participants Who Passed the Manipulation Check (Full Models)}
\label{tab:si-robustness-check1-polarization}
\centering
\begin{adjustbox}{width=\textwidth}
\begin{tabular}{llllllll}
\toprule
& Ingroup Warmth & Outgroup Warmth & Affective Polarization & Ingroup Issue Distance & Outgroup Issue Distance & Perceived Issue Polarization & Attitude Polarization \\
\midrule
Intercept & \num{8.797}*** & \num{4.598}** & \num{4.119}+ & \num{7.835}*** & \num{18.271}*** & \num{12.581}*** & \num{6.172}*** \\
& (\num{1.458}) & (\num{1.626}) & (\num{2.255}) & (\num{2.156}) & (\num{3.574}) & (\num{3.310}) & (\num{1.707}) \\
& [\num{5.938}, \num{11.656}] & [\num{1.409}, \num{7.787}] & [\num{-0.305}, \num{8.543}] & [\num{3.605}, \num{12.064}] & [\num{11.260}, \num{25.281}] & [\num{6.087}, \num{19.074}] & [\num{2.823}, \num{9.521}] \\
Ingroup-Disagree & \num{-2.593}*** & \num{0.504} & \num{-3.088}** & \num{6.248}*** & \num{-0.672} & \num{-4.126}** & \num{-1.844}+ \\
& (\num{0.668}) & (\num{0.761}) & (\num{1.014}) & (\num{1.135}) & (\num{1.389}) & (\num{1.400}) & (\num{1.021}) \\
& [\num{-3.904}, \num{-1.282}] & [\num{-0.988}, \num{1.997}] & [\num{-5.078}, \num{-1.099}] & [\num{4.022}, \num{8.474}] & [\num{-3.397}, \num{2.054}] & [\num{-6.873}, \num{-1.380}] & [\num{-3.848}, \num{0.159}] \\
Outgroup-Agree & \num{-1.193}* & \num{4.953}*** & \num{-6.147}*** & \num{0.992} & \num{-13.756}*** & \num{-11.356}*** & \num{0.682} \\
\end{tabular}
\end{adjustbox}
\end{table}

\begin{table}
\caption{Treatment Effects on Discussion-related Outcomes Among Participants Who Passed the Manipulation Check (Full Models)}
\label{tab:si-robustness-check1-discussion}
\centering
\begin{adjustbox}{width=0.7\textwidth}
\begin{tabular}{llll}
\toprule
& Future Willingness & Discussion Satisfaction & Discussion Quality \\
\midrule
Intercept & \num{3.826}*** & \num{3.357}*** & \num{1.585}** \\
& (\num{0.176}) & (\num{0.137}) & (\num{0.506}) \\
& [\num{3.480}, \num{4.172}] & [\num{3.089}, \num{3.625}] & [\num{0.591}, \num{2.578}] \\
Ingroup-Disagree & \num{-0.357}*** & \num{-1.135}*** & \num{-0.196} \\
& (\num{0.076}) & (\num{0.083}) & (\num{0.217}) \\
& [\num{-0.506}, \num{-0.207}] & [\num{-1.298}, \num{-0.971}] & [\num{-0.622}, \num{0.229}] \\
Outgroup-Agree & \num{-0.635}*** & \num{-0.208}*** & \num{0.387}* \\
\end{tabular}
\end{adjustbox}
\end{table}

\begin{table}
\caption{Treatment Effects on Political Polarization Outcomes Excluding Pilot-Phase Participants (Full Models)}
\label{tab:si-robustness-check2-polarization}
\centering
\begin{adjustbox}{width=\textwidth}
\begin{tabular}{llllllll}
\toprule
& Ingroup Warmth & Outgroup Warmth & Affective Polarization & Ingroup Issue Distance & Outgroup Issue Distance & Perceived Issue Polarization & Attitude Polarization \\
\midrule
Intercept & \num{8.873}*** & \num{5.655}*** & \num{3.391}+ & \num{7.959}*** & \num{14.570}*** & \num{11.772}*** & \num{6.425}*** \\
& (\num{1.230}) & (\num{1.500}) & (\num{1.993}) & (\num{1.933}) & (\num{2.492}) & (\num{2.689}) & (\num{1.787}) \\
& [\num{6.461}, \num{11.285}] & [\num{2.713}, \num{8.596}] & [\num{-0.517}, \num{7.299}] & [\num{4.169}, \num{11.749}] & [\num{9.683}, \num{19.458}] & [\num{6.498}, \num{17.046}] & [\num{2.921}, \num{9.929}] \\
Ingroup-Disagree & \num{-2.181}*** & \num{1.039} & \num{-3.209}*** & \num{5.362}*** & \num{-2.393}* & \num{-4.679}*** & \num{-1.287} \\
& (\num{0.518}) & (\num{0.665}) & (\num{0.845}) & (\num{0.913}) & (\num{1.137}) & (\num{1.113}) & (\num{0.873}) \\
& [\num{-3.197}, \num{-1.165}] & [\num{-0.266}, \num{2.343}] & [\num{-4.866}, \num{-1.552}] & [\num{3.572}, \num{7.153}] & [\num{-4.623}, \num{-0.164}] & [\num{-6.862}, \num{-2.495}] & [\num{-3.000}, \num{0.426}] \\
Outgroup-Agree & \num{-1.039}* & \num{4.551}*** & \num{-5.566}*** & \num{0.982} & \num{-11.221}*** & \num{-9.753}*** & \num{1.339} \\
\end{tabular}
\end{adjustbox}
\end{table}

\begin{table}
\caption{Treatment Effects on Discussion-related Outcomes Excluding Pilot-Phase Participants (Full Models)}
\label{tab:si-robustness-check2-discussion}
\centering
\begin{adjustbox}{width=0.7\textwidth}
\begin{tabular}{llll}
\toprule
& Future Willingness & Discussion Satisfaction & Discussion Quality \\
\midrule
Intercept & \num{3.811}*** & \num{3.252}*** & \num{1.335}*** \\
& (\num{0.164}) & (\num{0.157}) & (\num{0.358}) \\
& [\num{3.490}, \num{4.132}] & [\num{2.944}, \num{3.559}] & [\num{0.632}, \num{2.038}] \\
Ingroup-Disagree & \num{-0.219}*** & \num{-0.812}*** & \num{0.065} \\
& (\num{0.062}) & (\num{0.066}) & (\num{0.187}) \\
& [\num{-0.340}, \num{-0.099}] & [\num{-0.941}, \num{-0.683}] & [\num{-0.303}, \num{0.433}] \\
Outgroup-Agree & \num{-0.604}*** & \num{-0.269}*** & \num{0.369}* \\
\end{tabular}
\end{adjustbox}
\end{table}

\begin{table}
\caption{Mixed Effects Model}
\label{tab:si-mixed-effects-model}
\centering
\begin{adjustbox}{width=0.5\textwidth}
\begin{tabular}{lll}
\toprule
& Affective Polarization & Perceived Issue Polarization \\
\midrule
Intercept & \num{61.357}*** & \num{3.468} \\
& (\num{2.279}) & (\num{2.571}) \\
& [\num{56.889}, \num{65.825}] & [\num{-1.572}, \num{8.509}] \\
Ingroup-Disagree & \num{-0.688} & \num{0.842} \\
& (\num{1.223}) & (\num{1.262}) \\
& [\num{-3.085}, \num{1.709}] & [\num{-1.632}, \num{3.315}] \\
Outgroup-Agree & \num{-0.729} & \num{0.537} \\
\end{tabular}
\end{adjustbox}
\end{table}


\begin{table}
\caption{Heterogeneous Treatment Effects on Political Polarization Outcomes by Topic (Full Models)}
\label{tab:si-hte-topic}
\centering
\begin{adjustbox}{width=\textwidth}
\begin{tabular}{llllllll}
\toprule
& Ingroup Warmth & Outgroup Warmth & Affective Polarization & Ingroup Issue Distance & Outgroup Issue Distance & Perceived Issue Polarization & Attitude Polarization \\
\midrule
Intercept & \num{8.008}*** & \num{5.899}*** & \num{1.998} & \num{6.647}*** & \num{17.301}*** & \num{13.730}*** & \num{8.846}*** \\
& (\num{1.264}) & (\num{1.634}) & (\num{2.032}) & (\num{1.861}) & (\num{2.794}) & (\num{2.941}) & (\num{1.829}) \\
& [\num{5.529}, \num{10.488}] & [\num{2.694}, \num{9.103}] & [\num{-1.987}, \num{5.982}] & [\num{2.996}, \num{10.297}] & [\num{11.822}, \num{22.780}] & [\num{7.962}, \num{19.497}] & [\num{5.259}, \num{12.434}] \\
Ingroup-Disagree & \num{-2.944}*** & \num{0.638} & \num{-3.538}*** & \num{5.921}*** & \num{-3.217}* & \num{-7.123}*** & \num{-2.759}** \\
& (\num{0.643}) & (\num{0.822}) & (\num{1.069}) & (\num{1.105}) & (\num{1.378}) & (\num{1.417}) & (\num{1.012}) \\
& [\num{-4.205}, \num{-1.682}] & [\num{-0.975}, \num{2.250}] & [\num{-5.634}, \num{-1.442}] & [\num{3.754}, \num{8.088}] & [\num{-5.920}, \num{-0.513}] & [\num{-9.903}, \num{-4.344}] & [\num{-4.744}, \num{-0.774}] \\
Outgroup-Agree & \num{-0.584} & \num{4.281}*** & \num{-4.828}*** & \num{2.016}* & \num{-11.896}*** & \num{-10.628}*** & \num{-0.718} \\
\end{tabular}
\end{adjustbox}
\end{table}

\begin{table}
\caption{Heterogeneous Treatment Effects on Political Polarization Outcomes by Partisanship (Full Models)}
\label{tab:si-hte-pid}
\centering
\begin{adjustbox}{width=\textwidth}
\begin{tabular}{llllllll}
\toprule
& Ingroup Warmth & Outgroup Warmth & Affective Polarization & Ingroup Issue Distance & Outgroup Issue Distance & Perceived Issue Polarization & Attitude Polarization \\
\midrule
Intercept & \num{8.664}*** & \num{4.821}** & \num{3.884}+ & \num{7.013}*** & \num{15.427}*** & \num{11.720}*** & \num{7.224}*** \\
& (\num{1.289}) & (\num{1.554}) & (\num{2.034}) & (\num{1.713}) & (\num{2.637}) & (\num{2.776}) & (\num{1.881}) \\
& [\num{6.137}, \num{11.192}] & [\num{1.774}, \num{7.867}] & [\num{-0.105}, \num{7.873}] & [\num{3.654}, \num{10.372}] & [\num{10.256}, \num{20.599}] & [\num{6.276}, \num{17.163}] & [\num{3.536}, \num{10.912}] \\
Ingroup-Disagree & \num{-3.225}*** & \num{1.847}* & \num{-5.050}*** & \num{5.171}*** & \num{-2.749}+ & \num{-5.711}*** & \num{-2.605}* \\
& (\num{0.714}) & (\num{0.904}) & (\num{1.144}) & (\num{1.109}) & (\num{1.489}) & (\num{1.473}) & (\num{1.133}) \\
& [\num{-4.624}, \num{-1.825}] & [\num{0.074}, \num{3.621}] & [\num{-7.293}, \num{-2.807}] & [\num{2.996}, \num{7.347}] & [\num{-5.670}, \num{0.171}] & [\num{-8.599}, \num{-2.822}] & [\num{-4.827}, \num{-0.383}] \\
Outgroup-Agree & \num{-1.085}+ & \num{5.682}*** & \num{-6.740}*** & \num{-0.205} & \num{-11.791}*** & \num{-10.181}*** & \num{0.757} \\
\end{tabular}
\end{adjustbox}
\end{table}

\clearpage

\section{Replication Materials}
\label{sec:si-replication}

Additional replication materials (replication code, aggregated data, etc.) are available on GitHub at~\url{https://github.com/DO-WON/AI_Conv}

\subsection{Survey Questions}
\label{sec:si-questionnaire}

\subsubsection*{Party Identification}

\begin{enumerate}
    \item Generally speaking, do you usually think of yourself as a Republican, a Democrat, an Independent, or something else?\\
    \emph{Options: Democrat; Republican; Independent; Other; Not sure}

    \item[\phantom{0}] \emph{If Democrat:} Would you call yourself a strong Democrat or not a very strong Democrat?\\
    \emph{Options: Strong Democrat; Not very strong Democrat}

    \item[\phantom{0}] \emph{If Republican:} Would you call yourself a strong Republican or not a very strong Republican?\\
    \emph{Options: Strong Republican; Not very strong Republican}

    \item[\phantom{0}] \emph{If Independent or Other:} Do you think of yourself as closer to the Republican Party or to the Democratic Party?\\
    \emph{Options: Closer to the Republican Party; Closer to the Democratic Party; Neither}
\end{enumerate}

\subsubsection*{Political Ideology}

\begin{enumerate}
    \item When it comes to politics, how would you describe yourself—as liberal, conservative, or neither liberal nor conservative?\\
    \emph{Options: Very conservative; Somewhat conservative; Slightly conservative; Moderate / middle of the road; Slightly liberal; Somewhat liberal; Very liberal; Not sure}
\end{enumerate}

\subsubsection*{Affective Polarization (Feeling Thermometers)}

\begin{enumerate}
    \item How would you rate your feelings toward \textbf{Republicans}?\\
    \emph{Slider: 0 (Extremely cold or unfavorable) to 100 (Extremely warm or favorable)}

    \item How would you rate your feelings toward \textbf{Democrats}?\\
    \emph{Slider: 0 (Extremely cold or unfavorable) to 100 (Extremely warm or favorable)}
\end{enumerate}

\subsubsection*{Trust in AI}

\begin{enumerate}
    \item Many AI systems like ChatGPT are able to create summaries of long documents or conversations. Do you trust these AI systems to provide a fair and accurate representation of what someone thinks or says?\\
    \emph{Options: Yes; No}
\end{enumerate}

\subsubsection*{Issue Selection and Opinion Statement}

\begin{enumerate}
    \item Thinking about the next U.S. Presidential Election, which one of the following issues is most important to you in deciding which political candidate you will support?\\
    \emph{Options: Gender inequalities; Racial inequalities; Gun policy; Foreign policy; Economy; Immigration; LGBTQ+ rights; Health care; Abortion; Climate change; Other}

    \item Reflecting on the issue you selected, what is your position on it? Please write at least 100 characters (about 2--3 sentences) explaining your view and the reasons behind it.\\
    \emph{[Open-ended text entry]}
\end{enumerate}

\subsubsection*{Opinion Summary Validation}

After participants submitted their open-ended response, an AI-generated summary was presented to them. Below are the survey items used:

\begin{enumerate}
    \item Does this capture your view well enough?\\
    \emph{Options: Yes; No}

    \item[\phantom{0}] \emph{If No:} Please revise the summary to better reflect your view. You can edit, rephrase, or rewrite it as needed.\\
    \emph{[Editable text field pre-filled with AI summary]}
\end{enumerate}

\subsubsection*{Pre-Treatment Agreement and Perceived Issue Polarization}

\begin{enumerate}
    \item How strongly do you agree or disagree with the statement?
    \emph{Slider: 0 (Strongly disagree) to 100 (Strongly agree)}

    \item How do you think \textbf{Democrats} would rate their agreement with the same statement?
    \emph{Slider: 0 (Strongly disagree) to 100 (Strongly agree)}

    \item How do you think \textbf{Republicans} would rate their agreement with the same statement?
    \emph{Slider: 0 (Strongly disagree) to 100 (Strongly agree)}
\end{enumerate}

\subsubsection*{Post-Treatment Agreement and Persuasion}

Post-treatment agreement was measured using the same three items above, administered again after the AI conversation.
Persuasion is computed as the change in distance from the neutral midpoint (50): post$-$pre when the pre-treatment stance is below 50, and pre$-$post when it is above 50, such that positive values indicate attitude moderation and negative values indicate attitude reinforcement. Below are the survey items used:

\begin{enumerate}
    \item How much do you agree or disagree with the statement now?
    \emph{Slider: 0 (Strongly disagree) to 100 (Strongly agree)}

    \item And how do you think \textbf{Democrats} would now rate their agreement with the same statement?
    \emph{Slider: 0 (Strongly disagree) to 100 (Strongly agree)}

    \item And how do you think \textbf{Republicans} would now rate their agreement with the same statement?
    \emph{Slider: 0 (Strongly disagree) to 100 (Strongly agree)}
\end{enumerate}

\clearpage
\bibliographystyle{abbrv}
\bibliography{main}

@article{franken2015eye,
  title={Eye-tracking study of reading speed from LCD displays: influence of type style and type size},
  author={Franken, Gregor and Podlesek, Anja and Mo{\v{z}}ina, Klementina},
  journal={Journal of eye movement research},
  volume={8},
  number={1},
  pages={3},
  year={2015},
  publisher={Bern Open Publishing}
}

@incollection{graf2016investigating,
  title={Investigating positive and negative intergroup contact: Rectifying a long-standing positivity bias in the literature},
  author={Graf, Sylvie and Paolini, Stefania},
  booktitle={Intergroup contact theory},
  pages={100--121},
  year={2016},
  publisher={Routledge}
}

@article{paolini2010negative,
  title={Negative intergroup contact makes group memberships salient: Explaining why intergroup conflict endures},
  author={Paolini, Stefania and Harwood, Jake and Rubin, Mark},
  journal={Personality and social Psychology bulletin},
  volume={36},
  number={12},
  pages={1723--1738},
  year={2010},
  publisher={Sage Publications Sage CA: Los Angeles, CA}
}

@article{blattner2023does,
  title={Does contact reduce affective polarization? Field evidence from Germany},
  author={Blattner, Adrian and Koenen, Martin},
  journal={Field Evidence from Germany (July 12, 2023)},
  year={2023}
}

@article{balietti2021reducing,
  title={Reducing opinion polarization: Effects of exposure to similar people with differing political views},
  author={Balietti, Stefano and Getoor, Lise and Goldstein, Daniel G and Watts, Duncan J},
  journal={Proceedings of the National Academy of Sciences},
  volume={118},
  number={52},
  pages={e2112552118},
  year={2021},
  publisher={National Academy of Sciences}
}

@misc{braghieri2024talking,
  title={Talking across the Aisle},
  author={Braghieri, Luca and Schwardmann, Peter and Tripodi, Egon},
  url={https://www.egontripodi.com/papers/talkingacross.pdf},
  year={2024}
}

@article{fang2025person,
  title={How in-person conversations shape political polarization: Quasi-experimental evidence from a nationwide initiative},
  author={Fang, Ximeng and Heuser, Sven and St{\"o}tzer, Lasse S},
  journal={Journal of Public Economics},
  volume={242},
  pages={105309},
  year={2025},
  publisher={Elsevier}
}

@article{you2024ingroupnorms,
    author = {You, Zi Ting and Lee, Spike W S},
    title = {Explanations of and interventions against affective polarization cannot afford to ignore the power of ingroup norm perception},
    journal = {PNAS Nexus},
    volume = {3},
    number = {10},
    pages = {pgae286},
    year = {2024},
    month = {10},
    issn = {2752-6542},
    doi = {10.1093/pnasnexus/pgae286},
    url = {https://doi.org/10.1093/pnasnexus/pgae286},
    eprint = {https://academic.oup.com/pnasnexus/article-pdf/3/10/pgae286/59961211/pgae286.pdf},
}

@article{stagnaro2025factual,
  title={Factual knowledge can reduce attitude polarization},
  author={Stagnaro, Michael Nicholas and Amsalem, Eran},
  journal={Nature Communications},
  volume={16},
  pages={3809},
  year={2025},
  doi={10.1038/s41467-025-58697-3}
}

@article{bor2023quantifying,
  title={Quantifying polarization across political groups on key policy issues using sentiment analysis},
  author={Bor, Dennies and Lee, Benjamin Seiyon and Oughton, Edward J},
  journal={arXiv preprint arXiv:2302.07775},
  year={2023}
}

@article{jung2025varieties,
  title={Varieties of values: Moral values are uniquely divisive},
  author={Jung, Jae-Hee and Clifford, Scott},
  journal={American Political Science Review},
  volume={119},
  number={1},
  pages={462--478},
  year={2025},
  publisher={Cambridge University Press}
}

@article{dellaposta2015liberals,
  title={Why do liberals drink lattes?},
  author={DellaPosta, Daniel and Shi, Yongren and Macy, Michael},
  journal={American Journal of Sociology},
  volume={120},
  number={5},
  pages={1473--1511},
  year={2015},
  publisher={University of Chicago Press Chicago, IL}
}

@article{mason2015disrespectfully,
author = {Mason, Lilliana},
title = {“I Disrespectfully Agree”: The Differential Effects of Partisan Sorting on Social and Issue Polarization},
journal = {American Journal of Political Science},
volume = {59},
number = {1},
pages = {128-145},
doi = {https://doi.org/10.1111/ajps.12089},
url = {https://onlinelibrary.wiley.com/doi/abs/10.1111/ajps.12089},
eprint = {https://onlinelibrary.wiley.com/doi/pdf/10.1111/ajps.12089},
year = {2015}
}

@book{allport1954nature,
  author    = {Allport, Gordon W.},
  title     = {The Nature of Prejudice},
  year      = {1954},
  publisher = {Addison-Wesley},
  address   = {Reading, MA}
}

@book{heider2013psychology,
  title={The psychology of interpersonal relations},
  author={Heider, Fritz},
  year={2013},
  publisher={Psychology Press}
}

@article{huddy2015expressive,
  title={Expressive partisanship: Campaign involvement, political emotion, and partisan identity},
  author={Huddy, Leonie and Mason, Lilliana and Aar{\o}e, Lene},
  journal={American Political Science Review},
  volume={109},
  number={1},
  pages={1--17},
  year={2015},
  publisher={Cambridge University Press}
}

@article{mummolo2019demand,
  author  = {Mummolo, Jonathan and Peterson, Erik},
  title   = {Demand Effects in Survey Experiments: An Empirical Assessment},
  journal = {American Political Science Review},
  year    = {2019},
  volume  = {113},
  number  = {2},
  pages   = {517--529},
  doi     = {10.1017/S0003055418000837}
}

@article{deJong2024,
  author       = {de Jong, Jona F.},
  title        = {Cross‑partisan discussions reduced political polarization between UK voters, but less so when they disagreed},
  journal      = {Communications Psychology},
  year         = {2024},
  volume       = {2},
  number       = {1},
  pages        = {5},
  doi          = {10.1038/s44271-023-00051-8},
  url          = {https://doi.org/10.1038/s44271-023-00051-8}
}

@article{Ahler2018,
  author  = {Ahler, Douglas J. and Sood, Gaurav},
  title   = {The Parties in Our Heads: Misperceptions about Party Composition and Their Consequences},
  journal = {The Journal of Politics},
  year    = {2018},
  volume  = {80},
  number  = {3},
  pages   = {964--981},
  doi     = {10.1086/697253},
  url     = {https://doi.org/10.1086/697253}
}

@article{Voelkel2023,
  author = {Voelkel, J. G. and Chu, J. and Stagnaro, M. N. and Mernyk, J. S. and Redekopp, C. and Pink, S. L. and Druckman, J. N. and Rand, D. G. and Willer, R.},
  title = {Interventions reducing affective polarization do not necessarily improve anti-democratic attitudes},
  journal = {Nature Human Behaviour},
  year = {2023},
  volume = {7},
  number = {1},
  pages = {55--64},
  doi = {10.1038/s41562-022-01466-9}
}

@article{dias2022nature,
  author    = {Dias, Nicholas and Lelkes, Yphtach},
  title     = {The Nature of Affective Polarization: Disentangling Policy Disagreement from Partisan Identity},
  journal   = {American Journal of Political Science},
  volume    = {66},
  number    = {3},
  pages     = {775--790},
  year      = {2022},
  doi       = {10.1111/ajps.12628}
}

@article{goggin2019goes, 
    author = {Goggin, Stephen N. and Henderson, John A. and Theodoridis, Alexander G.}, 
    title = {What Goes with Red and Blue? Mapping Partisan and Ideological Associations in the Minds of Voters}, 
    journal = {Political Behavior}, 
    year = {2019}, 
    doi = {10.1007/s11109-018-09525-6} 
}

@book{mason2018uncivil,
  author    = {Mason, Lilliana},
  title     = {Uncivil Agreement: How Politics Became Our Identity},
  publisher = {University of Chicago Press},
  address   = {Chicago},
  year      = {2018}
}

@article{barber2019does,
  author    = {Barber, Michael and Pope, Jeremy C.},
  title     = {Does Party Trump Ideology? {D}isentangling Party and Ideology in America},
  journal   = {American Political Science Review},
  volume    = {113},
  number    = {1},
  pages     = {38--54},
  year      = {2019},
  doi       = {10.1017/S0003055418000795}
}

@article{bartels2002beyond,
  author    = {Bartels, Larry M.},
  title     = {Beyond the Running Tally: Partisan Bias in Political Perceptions},
  journal   = {Political Behavior},
  volume    = {24},
  number    = {2},
  pages     = {117--150},
  year      = {2002},
  doi       = {10.1023/A:1021226224601}
}

@book{kinder2017neither,
  author    = {Kinder, Donald R. and Kalmoe, Nathan P.},
  title     = {Neither Liberal nor Conservative: Ideological Innocence in the American Public},
  publisher = {University of Chicago Press},
  address   = {Chicago},
  year      = {2017}
}

@book{festinger1957theory,
  author    = {Festinger, Leon},
  title     = {A Theory of Cognitive Dissonance},
  year      = {1957},
  publisher = {Stanford University Press},
  address   = {Stanford, CA}
}

@article{pettigrew2006meta,
  author  = {Pettigrew, Thomas F. and Tropp, Linda R.},
  title   = {A Meta-Analytic Test of Intergroup Contact Theory},
  journal = {Journal of Personality and Social Psychology},
  volume  = {90},
  number  = {5},
  pages   = {751--783},
  year    = {2006},
  doi     = {10.1037/0022-3514.90.5.751}
}

@article{paluck2021prejudice,
  title={Prejudice reduction: Progress and challenges},
  author={Paluck, Elizabeth Levy and Porat, Roni and Clark, Chelsey S and Green, Donald P},
  journal={Annual review of psychology},
  volume={72},
  number={1},
  pages={533--560},
  year={2021},
  publisher={Annual Reviews}
}

@article{paluck2019contact,
  author  = {Paluck, Elizabeth Levy and Green, Seth A. and Green, Donald P.},
  title   = {The Contact Hypothesis Re-Evaluated},
  journal = {Behavioural Public Policy},
  volume  = {3},
  number  = {2},
  pages   = {129--158},
  year    = {2019},
  doi     = {10.1017/bpp.2018.25}
}

@article{santoro2022promise,
  title={The promise and pitfalls of cross-partisan conversations for reducing affective polarization: Evidence from randomized experiments},
  author={Santoro, Erik and Broockman, David E},
  journal={Science advances},
  volume={8},
  number={25},
  pages={eabn5515},
  year={2022},
  publisher={American Association for the Advancement of Science}
}

@article{voelkel2024strengthening, 
author = {Jan G. Voelkel and Michael N. Stagnaro and James Y. Chu and Sophia L. Pink and Joseph S. Mernyk and Chrystal Redekopp and Isaias Ghezae and Matthew Cashman and Dhaval Adjodah and Levi G. Allen and L. Victor Allis and Gina Baleria and Nathan Ballantyne and Jay J. Van Bavel and Hayley Blunden and Alia Braley and Christopher J. Bryan and Jared B. Celniker and Mina Cikara and Margarett V. Clapper and Katherine Clayton and Hanne Collins and Evan DeFilippis and Macrina Dieffenbach and Kimberly C. Doell and Charles Dorison and Mylien Duong and Peter Felsman and Maya Fiorella and David Francis and Michael Franz and Roman A. Gallardo and Sara Gifford and Daniela Goya-Tocchetto and Kurt Gray and Joe Green and Joshua Greene and Mertcan Güngör and Matthew Hall and Cameron A. Hecht and Ali Javeed and John T. Jost and Aaron C. Kay and Nick R. Kay and Brandyn Keating and John Michael Kelly and James R. G. Kirk and Malka Kopell and Nour Kteily and Emily Kubin and Jeffrey Lees and Gabriel Lenz and Matthew Levendusky and Rebecca Littman and Kara Luo and Aaron Lyles and Ben Lyons and Wayde Marsh and James Martherus and Lauren Alpert Maurer and Caroline Mehl and Julia Minson and Molly Moore and Samantha L. Moore-Berg and Michael H. Pasek and Alex Pentland and Curtis Puryear and Hossein Rahnama and Steve Rathje and Jay Rosato and Maytal Saar-Tsechansky and Luiza Almeida Santos and Colleen M. Seifert and Azim Shariff and Otto Simonsson and Shiri Spitz Siddiqi and Daniel F. Stone and Palma Strand and Michael Tomz and David S. Yeager and Erez Yoeli and Jamil Zaki and James N. Druckman and David G. Rand and Robb Willer},
title = {Megastudy testing 25 treatments to reduce antidemocratic attitudes and partisan animosity},
journal = {Science},
volume = {386},
number = {6719},
pages = {eadh4764},
year = {2024},
doi = {10.1126/science.adh4764},
URL = {https://www.science.org/doi/abs/10.1126/science.adh4764},
eprint = {https://www.science.org/doi/pdf/10.1126/science.adh4764}
}

@article{costello_durably_2024,
    title = {Durably reducing conspiracy beliefs through dialogues with {AI}},
    volume = {385},
    issn = {0036-8075, 1095-9203},
    url = {https://www.science.org/doi/10.1126/science.adq1814},
    doi = {10.1126/science.adq1814},
    language = {en},
    number = {6714},
    urldate = {2025-07-15},
    journal = {Science},
    author = {Costello, Thomas H. and Pennycook, Gordon and Rand, David G.},
    month = sep,
    year = {2024},
    note = {Publisher: American Association for the Advancement of Science (AAAS)},
}

@article{velez_confronting_2024,
    title = {Confronting {Core} {Issues}: {A} {Critical} {Assessment} of {Attitude} {Polarization} {Using} {Tailored} {Experiments}},
    copyright = {http://creativecommons.org/licenses/by/4.0},
    issn = {0003-0554, 1537-5943},
    shorttitle = {Confronting {Core} {Issues}},
    url = {https://www.cambridge.org/core/product/identifier/S0003055424000819/type/journal_article},
    doi = {10.1017/S0003055424000819},
    language = {en},
    urldate = {2025-02-25},
    journal = {American Political Science Review},
    author = {Velez, Yamil Ricardo and Liu, Patrick},
    month = aug,
    year = {2024},
    pages = {1--18},
}

@article{iyengar2015fear,
  title={Fear and loathing across party lines: New evidence on group polarization},
  author={Iyengar, Shanto and Westwood, Sean J},
  journal={American journal of political science},
  volume={59},
  number={3},
  pages={690--707},
  year={2015},
  publisher={Wiley Online Library}
}

@article{graham2020democracy,
  title={Democracy in America? Partisanship, polarization, and the robustness of support for democracy in the United States},
  author={Graham, Matthew H and Svolik, Milan W},
  journal={American Political Science Review},
  volume={114},
  number={2},
  pages={392--409},
  year={2020},
  publisher={Cambridge University Press}
}

@article{piazza2023political,
  title={Political polarization and political violence},
  author={Piazza, James A},
  journal={Security Studies},
  volume={32},
  number={3},
  pages={476--504},
  year={2023},
  publisher={Taylor \& Francis}
}

@article{huber2017political,
  title={Political homophily in social relationships: Evidence from online dating behavior},
  author={Huber, Gregory A and Malhotra, Neil},
  journal={The Journal of Politics},
  volume={79},
  number={1},
  pages={269--283},
  year={2017},
  publisher={University of Chicago Press Chicago, IL}
}

@article{lin2025persuading,
  title={Persuading voters using human--artificial intelligence dialogues},
  author={Lin, Hause and Czarnek, Gabriela and Lewis, Benjamin and White, Joshua P and Berinsky, Adam J and Costello, Thomas and Pennycook, Gordon and Rand, David G},
  journal={Nature},
  pages={1--8},
  year={2025},
  publisher={Nature Publishing Group UK London}
}

@article{boissin2025aiperceived,
    author = {Boissin, Esther and Costello, Thomas H and Spinoza-Martín, Daniel and Rand, David G and Pennycook, Gordon},
    title = {Dialogues with large language models reduce conspiracy beliefs even when the AI is perceived as human},
    journal = {PNAS Nexus},
    volume = {4},
    number = {11},
    pages = {pgaf325},
    year = {2025},
    month = {10},
    issn = {2752-6542},
    doi = {10.1093/pnasnexus/pgaf325},
    url = {https://doi.org/10.1093/pnasnexus/pgaf325},
    eprint = {https://academic.oup.com/pnasnexus/article-pdf/4/11/pgaf325/64700471/pgaf325.pdf},
}

@article{rathje2025sycophantic,
  title={Sycophantic AI increases attitude extremity and overconfidence},
  author={Rathje, Steve and Ye, Meryl and Globig, Laura and Pillai, Raunak and de Mello, Victoria and Van Bavel, Jay},
  year={2025},
  publisher={OSF},
  journal={OSF},
}

@article{sharma2023towards,
  title={Towards understanding sycophancy in language models},
  author={Sharma, Mrinank and Tong, Meg and Korbak, Tomasz and Duvenaud, David and Askell, Amanda and Bowman, Samuel R and Cheng, Newton and Durmus, Esin and Hatfield-Dodds, Zac and Johnston, Scott R and others},
  journal={arXiv preprint arXiv:2310.13548},
  year={2023}
}

@article{
holliday2025depolarization,
author = {Derek E. Holliday  and Yphtach Lelkes  and Sean J. Westwood },
title = {Why depolarization is hard: Evaluating attempts to decrease partisan animosity in America},
journal = {Proceedings of the National Academy of Sciences},
volume = {122},
number = {39},
pages = {e2508827122},
year = {2025},
doi = {10.1073/pnas.2508827122},
URL = {https://www.pnas.org/doi/abs/10.1073/pnas.2508827122},
eprint = {https://www.pnas.org/doi/pdf/10.1073/pnas.2508827122},
}

@article{glickman2025human,
  title={How human--AI feedback loops alter human perceptual, emotional and social judgements},
  author={Glickman, Moshe and Sharot, Tali},
  journal={Nature Human Behaviour},
  volume={9},
  number={2},
  pages={345--359},
  year={2025},
  publisher={Nature Publishing Group UK London}
}

@article{strandberg2019discussions,
  title={Do discussions in like-minded groups necessarily lead to more extreme opinions? Deliberative democracy and group polarization},
  author={Strandberg, Kim and Himmelroos, Staffan and Gr{\"o}nlund, Kimmo},
  journal={International Political Science Review},
  volume={40},
  number={1},
  pages={41--57},
  year={2019},
  publisher={Sage publications Sage UK: London, England}
}

@article{fisher2025political,
  title={Political Neutrality in AI Is Impossible-But Here Is How to Approximate It},
  author={Fisher, Jillian and Appel, Ruth E and Park, Chan Young and Potter, Yujin and Jiang, Liwei and Sorensen, Taylor and Feng, Shangbin and Tsvetkov, Yulia and Roberts, Margaret E and Pan, Jennifer and others},
  journal={arXiv preprint arXiv:2503.05728},
  year={2025}
}

@article{
argyle2023leveraging,
author = {Lisa P. Argyle  and Christopher A. Bail  and Ethan C. Busby  and Joshua R. Gubler  and Thomas Howe  and Christopher Rytting  and Taylor Sorensen  and David Wingate },
title = {Leveraging AI for democratic discourse: Chat interventions can improve online political conversations at scale},
journal = {Proceedings of the National Academy of Sciences},
volume = {120},
number = {41},
pages = {e2311627120},
year = {2023},
doi = {10.1073/pnas.2311627120},
URL = {https://www.pnas.org/doi/abs/10.1073/pnas.2311627120},
eprint = {https://www.pnas.org/doi/pdf/10.1073/pnas.2311627120}
}

@article{
piccardi2025reranking,
author = {Tiziano Piccardi  and Martin Saveski  and Chenyan Jia  and Jeffrey Hancock  and Jeanne L. Tsai  and Michael S. Bernstein },
title = {Reranking partisan animosity in algorithmic social media feeds alters affective polarization},
journal = {Science},
volume = {390},
number = {6776},
pages = {eadu5584},
year = {2025},
doi = {10.1126/science.adu5584},
URL = {https://www.science.org/doi/abs/10.1126/science.adu5584},
eprint = {https://www.science.org/doi/pdf/10.1126/science.adu5584}
}

@article{levendusky2016mis,
  title={(Mis) perceptions of partisan polarization in the American public},
  author={Levendusky, Matthew S and Malhotra, Neil},
  journal={Public Opinion Quarterly},
  volume={80},
  number={S1},
  pages={378--391},
  year={2016},
  publisher={Oxford University Press US}
}

@article{lees2020inaccurate,
  title={Inaccurate group meta-perceptions drive negative out-group attributions in competitive contexts},
  author={Lees, Jeffrey and Cikara, Mina},
  journal={Nature human behaviour},
  volume={4},
  number={3},
  pages={279--286},
  year={2020},
  publisher={Nature Publishing Group UK London}
}

@article{rossiter2024cross,
  title={Cross-partisan conversation reduced affective polarization for republicans and democrats even after the contentious 2020 election},
  author={Rossiter, Erin L and Carlson, Taylor N},
  journal={The Journal of Politics},
  volume={86},
  number={4},
  pages={1608--1612},
  year={2024},
  publisher={The University of Chicago Press Chicago, IL}
}

@article{pettigrew1998intergroup,
  title={Intergroup contact theory},
  author={Pettigrew, Thomas F},
  journal={Annual review of psychology},
  volume={49},
  number={1},
  pages={65--85},
  year={1998},
  publisher={Annual Reviews 4139 El Camino Way, PO Box 10139, Palo Alto, CA 94303-0139, USA}
}

@article{EndersEtAl2019,
  title={The differential effects of actual and perceived polarization},
  author={Enders, Adam M and Armaly, Miles T},
  journal={Political Behavior},
  volume={41},
  number={3},
  pages={815--839},
  year={2019},
  publisher={Springer}
}

@article{barlow2012contact,
  title={The contact caveat: Negative contact predicts increased prejudice more than positive contact predicts reduced prejudice},
  author={Barlow, Fiona Kate and Paolini, Stefania and Pedersen, Anne and Hornsey, Matthew J and Radke, Helena RM and Harwood, Jake and Rubin, Mark and Sibley, Chris G},
  journal={Personality and social Psychology bulletin},
  volume={38},
  number={12},
  pages={1629--1643},
  year={2012},
  publisher={Sage Publications Sage CA: Los Angeles, CA}
}

@article{tessler2024ai,
author = {Michael Henry Tessler  and Michiel A. Bakker  and Daniel Jarrett  and Hannah Sheahan  and Martin J. Chadwick  and Raphael Koster  and Georgina Evans  and Lucy Campbell-Gillingham  and Tantum Collins  and David C. Parkes  and Matthew Botvinick  and Christopher Summerfield },
title = {AI can help humans find common ground in democratic deliberation},
journal = {Science},
volume = {386},
number = {6719},
pages = {eadq2852},
year = {2024},
doi = {10.1126/science.adq2852},
URL = {https://www.science.org/doi/abs/10.1126/science.adq2852},
eprint = {https://www.science.org/doi/pdf/10.1126/science.adq2852}
}

@inproceedings{hughes2024pursuit,
author = {Hughes, Margaret and Roy, Brandon C. and Roy, Deb},
title = {In Pursuit of Constructive Communication: Designing Tools to Support Development of Constructive Communication Metrics},
year = {2024},
isbn = {9798400706325},
publisher = {Association for Computing Machinery},
address = {New York, NY, USA},
url = {https://doi.org/10.1145/3656156.3663720},
doi = {10.1145/3656156.3663720},
booktitle = {Companion Publication of the 2024 ACM Designing Interactive Systems Conference},
pages = {121–124},
numpages = {4},
location = {IT University of Copenhagen, Denmark},
series = {DIS '24 Companion}
}

@inproceedings{bao2021alrightconversations,
  title={Conversations gone alright: Quantifying and predicting prosocial outcomes in online conversations},
  author={Bao, Jiajun and Wu, Junjie and Zhang, Yiming and Chandrasekharan, Eshwar and Jurgens, David},
  booktitle={Proceedings of the Web Conference 2021},
  pages={1134--1145},
  year={2021}
}

@article{yeomans2020conversational,
  title={Conversational receptiveness: Improving engagement with opposing views},
  author={Yeomans, Michael and Minson, Julia and Collins, Hanne and Chen, Frances and Gino, Francesca},
  journal={Organizational Behavior and Human Decision Processes},
  volume={160},
  pages={131--148},
  year={2020},
  publisher={Elsevier}
}

@article{seckin2025identifying,
  title={Identifying Constructive Conflict in Online Discussions through Controversial yet Toxicity Resilient Posts},
  author={Seckin, Ozgur Can and Truong, Bao Tran and Flammini, Alessandro and Menczer, Filippo},
  journal={arXiv preprint arXiv:2509.18303},
  year={2025}
}

@article{zhang2018awryconversations,
  title={Conversations gone awry: Detecting early signs of conversational failure},
  author={Zhang, Justine and Chang, Jonathan P and Danescu-Niculescu-Mizil, Cristian and Dixon, Lucas and Hua, Yiqing and Thain, Nithum and Taraborelli, Dario},
  journal={arXiv preprint arXiv:1805.05345},
  year={2018}
}

@inproceedings{crossley2011text,
  title={Text coherence and judgments of essay quality: Models of quality and coherence},
  author={Crossley, Scott and McNamara, Danielle},
  booktitle={Proceedings of the Annual Meeting of the Cognitive Science Society},
  volume={33},
  year={2011}
}

@article{han2018you,
  title={How do you perceive this author? Understanding and modeling authors’ communication quality in social media},
  author={Han, Kyungsik},
  journal={PloS one},
  volume={13},
  number={2},
  pages={e0192061},
  year={2018},
  publisher={Public Library of Science San Francisco, CA USA}
}

@inproceedings{hutto2014vader,
  title={Vader: A parsimonious rule-based model for sentiment analysis of social media text},
  author={Hutto, Clayton and Gilbert, Eric},
  booktitle={Proceedings of the International AAAI Conference on Web and Social Media},
  volume={8},
  pages={216--225},
  year={2014}
}

@article{chen2024online,
  title={Are online users influenced by what other users say? Meta-analyzing the cognitive, emotional, and behavioral impact of online comment valence},
  author={Chen, Junhan and Xia, Shilin},
  journal={Cyberpsychology: Journal of Psychosocial Research on Cyberspace},
  volume={18},
  number={5},
  year={2024}
}

@article{
dryzek2019crisis,
author = {John S. Dryzek  and André Bächtiger  and Simone Chambers  and Joshua Cohen  and James N. Druckman  and Andrea Felicetti  and James S. Fishkin  and David M. Farrell  and Archon Fung  and Amy Gutmann  and Hélène Landemore  and Jane Mansbridge  and Sofie Marien  and Michael A. Neblo  and Simon Niemeyer  and Maija Setälä  and Rune Slothuus  and Jane Suiter  and Dennis Thompson  and Mark E. Warren },
title = {The crisis of democracy and the science of deliberation},
journal = {Science},
volume = {363},
number = {6432},
pages = {1144-1146},
year = {2019},
doi = {10.1126/science.aaw2694},
URL = {https://www.science.org/doi/abs/10.1126/science.aaw2694},
eprint = {https://www.science.org/doi/pdf/10.1126/science.aaw2694}
}

@ARTICLE{caluwaerts2023deliberation,
AUTHOR={Caluwaerts, Didier  and Bernaerts, Kamil  and Kesberg, Rebekka  and Smets, Lien  and Spruyt, Bram },
TITLE={Deliberation and polarization: a multi-disciplinary review},
JOURNAL={Frontiers in Political Science},
VOLUME={Volume 5 - 2023},
YEAR={2023},
URL={https://www.frontiersin.org/journals/political-science/articles/10.3389/fpos.2023.1127372},
DOI={10.3389/fpos.2023.1127372},
ISSN={2673-3145}
}

@incollection{carlson2020talking,
    author = {Carlson, Taylor N. and Abrajano, Marisa and Bedolla, Lisa García},
    isbn = {9780190082116},
    title = {The Composition and Determinants of Political Discussion Networks},
    booktitle = {Talking Politics: Political Discussion Networks and the New American Electorate},
    publisher = {Oxford University Press},
    year = {2020},
    month = {05}
}

@article{cinelli2021echo,
author = {Matteo Cinelli  and Gianmarco De Francisci Morales  and Alessandro Galeazzi  and Walter Quattrociocchi  and Michele Starnini },
title = {The echo chamber effect on social media},
journal = {Proceedings of the National Academy of Sciences},
volume = {118},
number = {9},
pages = {e2023301118},
year = {2021},
doi = {10.1073/pnas.2023301118},
URL = {https://www.pnas.org/doi/abs/10.1073/pnas.2023301118},
eprint = {https://www.pnas.org/doi/pdf/10.1073/pnas.2023301118}
}

@article{mcpherson2001birds,
  title={Birds of a feather: Homophily in social networks},
  author={McPherson, Miller and Smith-Lovin, Lynn and Cook, James M},
  journal={Annual review of sociology},
  volume={27},
  number={1},
  pages={415--444},
  year={2001},
  publisher={Annual Reviews 4139 El Camino Way, PO Box 10139, Palo Alto, CA 94303-0139, USA}
}

@article{boxell2024cross,
    author = {Boxell, Levi and Gentzkow, Matthew and Shapiro, Jesse M.},
    title = {Cross-Country Trends in Affective Polarization},
    journal = {The Review of Economics and Statistics},
    volume = {106},
    number = {2},
    pages = {557--565},
    year = {2024},
    month = {03},
    issn = {0034-6535},
    doi = {10.1162/rest_a_01160},
}

@misc{pew2019topic,
    author = {Doherty, Carroll and Kiley, Jocelyn and Tyson, Alec and Johnson, Bridget},
    title = {Public Highly Critical of State of Political Discourse in the U.S},
    howpublished = {Online at https://www.pewresearch.org/politics/2019/06/19/the-publics-level-of-comfort-talking-politics-and-trump},
    month = {Jun},
    year = {2019},
    note = {Pew Research},
}

@Article{velez2025chatbot,
  author    = {Velez, Yamil R. and Green, Donald P. and Sevi, Semra},
  journal   = {Proceedings of the National Academy of Sciences},
  title     = {Chatbot {Voting} {Advice} {Applications} inform but seldom sway young unaligned voters},
  year      = {2025},
  month     = dec,
  number    = {50},
  pages     = {e2515516122},
  volume    = {122},
  doi       = {10.1073/pnas.2515516122},
  publisher = {Proceedings of the National Academy of Sciences},
  url       = {https://www.pnas.org/doi/10.1073/pnas.2515516122},
  urldate   = {2026-03-04},
}

@Article{schroeder2026how,
  author    = {Schroeder, Daniel Thilo and Cha, Meeyoung and Baronchelli, Andrea and Bostrom, Nick and Christakis, Nicholas A. and Garcia, David and Goldenberg, Amit and Kyrychenko, Yara and Leyton-Brown, Kevin and Lutz, Nina and Marcus, Gary and Menczer, Filippo and Pennycook, Gordon and Rand, David G. and Ressa, Maria and Schweitzer, Frank and Song, Dawn and Summerfield, Christopher and Tang, Audrey and Van Bavel, Jay J. and van der Linden, Sander and Kunst, Jonas R.},
  journal   = {Science},
  title     = {How malicious {AI} swarms can threaten democracy},
  year      = {2026},
  month     = jan,
  number    = {6783},
  pages     = {354--357},
  volume    = {391},
  doi       = {10.1126/science.adz1697},
  publisher = {American Association for the Advancement of Science},
  url       = {https://www.science.org/doi/10.1126/science.adz1697},
  urldate   = {2026-03-04},
}

@article{benjamini2006adaptive,
    author = {Benjamini, Yoav and Krieger, Abba M. and Yekutieli, Daniel},
    title = {Adaptive linear step-up procedures that control the false discovery rate},
    journal = {Biometrika},
    volume = {93},
    number = {3},
    pages = {491-507},
    year = {2006},
    month = {09},
    issn = {0006-3444},
    doi = {10.1093/biomet/93.3.491},
    url = {https://doi.org/10.1093/biomet/93.3.491},
    eprint = {https://academic.oup.com/biomet/article-pdf/93/3/491/1080958/933491.pdf},
}

@article{rahn1993effect,
 ISSN = {00925853, 15405907},
 URL = {http://www.jstor.org/stable/2111381},
 author = {Wendy M. Rahn},
 journal = {American Journal of Political Science},
 number = {2},
 pages = {472--496},
 publisher = {[Midwest Political Science Association, Wiley]},
 title = {The Role of Partisan Stereotypes in Information Processing about Political Candidates},
 urldate = {2026-06-12},
 volume = {37},
 year = {1993}
}

@article{fiske1990continuum,
title = {A Continuum of Impression Formation, from Category-Based to Individuating Processes: Influences of Information and Motivation on Attention and Interpretation},
editor = {Mark P. Zanna},
journal = {Advances in Experimental Social Psychology},
publisher = {Academic Press},
volume = {23},
pages = {1-74},
year = {1990},
issn = {0065-2601},
doi = {https://doi.org/10.1016/S0065-2601(08)60317-2},
url = {https://www.sciencedirect.com/science/article/pii/S0065260108603172},
author = {Susan T. Fiske and Steven L. Neuberg}
}

@article{jussim1987nature,
author = {Jussim, Lee and Coleman, Lerita and Lerch, Lauren},
year = {1987},
month = {03},
pages = {536-546},
title = {The Nature of Stereotypes: A Comparison and Integration of Three Theories},
volume = {52},
journal = {Journal of Personality and Social Psychology},
doi = {10.1037/0022-3514.52.3.536}
}

@article{
westwood2022current,
author = {Sean J. Westwood  and Justin Grimmer  and Matthew Tyler  and Clayton Nall },
title = {Current research overstates American support for political violence},
journal = {Proceedings of the National Academy of Sciences},
volume = {119},
number = {12},
pages = {e2116870119},
year = {2022},
doi = {10.1073/pnas.2116870119},
URL = {https://www.pnas.org/doi/abs/10.1073/pnas.2116870119},
eprint = {https://www.pnas.org/doi/pdf/10.1073/pnas.2116870119}
}

@article{huang2017doesn,
  title={It doesn’t hurt to ask: Question-asking increases liking.},
  author={Huang, Karen and Yeomans, Michael and Brooks, Alison Wood and Minson, Julia and Gino, Francesca},
  journal={Journal of personality and social psychology},
  volume={113},
  number={3},
  pages={430},
  year={2017},
  publisher={American Psychological Association}
}

@article{saltz2024re,
  title={Re-ranking news comments by constructiveness and curiosity significantly increases perceived respect, trustworthiness, and interest},
  author={Saltz, Emily and Jalan, Zaria and Acosta, Tin},
  journal={arXiv preprint arXiv:2404.05429},
  year={2024}
}

@article{anderson2014nasty,
  title={The “nasty effect:” Online incivility and risk perceptions of emerging technologies},
  author={Anderson, Ashley A and Brossard, Dominique and Scheufele, Dietram A and Xenos, Michael A and Ladwig, Peter},
  journal={Journal of computer-mediated communication},
  volume={19},
  number={3},
  pages={373--387},
  year={2014},
  publisher={Oxford University Press Oxford, UK}
}

@article{meltzer2015journalistic,
  title={Journalistic concern about uncivil political talk in digital news media: Responsibility, credibility, and academic influence},
  author={Meltzer, Kimberly},
  journal={The International Journal of Press/Politics},
  volume={20},
  number={1},
  pages={85--107},
  year={2015},
  publisher={SAGE Publications Sage CA: Los Angeles, CA}
}

@article{thorson2010credibility,
  title={Credibility in context: How uncivil online commentary affects news credibility},
  author={Thorson, Kjerstin and Vraga, Emily and Ekdale, Brian},
  journal={Mass Communication and Society},
  volume={13},
  number={3},
  pages={289--313},
  year={2010},
  publisher={Taylor \& Francis}
}

@article{rossini2022beyond,
  title={Beyond incivility: Understanding patterns of uncivil and intolerant discourse in online political talk},
  author={Rossini, Patr{\'\i}cia},
  journal={Communication Research},
  volume={49},
  number={3},
  pages={399--425},
  year={2022},
  publisher={SAGE Publications Sage CA: Los Angeles, CA}
}

@book{scudder2020beyond,
  title={Beyond empathy and inclusion: The challenge of listening in democratic deliberation},
  author={Scudder, Mary F},
  year={2020},
  publisher={Oxford University Press}
}

@article{stromer2007measuring,
  title={Measuring deliberation’s content: A coding scheme},
  author={Stromer-Galley, Jennifer},
  journal={Journal of Deliberative Democracy},
  volume={3},
  number={1},
  year={2007},
  publisher={University of Westminster Press}
}

@article{you2006gratitude,
  title={Gratitude and Prosocial Behavior},
  author={You, Helping When It Costs},
  journal={Psychological Science},
  volume={17},
  number={4},
  pages={319--325},
  year={2006}
}

@article{algoe2012find,
  title={Find, remind, and bind: The functions of gratitude in everyday relationships},
  author={Algoe, Sara B},
  journal={Social and personality psychology compass},
  volume={6},
  number={6},
  pages={455--469},
  year={2012},
  publisher={Wiley Online Library}
}

@article{crossley2016say,
  title={Say more and be more coherent: How text elaboration and cohesion can increase writing quality},
  author={Crossley, Scott A and McNamara, Danielle S},
  journal={Journal of Writing Research},
  volume={7},
  number={3},
  pages={351--370},
  year={2016}
}

@article{feng2023effects,
  title={Effects of rhetorical devices on audience responses with online videos: An augmented elaboration likelihood model},
  author={Feng, Guangchao Charles and Luo, Yiwen and Yu, Zhenwei and Wen, Jinlang},
  journal={PLoS One},
  volume={18},
  number={3},
  pages={e0282663},
  year={2023},
  publisher={Public Library of Science San Francisco, CA USA}
}

@article{johansson2008lexical,
  title={Lexical diversity and lexical density in speech and writing: A developmental perspective},
  author={Johansson, Victoria},
  journal={Working papers/Lund University, Department of Linguistics and Phonetics},
  volume={53},
  pages={61--79},
  year={2008}
}

@Misc{kim2025effects,
  author    = {Kim, Do Won and Can Seckin, Ozgur and Bhadani, Saumya and Flammini, Alessandro and Ciampaglia, Giovanni Luca and Truong, Bao Tran},
  title     = {The Effects of Outgroup Agreement and Ingroup Dissent on Political Polarization},
  year      = {2025},
  copyright = {CC-By Attribution 4.0 International},
  doi       = {10.17605/OSF.IO/7FV3S},
  publisher = {OSF Registries},
  url       = {https://doi.org/10.17605/OSF.IO/7FV3S},
}
\end{document}